\documentclass[11pt]{article}
\newcommand{\ccpaperdate}{November 2026}
\usepackage{arxiv}
\usepackage[utf8]{inputenc}
\usepackage[T1]{fontenc}
\usepackage{lmodern}
\usepackage{microtype}
\usepackage{amsmath,amssymb}
\usepackage{booktabs}
\usepackage{array}
\usepackage{tabularx}
\usepackage{longtable}
\usepackage{enumitem}
\usepackage{graphicx}
\usepackage{xcolor}
\usepackage{xurl}
\usepackage{hyperref}
\usepackage[nameinlink,noabbrev]{cleveref}
\usepackage{caption}
\usepackage{subcaption}
\usepackage{float}
\usepackage{flafter}
\usepackage{needspace}
\usepackage[section]{placeins}
\usepackage{xspace}
\usepackage{ragged2e}
\newcommand{\field}[1]{\mathsf{#1}}
\newcommand{\rid}{\ensuremath{\mathsf{rid}}}
\newcommand{\kid}{\ensuremath{\mathsf{kid}}}
\newcommand{\state}[1]{\texttt{#1}}
\newcommand{\code}[1]{\texttt{\detokenize{#1}}}
\newcolumntype{Y}{>{\raggedright\arraybackslash}X}
\newcolumntype{P}[1]{>{\raggedright\arraybackslash}p{#1}}
\renewcommand{\arraystretch}{1.12}
\hypersetup{
  pdftitle={Collaboration Contracts for Cross-Organizational Industrial Agents},
  pdfauthor={Zhongjie Wang, Mingyi Liu, Xiang He, Shuang Yu},
  pdfsubject={A lifecycle software model with formation and evolution protocols for collaboration among industrial agents across enterprises},
  pdfkeywords={industrial agents, collaboration contracts, multi-agent coordination, contract construction, runtime governance, accountability, experience reuse},
  colorlinks=true,
  linkcolor=blue!55!black,
  citecolor=blue!55!black,
  urlcolor=blue!55!black
}
\title{\textbf{Collaboration Contracts for Cross-Organizational Industrial Agents}}
\author{
  Zhongjie Wang, Mingyi Liu, Xiang He, Shuang Yu\\[0.45em]
  Faculty of Computing, Harbin Institute of Technology, China\\[0.25em]
  \texttt{\{rainy,liumy,hexiang,yushuang\}@hit.edu.cn}
}
\date{\ccpaperdate}
\setlist{itemsep=0.25em,topsep=0.35em}

\begin{document}
\maketitle
\begin{abstract}
Industrial procurement, logistics, and collaborative manufacturing require enterprises to coordinate outcomes while retaining private information and decision authority. We propose collaboration contracts (CC), which make cross-enterprise relationships first-class software entities with persistent identities, immutable versions, commitments, and evidence references. The model defines the effects of relationship operations and the dependencies used to compose relationships. Its engineering methods validate candidate terms, bind enterprise consent to exact content, and commit amendments atomically. The effective ledger determines which obligations an amendment must address. Separate evidence applications retain their sources and identify which obligations they support. These references also support accountability and qualified experience reuse. Reference implementations and controlled studies evaluate selected mechanisms and configurations. The clearest comparative result concerns ledger-derived disposition scope: correct normative outcomes reach 74\% with the guard and 32\% without it on targeted challenge cases. Other comparisons expose coordination and coverage limits, including a reverse result whose process attribution remains unresolved. A three-enterprise online instance provides a configuration-specific, agent-directed formation-to-closure path, with independent reconciliation of the implemented business predicates.
\end{abstract}
\section{Introduction}\label{sec:1}

Fulfilling an industrial order often involves several enterprises. A buyer specifies quantity, budget, and delivery dates; a supplier allocates capacity and delivers conforming goods; a logistics provider transports them; and a partner factory performs subsequent processing. Whether the goods enter production on time depends on how these arrangements fit together. Each enterprise must also serve other customers, allocate internal resources, and manage its own risks. Changes to quantities, transport schedules, or quality requirements propagate along the business chain. The participants must jointly determine which arrangements can change and which responsibilities remain in force.

Procurement, supplier, and logistics agents query information, negotiate terms, and invoke enterprise systems within the authority granted by their organizations. A procurement platform may coordinate their work, while each enterprise controls its own disclosures, resources, and decisions to undertake commitments.

\begin{figure}[t]
\centering
\includegraphics[width=\linewidth]{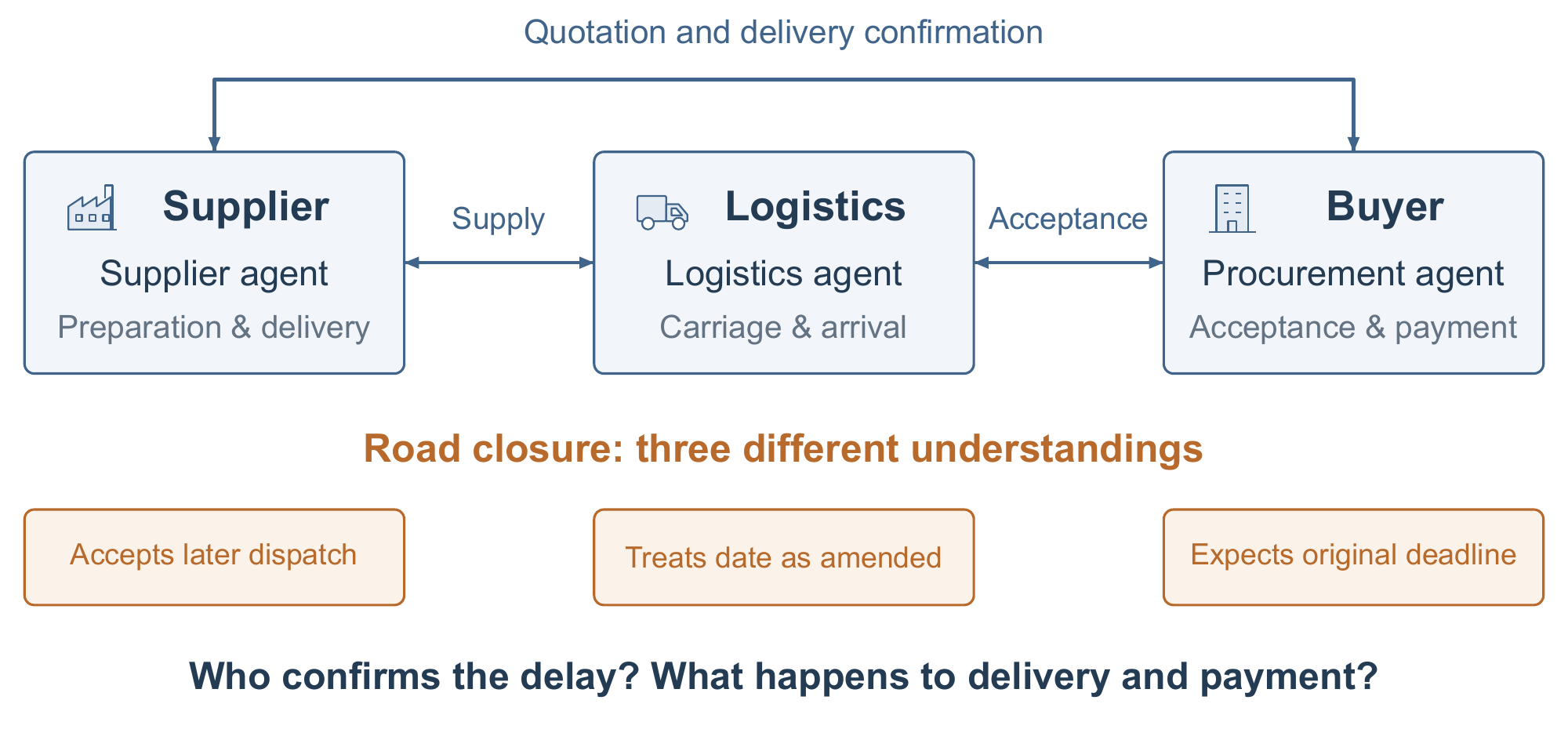}
\caption{Enterprise agents in a business collaboration and the challenges created by divergent understandings of an amendment.}\label{fig:1}
\end{figure}

Reduced supplier capacity, transport delays, or failed inspections can make the original plan infeasible. In an application-level integration, rules for the same relationship can be spread across application code, workflow scripts, interface configurations, and human agreements. After a road closure, logistics may record a revised delivery date while the buyer still assesses lateness against the original date (Figure~\ref{fig:1}). A shared revision must establish who approves the change, which obligations remain, and what grounds support later outcome judgments.

A recorded arrival may precede quality inspection, buyer acceptance, and the point when payment becomes due. Assessing a delay or quality dispute requires the applicable agreement and event records. Later use of a remedy depends on its applicability and the rights to the underlying information.

As agents propose and revise plans during execution, collaboration software must distinguish candidate terms, enterprise approvals, and effective agreements. We ask how to establish executable agreements and maintain clear responsibilities when information sharing is limited, decision authority remains with separate enterprises, and execution conditions continue to change.

We propose \emph{collaboration contracts} (CC) for cross-organizational industrial agents. A collaboration contract makes the collaboration relationship a first-class software entity with a persistent identity, explicit agreement versions, and operations for formation, revision, fulfillment, and closure. Enterprise systems supply capabilities for production, design and simulation, logistics, and inspection. The collaboration relationship specifies the terms under which enterprises use those capabilities together. Agents address that same relationship when they propose terms, confirm responsibilities, and submit evidence.

Enterprise interfaces retain authority over identity, permission, consent, factual attestation, and physical execution. Agent proposals enter the runtime for adjudication, and the runtime uses the authoritative inputs to determine their effect on the shared agreement.

Our contribution to software theory is an operational model of a persistent collaboration relationship. The model specifies the relationship state, the preconditions and effects of its operations, and the invariants required of valid transitions (Section~\ref{sec:4}). It ties authority to the agreed content, derives the scope of amendments from the effective obligation ledger, and identifies the grounds on which evidence can support fulfillment. Dependencies between versions or commitments support composition across enterprises that retain separate decision authority (Section~\ref{sec:7}). The same model gives experience reuse explicit grounds in observed outcomes and conditions for ownership and applicability (Section~\ref{subsec:6.3}).

Our contribution to software engineering is a set of methods for building and testing collaboration around that relationship state. Formation validates candidates assembled under selective disclosure and binds enterprise confirmations to exact content. Evolution checks the dispositions of outstanding obligations and commits the confirmed successor atomically. Monitoring and closure check fulfillment against authoritative facts and their version-specific uses. After a local change, explicit dependencies delimit the checks to run again. The reference interfaces expose these state changes, the grounds that support them, and reasons for rejection. Table~\ref{tab:evidence-map} maps the methods to their implemented and evaluated paths.

We evaluate selected mechanisms using reference implementations. Controlled simulations examine business mechanisms, paired online studies compare semantic configurations and implementations, and a targeted audit examines migration decisions. A three-enterprise online instance connects formation to verified closure. The strongest comparative evidence supports ledger-derived disposition scope on targeted challenge cases. The other studies identify constraints on coordination, action reachability, and evidence coverage.

The subsequent evaluation agenda extends coverage across scenarios, models, and implementations and examines long-running collaboration with enterprise systems (Section~\ref{subsec:11.4}).

The paper begins with procurement and manufacturing collaboration, then develops the model, protocols, lifecycle methods, and contract composition. Section~\ref{sec:8} maps these responsibilities to implementations. Sections~\ref{sec:9}--\ref{sec:10} present evaluation designs and results; Sections~\ref{sec:11}--\ref{sec:12} discuss their implications, coverage, and validity.

\section{Research Questions and Running Example}\label{sec:2}

\subsection{Procurement and delivery involving enterprise agents}\label{subsec:2.1}\label{sec:2.1}

An illustrative procurement case runs through the model and methods. The buyer seeks 100 units with a budget of 1,000 illustrative monetary units, and permits partial delivery of at least 80 units within the agreed window. Its procurement agent opens a provisional relationship, identifies the intended business goal, and invites a supplier agent and a logistics agent. The supplier can currently provide 90 units at a price of 10 per unit and may attest that it can commit to at least 90 units without revealing its exact inventory, costs, or other customers' orders. The logistics agent responds with a transport window. A participant may join, decline, request clarification, or return an alternative within its disclosure and action authority.

During formation, the agents exchange proposals and counterproposals until they produce a candidate covering 90 conforming units, transport, inspection, and payment after acceptance. Independent checks evaluate the candidate's structure, constraints, disclosed predicates, and action-time authority. Each required enterprise then confirms the same relationship, candidate version, and content hash. Only a successful activation registers the supplier's delivery commitment, the logistics provider's carriage commitment, and the buyer's conditional payment commitment. An invitation, candidate, or conversational ``accept'' creates no such obligations by itself. If a party declines and no alternative set of participants and terms is confirmed, the provisional relationship ends without an effective CC.

During execution, an approved source attests a road closure that threatens the original delivery date. Because the change affects shared deadlines and dependent rights, the logistics agent initiates an evolution round. Affected agents assess the impact, propose and counterpropose revised dates, and may reject the change or request additional authority. The logistics agent may offer expedited carriage if the buyer pays an additional fee. The buyer can report that confirmation is pending while its enterprise decides; this status supplies no consent. Once an authoritative response is available, the buyer may confirm the exact candidate or decline it. CC observes these external positions and their grounds; each enterprise determines how to reach its own decision. A successor candidate must dispose of delivery, carriage, and the still-untriggered payment commitment, separately validate any new fee obligation, and obtain the required enterprise confirmations. Until that candidate is validated and atomically activated, the original version remains effective. Without this rule, logistics may consider the date amended while the buyer still expects the original deadline and the supplier continues using the original payment terms.

An enterprise-side change does not always trigger evolution. Switching equipment or route internally can remain an execution choice when the effective commitments are still satisfiable. An option already authorized by the effective terms follows its prescribed conditions and notifications. A change to quantity, deadline, acceptance, payment, responsibility, or another party's rights beyond those terms enters the shared coordination loop. Arrival, inspection, and receipt records after rescheduling support outcome judgments. After closure, a remedy may inform later orders only when provenance, ownership, time limits, evidence, and business conditions qualify its reuse. The example therefore spans formation, execution, evolution, closure, and subsequent use.

\subsection{Business challenges from negotiation to exception handling}\label{subsec:2.2}\label{sec:2.2}

Limited disclosure makes evidence of feasibility different from access to an enterprise's full state. The buyer must know which decisions a disclosed predicate supports and when additional information or authority is needed. This creates a coordination problem even when every participant follows its own enterprise policy.

Negotiation messages must refer to business outcomes that the parties can undertake. An ``accept'' reply still needs to identify the goods, deadline, acceptance and payment conditions, and the sender's authority. Enterprises retain these obligations across agent restarts and changes of personnel or models. Software must therefore connect conversational expressions to a persistent, queryable agreement.

Local changes can affect rights and dependencies beyond the initiating enterprise. The challenge is to distinguish internal execution choices from amendments to the shared agreement, then account for obligations that persist through the change. Untriggered conditional commitments make this continuity requirement especially easy to overlook.

Outcome recognition, responsibility, and experience reuse require different judgments over the same collaboration history. Each must retain its applicable agreement and supporting grounds while adding its own factual, causal, or qualification checks.

\subsection{From business challenges to research questions}\label{subsec:2.3}\label{sec:2.3}

These challenges yield four lifecycle research questions, MRQ-A through MRQ-D, addressing formation, dynamic fulfillment, evidence-supported accountability and composition, and experience qualification and reuse. They organize the collaboration-contract framework. Section~\ref{sec:9} specifies which mechanisms the current evaluations cover.

\textbf{MRQ-A: Formation and validity.} When enterprises separately control information, resources, and authority, how can their agents initiate collaboration, negotiate a common candidate, and activate a shared agreement that each party is authorized and able to undertake under limited disclosure?

\textbf{MRQ-B: Dynamic fulfillment and evolution.} When execution conditions change, how can affected agents coordinate a lawful successor, continuously account for existing obligations, and obtain the grounds needed to continue, conclude, or terminate collaboration?

\textbf{MRQ-C: Evidence, accountability, and composition.} How can effective terms and event evidence support interpretation of outcomes and responsibility, and determine the effects of a local change on other collaboration relationships?

\textbf{MRQ-D: Experience qualification and reuse.} Which lessons from completed collaborations and concluded negotiations may inform a new negotiation, and how should applicability, ownership, time limits, and evidential support govern their use?

\begingroup
\small
\setlength{\tabcolsep}{4pt}
\renewcommand{\arraystretch}{1.18}
\setcounter{table}{0}
\begin{table}[htbp]
\caption{Information and normative distinctions that collaboration contracts must preserve.}\label{tab:1}
\centering
\small
\begin{tabular}{@{}>{\RaggedRight\arraybackslash}p{0.27\dimexpr\linewidth-4\tabcolsep\relax}>{\RaggedRight\arraybackslash}p{0.35\dimexpr\linewidth-4\tabcolsep\relax}>{\RaggedRight\arraybackslash}p{0.38\dimexpr\linewidth-4\tabcolsep\relax}@{}}
\toprule
\textbf{Distinction} & \textbf{Information that can differ despite the same visible outcome} & \textbf{Consequence for software judgment} \\
\midrule
Completion and fulfillment & Delivery authority, agreed quality, and evidence requirements & Task termination does not establish fulfillment \\
Generation and validity & Current permissions, budget, resources, and hard constraints & A plausible proposal may be unacceptable \\
Ex ante compliance and recovery & New conditions, required consent, and old-obligation dispositions & A lawful original agreement need not be recoverable \\
Logs and responsibility grounds & Effective terms, causal conditions, and excuse rules & The same delay may imply different responsibility \\
History and experience qualification & Ownership, applicability, provenance, and risk & Relevant history may be impermissible to reuse \\
Messages and normative effect & Current relationship/version and authority at action time & The same acceptance message may take effect or be rejected \\
\bottomrule
\end{tabular}
\end{table}
\endgroup

Table~\ref{tab:1} summarizes the information needed to answer these questions. Section~\ref{sec:3} positions the framework relative to existing mechanisms; Section~\ref{sec:4} defines its objects and runtime interpretation.

\section{Related Work and Positioning}\label{sec:3}

\subsection{Social commitments and normative multiagent systems}\label{subsec:3.1}\label{sec:3.1}

The Contract Net Protocol organizes distributed negotiation through announcements, bids, and awards~\cite{r1}. Social commitment models give obligations a debtor, creditor, antecedent, and consequent, and specify how commitments arise and change~\cite{r2,r3,r4,r5}. Commitment-based service architectures and interaction programming make these relations operational~\cite{r6,r7}. Telang and Singh use composable commitment patterns to specify cross-organizational business models and verify operational processes against them~\cite{telangbusiness}. In 2COMM, agents can adopt roles in programmable interaction artifacts, which maintain social state and monitor compliance~\cite{r8}. Constitutive and regulative specifications distinguish how social facts arise from whether behavior is permitted~\cite{r9}. Normative multiagent systems investigate autonomy and governance in open environments~\cite{r10,r11}.

Organizational approaches also make roles, norms, and interaction structures explicit. JaCaMo integrates agent, environment, and organization programming, and AMELI supplies execution infrastructure for electronic institutions~\cite{jacamo,ameli}. A relationship-centered account of agentic services distinguishes delegation contracts, which authorize an actor for one principal, from collaboration contracts among independently accountable parties~\cite{r60}. CC builds on these precedents by specifying how a persistent business relationship changes at runtime when the enterprises control their own disclosures, approvals, and factual attestations.

\subsection{Choreography, service agreements, and computable contracts}\label{subsec:3.2}\label{sec:3.2}

WS-CDL describes observable multiparty interactions and their ordering~\cite{r12}. BPMN choreography and collaboration conformance address realizability and local checks of global rules~\cite{r13,r14,r15}. Permissioned ledgers can retain shared execution records for BPMN choreographies~\cite{r58}. Temporal agreements and eContractual choreography also address coordination with private processes and selective visibility across organizations~\cite{r16,r17}. CC uses such interaction mechanisms to carry messages and assigns their effects on the shared agreement through the relationship, authority, and version they reference.

WS-Agreement specifies agreement objects, guarantee terms, and monitoring. Its negotiation extension can supersede an existing agreement with a renegotiated one~\cite{r18,r19}. Timed service protocols support compatibility and replaceability analysis when a service changes~\cite{protocolcompat}. Service-level agreements address service quality and violations~\cite{r20}, and global-contract generation combines functional and nonfunctional agreements for composite services~\cite{r57}. Smart legal contract templates and computable contracting connect legal text, operational parameters, and code~\cite{r21,r22,r23}. These approaches provide constructs for the terms and interactions that CC manages across effective versions.

B2B automation already supplies structured collaboration and transaction definitions. For example, ebXML BPSS describes business collaborations, document exchanges, and choreography~\cite{ebxml}. When a collaboration changes in flight, the participants also need to determine what the revised agreement does to commitments and evidence accumulated during execution.

\paragraph{Case-level comparison.}
In the road-closure case, an amended transport plan also affects an established payment obligation whose acceptance condition is still pending. The effective ledger determines the complete set of obligations to address, even if the proposal omits the payment. Each required enterprise confirms the digest covering both the revised terms and the disposition of existing commitments. When a receipt from an earlier version supports a successor obligation, a separate application records the authorized use and preserves the receipt's source. These rules define the effect of amendment as an operation on one shared relationship. Table~\ref{tab:case-comparison} relates the checks to existing mechanisms and the evaluation.

\begingroup
\scriptsize
\setlength{\tabcolsep}{2.5pt}
\renewcommand{\arraystretch}{1.15}
\setcounter{table}{1}
\begin{table}[!t]
\caption{Road-closure case: expressible elements, additional runtime checks, implementation entry points, and evidence.}\label{tab:case-comparison}
\centering
\begin{tabular}{@{}>{\RaggedRight\arraybackslash}p{0.16\dimexpr\linewidth-8\tabcolsep\relax}>{\RaggedRight\arraybackslash}p{0.20\dimexpr\linewidth-8\tabcolsep\relax}>{\RaggedRight\arraybackslash}p{0.28\dimexpr\linewidth-8\tabcolsep\relax}>{\RaggedRight\arraybackslash}p{0.17\dimexpr\linewidth-8\tabcolsep\relax}>{\RaggedRight\arraybackslash}p{0.19\dimexpr\linewidth-8\tabcolsep\relax}@{}}
\toprule
\textbf{Existing mechanism} & \textbf{Element already expressible} & \textbf{CC cross-stage responsibility} & \textbf{Reference entry point} & \textbf{Evidence in this paper} \\
\midrule
Commitment Machines / 2COMM & Conditional payment and debtor--creditor commitment evolution & Include established commitments with pending antecedents in the ledger-derived revision domain & Requirements query; migration commit & Scripted controls; O2 audit identifies a shared confirmation lock before terminal mechanism exposure \\
WS-Agreement and agreement versioning & Delivery guarantees and a relation between original and revised agreements & Derive the complete disposition domain from the effective ledger and require one $\mu$ entry for every outstanding registered obligation & Requirements query; migration commit & O1 main challenge cases: 74\% versus 32\%; targeted process traces are re-adjudicable \\
Provenance and usage control & Credential origin, permitted purpose, and usage policy & Preserve the original source; record each later application against \rid{}/version/\kid{} without rewriting origin & Evidence submission; closure query and commit & Kernel and bounded checks; evidence use is not isolated by an online ablation \\
TaaS trust-evidence envelope & Action manifest and issuer-attributed authority, provenance, status, and challenge records & Bind authoritative consent and evidence to the effective obligations and the requested change or closure & Enterprise evidence interface; evidence submission and closure & Source and application checks in the kernel; online coverage summarized in Table~\ref{tab:evidence-map} \\
Agent messages, tools, and retrieval & Proposal transport, tool invocation, and historical retrieval & Treat candidates as proposals; apply independent authority, validity, and experience-qualification decisions before normative effect or reuse & Acceptance, authority, and reuse gates & F2/F3/F6 adjacent-configuration differences under the evaluated stack \\
\bottomrule
\end{tabular}
\end{table}
\endgroup
\setcounter{table}{2}

\subsection{Design by contract, runtime verification, and industrial data sovereignty}\label{subsec:3.3}\label{sec:3.3}

Design by contract assigns responsibilities to callers and components through preconditions, postconditions, and invariants~\cite{r24}. Contracts for system design support composition and refinement through assumptions and guarantees~\cite{r25,r26,r27}. Runtime verification turns trace properties into monitors~\cite{r28,r29}, and a policy engine such as Cedar can decide whether an action is authorized~\cite{cedar}. CC gives these checks a common subject: the effective obligations of independently governed enterprises under a particular relationship version. An event-sourced store can reconstruct that state from recorded changes~\cite{eventsourcing}. The relationship specification determines which changes are authorized and how past events count toward current commitments.

ODRL represents permissions, prohibitions, duties, and remedies~\cite{r30}. Usage control and data-space protocols govern access, purpose, negotiation, and transfer~\cite{r31,r32}. Catena-X and Manufacturing-X work addresses industrial data sovereignty and policy representation~\cite{r33,r34}. Collaboration contracts obtain permitted judgments and evidence through enterprise policy interfaces. Data-use duties and business delivery commitments can be connected within one relationship while retaining their respective authoritative sources, purpose restrictions, and asset ownership.

Trustworthiness as a Service (TaaS) externalizes support for a bounded delegated action through an immutable manifest and issuer-attributed attestations~\cite{r62}. Such an envelope can supply authority, provenance, status, and challenge evidence to a CC enterprise interface. CC adds the relationship-level interpretation: which established obligation is affected, what disposition applies, and how an attestation is used under the effective version.

\subsection{Agent stacks, security, and experience memory}\label{subsec:3.4}\label{sec:3.4}

ReAct interleaves reasoning and acting~\cite{r52}. CAMEL, AutoGen, MetaGPT, and ChatDev organize multiagent collaboration through roles, conversations, and development workflows~\cite{r35,r36,r37,r38}. Guo et al. survey these approaches~\cite{r39}. A2A and MCP connect messages, tasks, tools, and context~\cite{r40,r41}. The experiments use message configurations from the cited A2A release together with project-specific business actions. SupplyChainAgent extends these agent-based approaches to supply-chain inquiry, negotiation, production, transactions, and fulfillment~\cite{r59}, making enterprise autonomy and cross-firm dependencies visible in an industrial simulation. CC supplies the business interpretation of the resulting proposals, approvals, amendments, and outcomes.

ToolEmu, AgentDojo, and CaMeL study tool risks, prompt injection, and separation of trusted control from untrusted input~\cite{r42,r43,r44}. Collaboration contracts likewise separate proposals from authoritative adjudication. RAG, Generative Agents, Reflexion, MemGPT, and A-Mem improve retrieval and memory organization~\cite{r45,r46,r47,r48,r49}. Motivated by documented contamination and evaluation-malpractice risks~\cite{r50,r51}, we use source isolation to constrain direct answer sharing in evaluation. Structural causal models and PROV provide foundations for accountability and provenance~\cite{r53,r54}, whose inputs are connected to effective obligations and evidence.

\paragraph{Recent agent contracts and execution infrastructure.}
Recent work makes several adjacent meanings of ``contract'' concrete. Agent Contracts regulates resource-bounded autonomous operation~\cite{agentcontracts}; RAC connects engineering requirements, artifact relations, observations, and repair under machine-checkable constraints~\cite{rac}. Evidence-to-effect infrastructure addresses authoritative observations, execution identity, consequential actions, and recovery for stateful agents~\cite{evidenceeffect}, while a recent authorization review organizes identity, delegation, enforcement, and provenance across principals~\cite{autharch}. CC shares these concerns at enterprise boundaries and focuses on the obligations that survive a jointly approved business amendment.

Database plan and evidence contracts bind admissible execution and supported claims~\cite{dbcontracts}; study contracts connect approved research choices to execution records and claim scope~\cite{studycontracts}. Both reinforce the separation between a proposed result and the evidence that warrants it. PANDA addresses decentralized orchestration among independently administered agents~\cite{panda}; executable self-governance in GovSim examines agents' adoption of rules under collective pressure~\cite{govsim}. These studies extend the relevant context from individual tool use to institutions and coordination. CC supplies a common relationship, version, commitment, and evidence vocabulary for industrial agreement-making; its distinctive empirical question is how the associated runtime checks affect business outcomes.

\begingroup
\small
\setlength{\tabcolsep}{4pt}
\renewcommand{\arraystretch}{1.18}
\renewcommand{\thetable}{2a}
\renewcommand{\theHtable}{2a}
\begin{table}[htbp]
\caption{Connections between neighboring mechanisms and collaboration contracts.}\label{tab:2}
\centering
\small
\begin{tabular}{@{}>{\RaggedRight\arraybackslash}p{0.27\dimexpr\linewidth-4\tabcolsep\relax}>{\RaggedRight\arraybackslash}p{0.35\dimexpr\linewidth-4\tabcolsep\relax}>{\RaggedRight\arraybackslash}p{0.38\dimexpr\linewidth-4\tabcolsep\relax}@{}}
\toprule
\textbf{Existing mechanism} & \textbf{Primary objects or responsibilities} & \textbf{Connection specified by collaboration contracts} \\
\midrule
Social commitments and normative MAS & Participant obligations, social states, and programmable interaction artifacts & Enterprise authority and disposition of commitments under each relationship version \\
Choreography and cross-organizational processes & Roles, message order, and local compliance & Connections from messages to normative effects and subsequent evidence \\
Service agreements and computable contracts & Guarantees, templates, negotiation, and superseding agreements & Complete obligation disposition and consent to the exact amendment \\
Service evolution and event sourcing & Protocol compatibility, replaceability, and state reconstruction & Admission of version changes and applicability of earlier facts to current obligations \\
Design by contract and runtime verification & Constraint checks, composition, and trace monitoring & Judgments traced to participants and rid/v/kid \\
Data sovereignty and usage control & Access, purpose, and disclosure policies & Retrievable connections between enterprise judgments and business commitments \\
Cross-domain trust-evidence profiles & Action-scoped manifests, issuer-attributed attestations, status, challenge, and recovery & Persistent relationship and obligation semantics that determine each attestation's normative scope and effect \\
Agent protocols and experience memory & Transport, tools, candidate generation, and retrieval & Runtime rules from candidates to effective obligations or qualified reuse \\
\bottomrule
\end{tabular}
\end{table}
\endgroup
\setcounter{table}{2}

\section{Collaboration-Contract Model and Runtime Semantics}\label{sec:4}

\paragraph{Reading the model against its implementation.}
Definitions D1--D5 map to the stores and interfaces in Section~\ref{subsec:8.1}; the guarded transitions map to the entry points in Section~\ref{subsec:8.2}. Table~\ref{tab:evidence-map} links each specification to its implemented subset and evidence.

\subsection{Responsibilities and terminology}\label{subsec:4.1}\label{sec:4.1}

A collaboration contract answers four questions about an enterprise collaboration: who participates, which agreement is currently recognized, which commitments remain to be performed, and what grounds support outcome judgments. ``Collaboration contract'' denotes the lifecycle model and methods. A \emph{contract version} is an immutable snapshot of agreed terms. A \emph{contract runtime} maintains agreements, checks proposals, and interprets events. The \emph{relationship-state interface} reads and writes the core state; the \emph{relationship layer} describes these responsibilities. These responsibilities may be implemented by a dedicated runtime or assembled from existing workflow, policy, provenance, and persistence components. The abstraction does not prescribe a cross-replica consensus, storage-replication, or global distributed-commit protocol; deployments supply those mechanisms.

The four core objects are relationship instance $R$, contract version $C^v$, atomic commitment $\kappa$, and evidence event $e$. They identify continuing collaboration, preserve approved terms, express who owes which outcome to whom, and connect results to their sources. Versions represent resource, temporal, safety, and disclosure conditions through rules or interface references. Each enterprise retains its authority over disclosure, permitted actions, and consent. Its interfaces supply scoped verdicts and confirmation credentials, which the runtime checks and references when it records an action's effect. Contract networks, accountability analysis, and experience assets inherit the core references to support multiple relationships and subsequent use.

\begin{figure}[H]
\centering
\includegraphics[width=\linewidth]{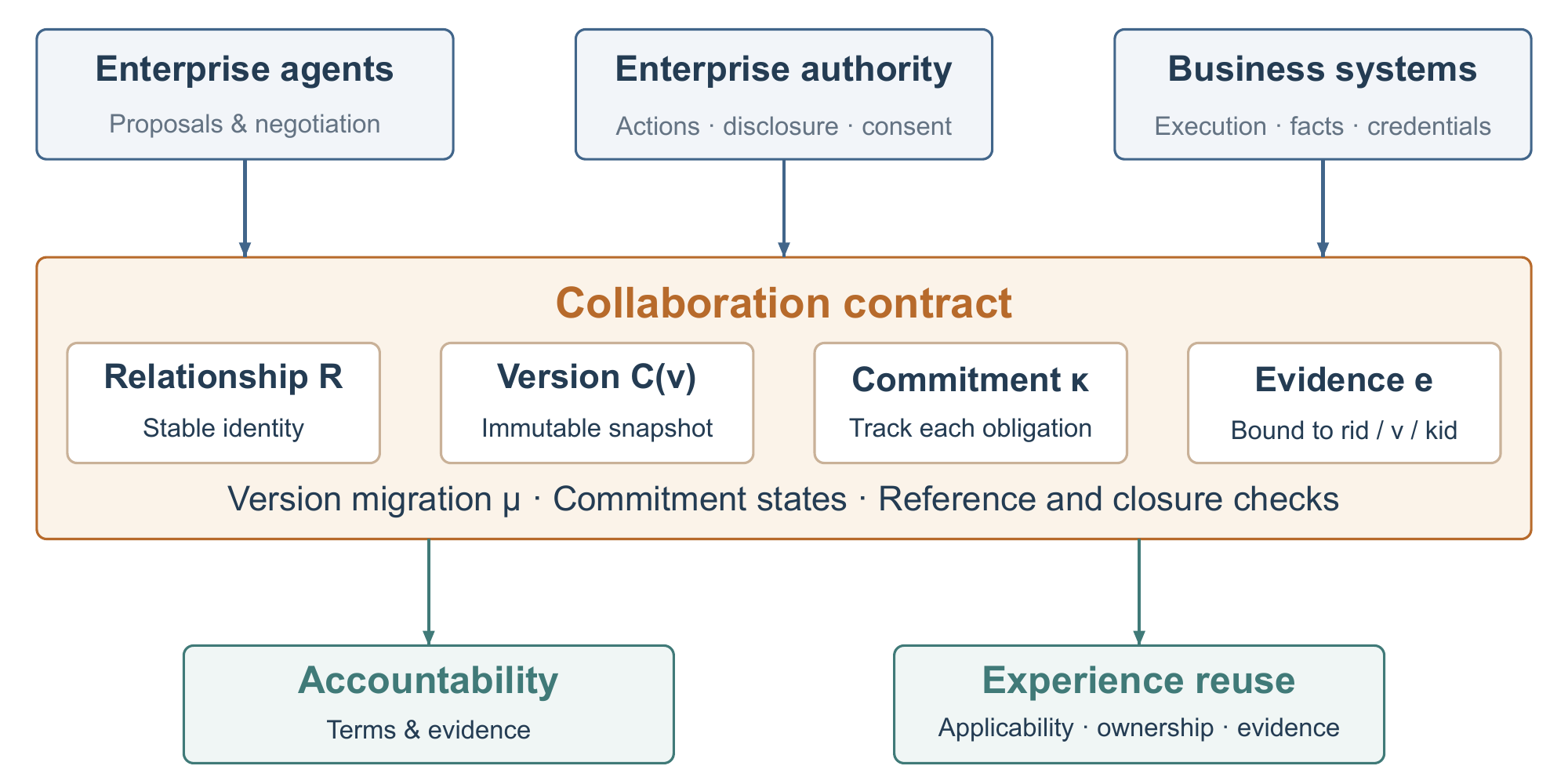}
\caption{Responsibilities and connections of collaboration contracts within a cross-organizational industrial-agent system.}\label{fig:2}
\end{figure}
\FloatBarrier

\subsection{Four core objects and the runtime projection}\label{subsec:4.2}\label{sec:4.2}

\textbf{Definition D1 (collaboration relationship).}
\begin{equation}
R=\langle \field{rid},\field{Participants},\field{Goal},\field{Scope}\rangle.
\end{equation}
The relationship identifier \rid{} remains stable. The remaining fields record the organizations and roles admitted to the broad collaboration scope, its goal, and its boundary. This relationship-level roster supports discovery and continuity; it does not replace the enterprise--role bindings frozen in each effective version. One collaboration can span multiple sessions, tasks, and agent processes while retaining the same \rid{}. A session bounds message interaction; an episode bounds an evaluation run.

\textbf{Definition D2 (agreement version).}
\begin{equation}
C^v=\langle\field{rid},v,\field{pred},B^v,K^v,\Phi^v,\Psi^v\rangle.
\end{equation}
Here $v$ is the version number, $\field{pred}$ references the predecessor within the same relationship, $B^v$ freezes the enterprise--role bindings for this version, and $K^v$ lists commitments. $\Phi^v$ stores hard constraints and external verdict references; $\Psi^v$ stores exception, remedy, exit, and evidence rules. A version is well bound only if every debtor and creditor reference in $K^v$ resolves through $B^v$ to an enterprise and contractual role authorized to occupy that position. Committed versions cannot be overwritten. Changing an enterprise participant or its contractual role requires a new version, but changing $B^v$ does not reassign an established commitment. Replacing an agent process or software carrier within the same enterprise and authority scope is an implementation change recorded through identity and carrier references; it does not by itself replace the enterprise party or amend the agreement.

\textbf{Definition D3 (atomic commitment).}
\begin{equation}
\begin{split}
\kappa=\langle &\field{kid},\field{debtor},\field{creditor},\field{antecedent},\field{consequent},\\
&\field{due},\field{authRef},\field{evReq},\field{remedy}\rangle.
\end{split}
\end{equation}
Here ``atomic'' means an individually identified debtor--creditor obligation; transaction atomicity is a separate property of activation and migration. The identifier \kid{} locates the commitment; the debtor owes the result and the creditor is entitled to it. The other fields specify the performance antecedent, required result, deadline, authorization ground, evidence requirements, and remedy. A commitment is established when the contract takes effect and necessary confirmations are complete. Its antecedent determines when performance is required. Thus ``pay after acceptance'' is established with the agreement, while acceptance triggers the payment requirement. Terms are fixed with the version; events update fulfillment and termination state.

\textbf{Definition D4 (evidence event).}
\begin{equation}
\begin{split}
e=\langle &\field{eid},\field{rid},v,\field{kid},\field{actor},\field{type},\field{object},\\
&t,\field{source},\field{integrity},\field{payloadHash}\rangle.
\end{split}
\end{equation}
An event references an original credential or an external verification verdict, identifying the relationship, version, and commitment it supports. Depending on event type, relationship-level events may use $\field{kid}=\bot$; commitment-level claims require complete references. An authorized source interface determines original credential provenance. A submission's destination relationship cannot establish that provenance by itself.

\textbf{Definition D5 (current collaboration state).} The runtime projection records the current version $v$, relationship phase $q$, the established-commitment ledger $K_a$, and evidence-support mapping $E_c$, as follows. $K_a$ retains the states of both outstanding and terminated commitments. Its outstanding subset includes established conditional commitments whose antecedents have not yet been triggered. The subscript $a$ denotes established obligations, not only obligations currently due for performance; $K_{\mathrm{out}}^v$ below selects the outstanding subset under version $v$. External interfaces determine authorization at the action time, and normative events preserve the verdict references.

Table~\ref{tab:3} summarizes four cross-stage correspondences. The core objects and normative runtime maintain the first three. Accountability and experience extensions inherit the core references and perform their own checks for the fourth. The same correspondences can be used to assess implementations in enterprise workflows or assembled modules.

\begin{equation}
\Sigma_R(t)=\langle\field{rid},v,q,K_a,E_c\rangle.
\end{equation}

\begingroup
\small
\setlength{\tabcolsep}{4pt}
\renewcommand{\arraystretch}{1.18}
\setcounter{table}{2}
\begin{table}[htbp]
\caption{Cross-stage semantic correspondences of collaboration contracts.}\label{tab:3}
\centering
\small
\begin{tabular}{@{}>{\RaggedRight\arraybackslash}p{0.19\dimexpr\linewidth-6\tabcolsep\relax}>{\RaggedRight\arraybackslash}p{0.28\dimexpr\linewidth-6\tabcolsep\relax}>{\RaggedRight\arraybackslash}p{0.29\dimexpr\linewidth-6\tabcolsep\relax}>{\RaggedRight\arraybackslash}p{0.24\dimexpr\linewidth-6\tabcolsep\relax}@{}}
\toprule
\textbf{Cross-stage connection} & \textbf{Correspondence to retain} & \textbf{Adjudication and implementation responsibility} & \textbf{Business error prevented} \\
\midrule
Proposal validation to acceptance and activation & Validation, confirmation, and activation concern the same content at rid/v & Core checks references and current version; authority is queried at action time & Treating a stale proposal or unauthorized reply as current agreement \\
Old version to new version & Every outstanding old commitment has a lawful $\mu$ disposition, including untriggered conditional commitments & Core derives the complete scope from the effective ledger and retains lineage and required confirmations & Omitting payment or another old obligation when rescheduling \\
Execution to fulfillment and closure & Outcomes and evidence bind to the supported rid/v/kid and prescribed requirements & External interfaces issue independently; core checks original origin and permitted use and records attachment separately & Treating arrival as full acceptance or old evidence as support for a new obligation \\
Closed records to responsibility and experience & Derived results inherit core references and meet their own grounds and qualification requirements & Extensions check causality, terms, ownership, purpose, and applicability & Equating provenance with correct attribution or unrestricted reuse \\
\bottomrule
\end{tabular}
\end{table}
\endgroup

A relationship can enter \state{Proposed}, \state{Negotiating}, \state{Active}, \state{Suspended}, \state{Renegotiating}, \state{Closed}, or \state{VerifiedClosed}. \state{Proposed} and \state{Negotiating} may use a provisional \rid{} to correlate invitations, disclosures, candidates, and confirmations, but have no effective contract version and establish no commitments. These process records bind the provisional \rid{} and candidate digest and carry a pending-version status; they do not masquerade as an effective $C^v$. \state{Active} begins only after a validated candidate receives the required exact-content confirmations and is committed. \state{Suspended} records a pause permitted by the agreement or an authoritative adjudication, with its scope and grounds retained through a state-update event. During \state{Renegotiating}, the previous version remains effective until a successor is atomically committed. Administrative or exit rules can produce \state{Closed}. \state{VerifiedClosed} requires an effective agreement, permitted terminal states or dispositions for all established and carried commitments, and satisfaction of designated evidence requirements. An ineffective proposal may be withdrawn or administratively ended. Even an effective relationship with no outstanding obligations retains its valid version, closure authorization, and closure event.

\FloatBarrier
\subsection{Normative transitions and external inputs}\label{subsec:4.3}\label{sec:4.3}

The interface receives a participant's intended action and the verdicts supplied by enterprise or fact interfaces, then decides whether to update the agreement. The relationship transition system is $\mathrm{RTS}=(S,A_R,\delta,s_0)$, where $S$ includes the projection, version store, and event store; $A_R$ is the set of normative actions; and $\delta(S,a,x)$ accepts a proposed action $a$ and external verdicts $x$, returning a new state or a rejection reason. Table~\ref{tab:4} groups actions into proposal, acceptance and activation, state and evidence update, version migration, and release and closure. Authorization grants, disclosure, and physical execution occur through external interfaces.

A rejection appends a diagnostic record while preserving effective obligations and the normative projection. An idempotent duplicate returns the existing event reference. Request processing, normative effect, factual verification, and subsequent business consequences are recorded separately. Enterprise systems enforce permissions for physical execution.

\begingroup
\small
\setlength{\tabcolsep}{4pt}
\renewcommand{\arraystretch}{1.18}
\setcounter{table}{3}
\begin{table}[H]
\caption{Preconditions and effects of five categories of normative transition.}\label{tab:4}
\centering
\small
\begin{tabular}{@{}>{\RaggedRight\arraybackslash}p{0.19\dimexpr\linewidth-4\tabcolsep\relax}>{\RaggedRight\arraybackslash}p{0.36\dimexpr\linewidth-4\tabcolsep\relax}>{\RaggedRight\arraybackslash}p{0.45\dimexpr\linewidth-4\tabcolsep\relax}@{}}
\toprule
\textbf{Transition class} & \textbf{Preconditions} & \textbf{Permitted effects} \\
\midrule
Formation proposals & Identifiable initiator, provisional relationship, invited roles, and candidate structure & Correlate invitations, participation responses, disclosures, and candidates without activating commitments \\
Confirmation and activation & No previously committed effective version under rid; canonical rid/version/binding/terms hash, action-time authority, hard constraints, and typed confirmations from the runtime-derived required set & Record confirmations; register all commitments only when the agreement becomes effective, including untriggered conditional commitments \\
State and evidence update & Trusted issuance or adjudication, original origin, antecedents, and bound references & Append state or verification events without changing terms \\
Version migration & Current predecessor, well-bound valid successor at commit time, typed confirmations from the runtime-derived change requirements, and complete lawful $\mu$ over $K_{\mathrm{out}}^v$ & Atomically commit new version and old-obligation dispositions \\
Release and closure & Release: authority and grounds; withdrawal or administrative ending: own rules; provable closure: effective agreement, complete obligation domain, and required evidence & Record release, withdrawal, or administrative ending according to request; record VerifiedClosed only when closure conditions hold \\
\bottomrule
\end{tabular}
\end{table}
\endgroup

Initial activation requires that no effective version has ever been committed for the provisional relationship, that authority and validity still hold at commit time, and that all required exact-content confirmations are present. Migration instead requires the declared predecessor to be current and a permitted disposition for every old outstanding obligation, committed atomically with the new version. Verifiable closure requires an effective agreement, an authorized request, the complete obligation domain, permitted terminal states, and designated evidence. Requests that fail their preconditions return \state{Reject} and preserve the current normative projection. Tables~\ref{tab:3}--\ref{tab:5} state the action effects, references, and disposition vocabulary used in this paper.

The coordination protocols distinguish candidate search from normative agreement. The canonical confirmation payload contains the relationship and version identifiers, version-specific enterprise--role bindings, all effective candidate terms and commitments, and, for evolution, the complete disposition mapping. Let $h(C)$ and $h(C,\mu)$ denote hashes over those canonical payloads. Any change to a hashed field invalidates candidate-specific confirmations. A confirmation credential identifies its issuing enterprise, confirmation role or subject, confirmation type, relationship and version, candidate digest, and validity interval. It therefore cannot serve as a blanket consent for unrelated roles or changes.

A typed confirmation requirement is $r_c=\langle p,\mathit{role},\mathit{type},\mathit{subject}\rangle$: enterprise $p$ must confirm the governed subject while acting in the stated contractual role and using the stated confirmation type. Let $\mathrm{ReqConf}(C,t)$ be the requirements that apply at time $t$ under the version-specific bindings, commitments, rights, and enterprise policies. The runtime independently derives or verifies this set. Agents may suggest confirmations, but their candidate cannot shrink the required set. The predicate $\mathrm{AuthorizedConfirm}(r_c,\rid,v,d,t)$ checks the issuing enterprise, role, type, subject, relationship and version, candidate digest $d$, authority, and validity at $t$. Initial activation is permitted only when
\begin{equation}
\begin{split}
&\mathrm{NoCommittedVersion}(\rid)\ \land\ \mathrm{Valid}(C^1,t_{\mathrm{commit}})\\
&\quad\land\ \forall r_c\in\mathrm{ReqConf}(C^1,t_{\mathrm{commit}}):
\mathrm{AuthorizedConfirm}(r_c,\rid,1,h(C^1),t_{\mathrm{commit}}).
\end{split}
\label{eq:formation-activation}
\end{equation}
$\mathrm{NoCommittedVersion}(\rid)$ means that no effective version has ever been committed under that relationship identifier. A closed relationship therefore cannot re-enter through initial activation. A later collaboration uses a new \rid{} or an explicit authorized reopening transition, which is outside the base kernel.

Negotiation strategies search for $C^1$; Equation~\ref{eq:formation-activation} determines its normative acceptance. Section~\ref{subsec:5.3} describes the strategy interface.

For evolution, let $K_{\mathrm{out}}^v$ be the obligations in the effective ledger that have not been lawfully terminated, and let $\mathrm{Affected}(C^v,C^{v+1},\mu)$ denote the enterprises whose bindings, rights, duties, or required grounds change. This set identifies the routing and impact scope. The runtime expands it into $\mathrm{ReqConf}_{\Delta}(C^v,C^{v+1},\mu,t)$, the typed confirmation requirements for the change. Both sets are derived or verified from the effective version, ledger, dependency rules, and enterprise policies; proposed sets are advisory and cannot omit a required enterprise, role, type, or subject. A successor can commit only when
\begin{equation}
\begin{split}
&\mathrm{Current}(C^v)\land \mathrm{SuccessorOf}(C^{v+1},C^v)\\
&\quad\land\mathrm{Valid}(C^{v+1},t_{\mathrm{commit}})
\land \mathrm{Complete}(\mu,K_{\mathrm{out}}^v)\\
&\quad\land\ \forall r_c\in\mathrm{ReqConf}_{\Delta}(C^v,C^{v+1},\mu,t_{\mathrm{commit}}):\\
&\qquad\mathrm{AuthorizedConfirm}(r_c,\rid,v{+}1,h(C^{v+1},\mu),t_{\mathrm{commit}}),
\end{split}
\label{eq:evolution-activation}
\end{equation}
Here $\mathrm{SuccessorOf}(C^{v+1},C^v)$ requires the same \rid{}, the next version number, and $C^{v+1}.\field{pred}=C^v$. Satisfying the predicate is followed by an atomic migration write. Failure to find or confirm a successor leaves $C^v$ effective. These predicates define agreement on an exact candidate; they do not guarantee that a negotiation strategy will find one or terminate within a bounded number of rounds. Replica agreement and distributed commit are deployment mechanisms beneath the logical atomic-write interface described in Section~\ref{subsec:4.1}.

A malformed, unauthorized, or rejected confirmation request leaves the enterprise's ability to review a revised candidate intact, unless an explicit enterprise policy states otherwise. Fresh confirmation binds the revised digest and supersedes any incompatible earlier confirmation. This supports correction while retaining the history of enterprise decisions.

\subsection{Current validity and future feasibility}\label{subsec:4.4}\label{sec:4.4}

$\mathrm{Valid}$ determines whether a proposal can currently be committed. The implementation checks structure and references, action authorization and disclosure scope, encodable hard constraints, and exception and evidence specifications. $\mathrm{WellFormed}$ includes $\mathrm{WellBound}(C^v)$: every debtor and creditor reference must resolve through the version-specific enterprise--role bindings in $B^v$. Authorization to establish those bindings is checked at activation or migration time. A later permission change may block a new action, but it does not erase an established obligation; the obligation must instead follow an applicable exit, excuse, breach, release, or migration rule. Appropriate interfaces supply authorization and private facts; domain models and constraint solvers handle resources and timing. A large language model (LLM) cannot replace unknown facts with affirmative verdicts. An incomplete required check results in rejection, a pending judgment, or a request for additional permitted information.

$\mathrm{Feasible}$ is a semantic condition on world dynamics: an allowed future trajectory must satisfy the goal and hard commitments that have not been excused. In the controlled experiments, solvers or simulations compute feasibility within finite parameterized worlds. Open dynamic environments require domain-specific models. A contract that is currently valid can still fail following a later fault.

\begin{equation}
\begin{split}
\mathrm{Valid}(C^v,t)\ \Longleftrightarrow\ &\mathrm{WellFormed}\land\mathrm{AuthOK}(t)\\
&\land\mathrm{HardConstraintsOK}(t)\land\mathrm{EvidenceSpecOK}.
\end{split}
\end{equation}

Correspondingly, $\mathrm{Feasible}(C^v,W_t)$ means that an allowed trajectory $\pi$ exists whose execution satisfies the goal and effective hard commitments. $W_t$ describes environment semantics. The evaluated methods do not read evaluator-private ground truth; they make judgments using accessible interfaces and predicates within their designated scope.

\subsection{Versioned disposition and continuity of obligations}\label{subsec:4.5}\label{sec:4.5}

Rescheduling after a road closure must account for existing delivery, transportation, and payment responsibilities. A scheduler proposes an execution plan; a disposition mapping $\mu$ specifies how each old obligation continues or terminates. Its checking domain comes from all old commitments that have not been validly terminated, including untriggered conditional commitments. A proposal cannot reduce that domain by shortening its own list. Each disposition requires applicable terms, authority, facts, or consent. The new version, $\mu$, and migration event are committed in one transaction. Missing or impermissible dispositions cause rejection.

For example, payment is already committed before acceptance. Its disposition after the road closure must either carry it forward or provide valid termination grounds. Agreement activation governs when a commitment is established; the performance antecedent governs when payment is required. Closure covers all established and carried commitments. Reducing the new version's list of required items cannot remove an old obligation from checking.

The model's \code{carry} continues an obligation with unchanged substantive content and debtor--creditor parties. It retains explicit lineage if its representation uses a new identifier; the base kernel uses the same \kid{} and ledger state across versions. Changes to quantity, amount, parties, or other substantive terms require grounded disposition of the old claim and separately authorized establishment of its successor. Original \rid{}/$v$/\kid{}, grounds, and event times remain in the record. The base disposition vocabulary is \code{carry}, \code{discharge}, \code{excuse}, \code{breach}, and \code{waive}; the last two correspond to \state{Violated} and \state{Released}.

\begin{figure}[H]
\centering
\includegraphics[width=\linewidth]{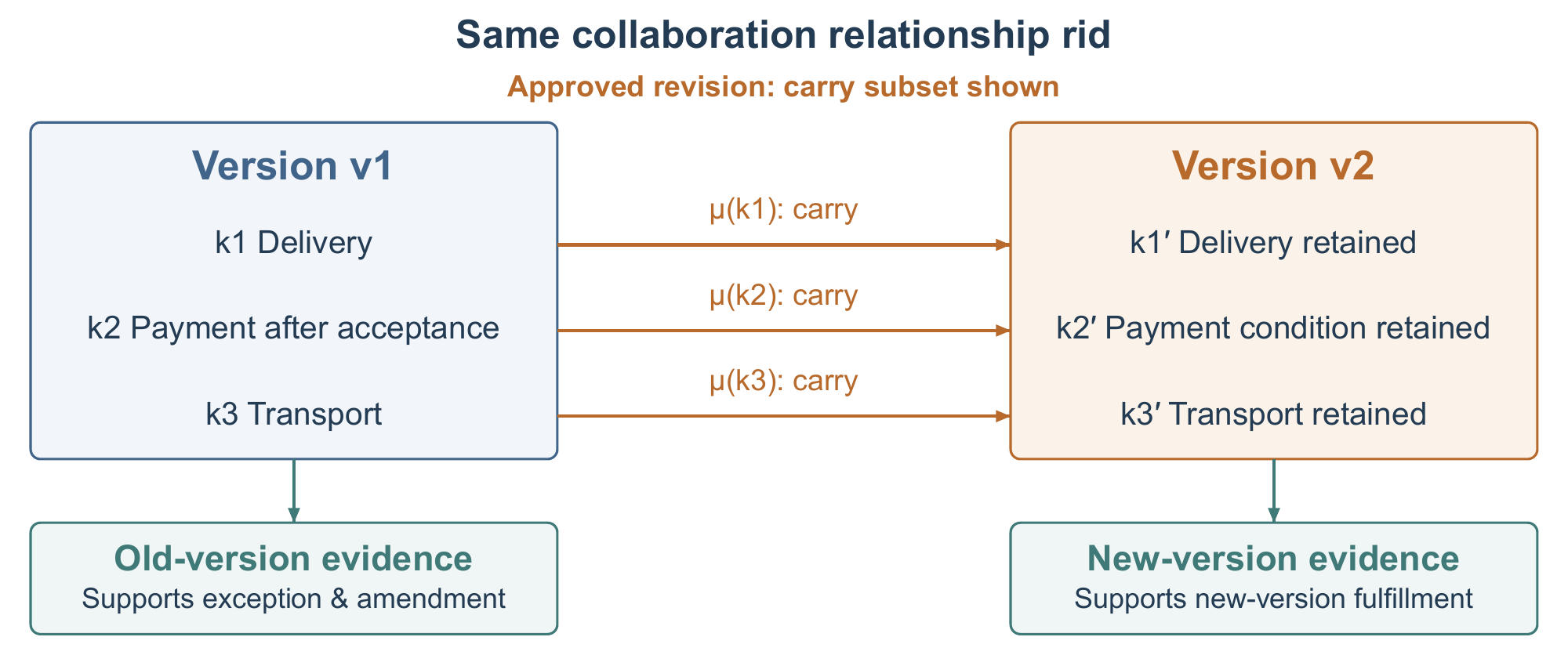}
\caption{Unchanged obligations carried into an approved version: lineage and the distinct uses of old and new evidence. Changed obligations require separate disposition and successor establishment.}\label{fig:3}
\end{figure}

In Figure~\ref{fig:3}, $k'_1$, $k'_2$, and $k'_3$ denote positions of the same obligations in the new version. The base kernel represents this continuation using unchanged identifiers. Section~\ref{subsec:8.2} distinguishes that implementation from the later successor transaction.

\begin{equation}
\mu^{v\rightarrow v+1}:K_{\mathrm{out}}^v\longrightarrow\mathcal D,
\qquad \mathcal D=\{\langle d,\ell,g\rangle\mid d\in D\}.
\end{equation}

Here $D=\{\mathsf{carry},\mathsf{discharge},\mathsf{excuse},\mathsf{breach},\mathsf{waive}\}$, $\ell$ is a continuation reference for \code{carry} (and $\bot$ otherwise), and $g$ contains the required grounds. Thus every entry has a disposition type, an applicable target reference, and adjudicable support. Table~\ref{tab:5} gives the grounds for each type.

The later successor implementation composes old-claim disposition, new-obligation establishment, and version switching in one transaction. Its historical status \code{superseded} records that replacement and points to the new obligation and disposition grounds; it does not assert that the old obligation was performed. Section~\ref{subsec:10.11} reports this extension, including explicit treatment of prior lateness and credit for prior performance.

\begingroup
\small
\setlength{\tabcolsep}{4pt}
\renewcommand{\arraystretch}{1.18}
\setcounter{table}{4}
\begin{table}[htbp]
\caption{Disposition of old obligations and the grounds that must be retained.}\label{tab:5}
\centering
\small
\begin{tabular}{@{}>{\RaggedRight\arraybackslash}p{0.27\dimexpr\linewidth-4\tabcolsep\relax}>{\RaggedRight\arraybackslash}p{0.35\dimexpr\linewidth-4\tabcolsep\relax}>{\RaggedRight\arraybackslash}p{0.38\dimexpr\linewidth-4\tabcolsep\relax}@{}}
\toprule
\textbf{Disposition} & \textbf{Meaning} & \textbf{Grounds to retain} \\
\midrule
carry & Continue an unchanged obligation into the new version & Old/new versions, unchanged content and parties, and obligation lineage; base kernel retains the same kid \\
discharge & Terminate on the basis of an achieved outcome & Performance result and corresponding verification references \\
excuse & Excuse under an exception rule & Applicable terms, events, and adjudication \\
breach & Record violation of an obligation & Effective obligation, deviation, and disposition grounds \\
waive & Release by an authorized participant & Release authority and valid confirmation \\
\bottomrule
\end{tabular}
\end{table}
\endgroup

\code{discharge} does not automatically establish \state{VerifiedFulfilled}. Permitted terminal states, factual outcomes, and evidential support are checked separately. Later revocation of a permission also does not erase an established commitment. It may prohibit a new performance action, but the old obligation must receive a disposition under the agreed exit, excuse, breach, or migration rules.

Commitment $k_2$ in Figure~\ref{fig:3} is the payment-after-acceptance obligation.

\Needspace{22\baselineskip}
\subsection{Four invariants and two supporting arguments}\label{subsec:4.6}\label{sec:4.6}

Table~\ref{tab:6} states the four invariants. P1 is an information-separation observation; P2 gives conditional preservation under trustworthy inputs and complete guards. E0 tests bounded implementation paths.

\begingroup
\small
\setlength{\tabcolsep}{4pt}
\renewcommand{\arraystretch}{1.18}
\setcounter{table}{5}
\begin{table}[H]
\caption{Four requirements for successful normative writes.}\label{tab:6}
\centering
\small
\begin{tabular}{@{}>{\RaggedRight\arraybackslash}p{0.26\dimexpr\linewidth-2\tabcolsep\relax}>{\RaggedRight\arraybackslash}p{0.74\dimexpr\linewidth-2\tabcolsep\relax}@{}}
\toprule
\textbf{Invariant} & \textbf{Requirement for successful normative writes} \\
\midrule
I1: Authority preservation & Formation, release, excuse, and migration use action-time authority and every required role/type/subject confirmation bound to the exact relationship, version, candidate digest, and valid time \\
I2: Obligation continuity & Every outstanding old commitment has a lawful disposition and supporting grounds, including untriggered conditional commitments; retain lineage \\
I3: Reference completeness & Normative claims of fulfillment, breach, excuse, release, and evidence closure cite corresponding rid/v/kid; relationship-level events follow D4 \\
I4: Evidence closure & VerifiedClosed references an effective relationship and agreement and covers lawful dispositions and prescribed support for all established and carried commitments \\
\bottomrule
\end{tabular}
\end{table}
\endgroup

\Needspace{13\baselineskip}
\textbf{Observation P1 (information insufficiency under a reduced view).} Let $s$ be a complete state containing the external context needed for a judgment, $\pi_X$ the entire accessible view after removing a class of information, and $J$ the target judgment under declared rules. If states $s_1$ and $s_2$ satisfy the following condition, no deterministic function $f_X$ depending only on $\pi_X$ can judge both states correctly.

\begin{equation}
\pi_X(s_1)=\pi_X(s_2)\quad\land\quad J(s_1)\ne J(s_2).
\end{equation}

\begin{equation}
\neg\exists f_X\ \forall i\in\{1,2\}:\quad f_X(\pi_X(s_i))=J(s_i).
\end{equation}

\emph{Argument.} The identical input view produces the same deterministic output, whereas the target judgments differ; at least one judgment is wrong. Witnesses N1--N6 each identify the removed information, identical view, and differing judgments. The view includes permitted additional queries and their replies. If another representation or query restores the information, the observation's premise must be reassessed. P1 identifies information required by the designated judgment; object names and software packaging may differ.

\textbf{Preservation argument P2 (conditional on complete guards).} Let $I(S)$ mean that I1--I4 hold for the core normative state and its version and event-reference stores. Assume that the initial state satisfies $I$ and that external authorization, typed confirmations, and factual verdicts are trustworthy and normalized by issuer or actor, action, confirmation role, type and subject, relationship and version, candidate digest, and valid time. The core runtime completely registers established and carried commitments and their lineage, derives migration and closure checking domains from those records, and checks references, dispositions, and required grounds. All normative writes pass the transitions and guards of Tables~\ref{tab:4}--\ref{tab:6}; versions, $\mu$, and normative events commit as serialized atomic transitions of one logical relationship state. Replicated implementations must realize that serialization and preserve event history. Denote these external and internal implementation conditions by $H$. For a successfully committed transition:

\begin{equation}
I(S)\land H(S,a,x)\land[\delta(S,a,x)=S']\quad\Longrightarrow\quad I(S').
\end{equation}

\emph{Preservation argument.} Rejection and idempotent returns preserve the normative projection. Acceptance and activation check authority and necessary confirmation and register commitments. State and evidence updates check factual sources and references. Migration checks every outstanding old obligation's disposition and retains lineage. Release records an authorized actor and disposition grounds. Verifiable closure checks an effective agreement, the complete obligation domain, permitted terminal states, and designated evidence. Under $H$, each action preserves I1--I4; induction yields preservation for finite sequences. Section~\ref{sec:10.6} reports implementation checks of these paths.

$H$ includes substantive deployment conditions. If a consent interface directly trusts a proposal field stating ``no consent needed,'' it bypasses independent authority. Complete references alone cannot then establish preservation of I1. Interfaces must check issuer, role, confirmation type and subject, scope, action, relationship and version, candidate digest, and valid time. P2 concerns conditional safety of normative writes; factual truth, physical overreach, causal-model correctness, and availability require corresponding enterprise mechanisms. Section~\ref{sec:8.5} defines this threat boundary.

\section{Coordination Protocols and Lifecycle Operation}\label{sec:5}

\begin{figure}[htbp]
\centering
\includegraphics[width=0.90\linewidth]{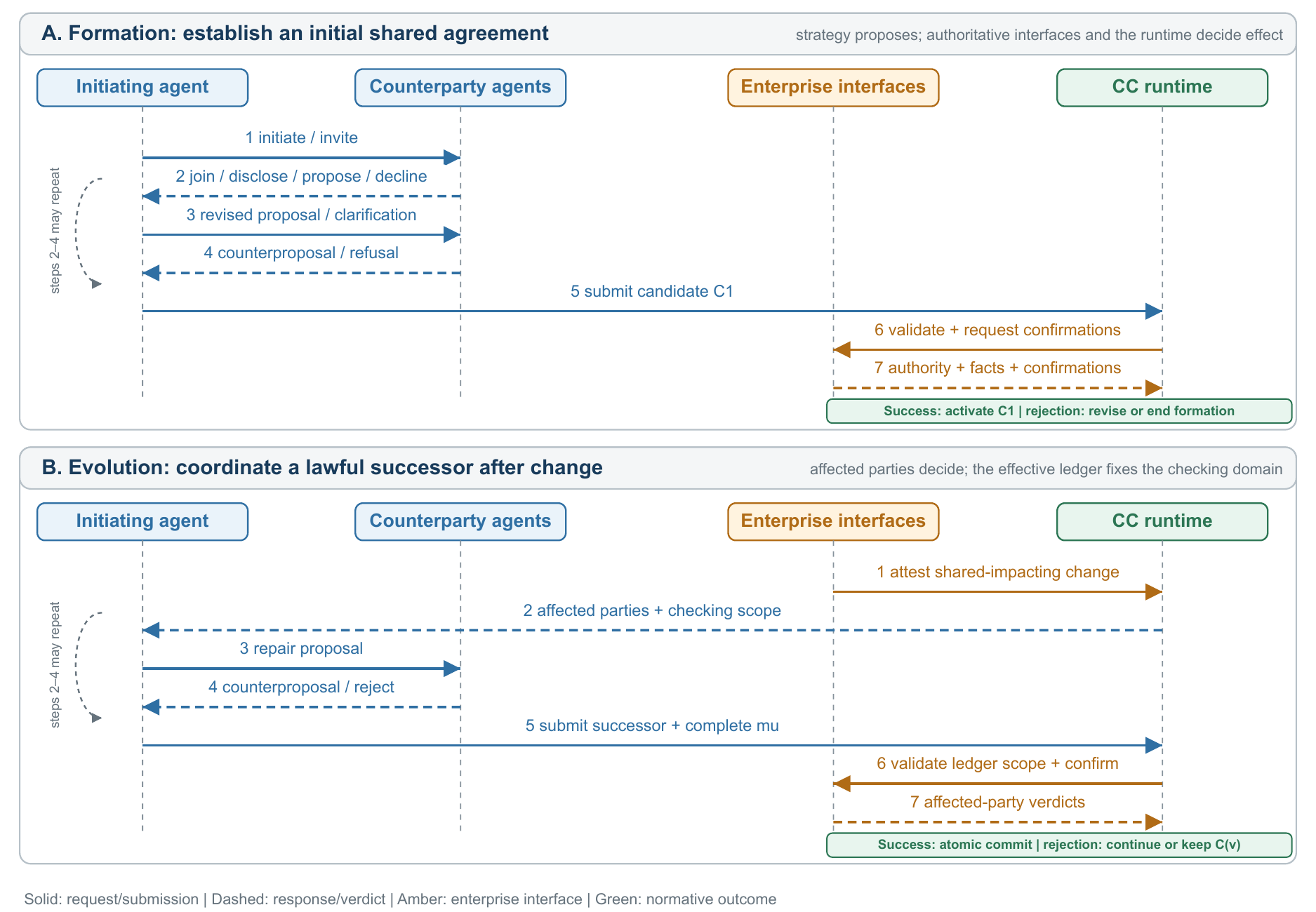}
\caption{Two coordination protocols within the CC lifecycle. Candidate-generation strategies may vary; exact-content confirmation, validation, rejection semantics, and normative writes follow the CC rules. Dashed arrows denote replies or returned verdicts; solid arrows denote requests or submissions. Steps 2--4 may repeat.}\label{fig:4}
\end{figure}

Two coordination protocols organize shared decisions: formation establishes the first effective agreement, and evolution establishes a lawful successor. Figure~\ref{fig:4} shows their common pattern. Blue interactions search for a candidate, amber interactions obtain enterprise-grounded verdicts, and green writes create normative effect. Other lifecycle operations use the resulting relationship and version references.

\subsection{Collaboration-contract formation protocol}\label{subsec:5.1}\label{sec:5.1}

Formation begins with an initiating agent and a business goal; it does not assume that a complete participant set or agreed terms already exist. The protocol supplies a common interaction skeleton:

\begin{enumerate}[leftmargin=2.3em,label=\textbf{FP\arabic*.}]
\item \textbf{Initiate and invite.} The initiator opens a provisional relationship, identifies the goal and intended roles, and invites candidate counterparties.
\item \textbf{Join, decline, or redirect.} An invited agent may join, decline, identify an alternative authorized enterprise--role binding, or request clarification. These responses do not establish commitments.
\item \textbf{Disclose and negotiate.} Participants exchange permitted predicates, proposals, counterproposals, refusals, and requests for additional enterprise authority.
\item \textbf{Construct and validate.} A candidate $C^1$ binds participants, terms, atomic commitments, constraints, exceptions, and evidence requirements to the provisional \rid{}. Independent adjudicators check its structure, current authority, disclosed grounds, and business constraints.
\item \textbf{Confirm exact content.} The runtime derives the required confirmer set from the version-specific bindings, commitments, rights, and policies. Every typed requirement in $\mathrm{ReqConf}(C^1,t_{\mathrm{commit}})$ is satisfied through the authoritative interface of its enterprise for the same canonical payload, role, confirmation type, and governed subject. A changed candidate invalidates earlier candidate-specific confirmations.
\Needspace{7\baselineskip}
\item \textbf{Activate or end formation.} If no effective version has ever been committed under the provisional \rid{} and validity still holds at commit time, the runtime atomically stores $C^1$, registers all commitments, appends the activation event, and enters \state{Active}. If validation or required confirmation fails, the participants may revise the candidate; if negotiation ends, no effective CC is created.
\end{enumerate}

\paragraph{Preparation and unsuccessful formation.}
Before activation, an enterprise may hold capacity, arrange an inspection, or reserve transport under its own authority or a separate agreement. The responsible enterprise records that action, its expiry, cost, and release conditions. If the main negotiation ends, FP6 records its outcome and notifies the responsible parties; their interfaces release or settle preparations under those existing grounds. For example, a supplier's temporary capacity hold may expire automatically, while a separately ordered inspection may still require payment. Ending formation creates no commitments under the proposed CC and does not cancel obligations established elsewhere. Preparation that requires a counterparty's commitment must first obtain an independently effective basis.

Agents generate commitment candidates using requirements, available experience, and counterpart disclosures. Enterprise interfaces control what information may be disclosed. A disclosure may provide an exact value, an interval, or an authorized predicate statement. A predicate answers a decision-relevant question, such as whether a quantity can be committed or a certification remains valid, without requiring disclosure of the enterprise's complete internal state.

Disclosure granularity and validity checking operate together. An allowed predicate statement can support a threshold constraint. If an interval crosses a decision boundary, the runtime requests refinement or retains an unresolved judgment. In the current experiments, a trusted simulated interface issues these predicates. Cryptographic privacy guarantees and issuer authentication are conditions for enterprise integration.

Candidate construction distinguishes targets, hard minimum requirements, and negotiable fields, as illustrated by the quantity choices in Section~\ref{subsec:2.1}. An early infeasibility judgment can prevent validation and repair from being reached. Replacing a declining participant or revising terms produces a new candidate requiring validation and confirmation.

\subsection{Inductive proposals and deductive adjudication}\label{subsec:5.2}\label{sec:5.2}

The implementation separates generation from adjudication. First, it parses an agent proposal into candidate participants, obligations, deadlines, and terms. Second, it checks identity, structure, authority, and external predicates. Third, it submits encodable resource, quantity, budget, and timing constraints to CP-SAT or the corresponding rule module. Fourth, it checks that exceptions, responsibility, and evidence requirements have explicit interpretations. A proposal proceeds to commitment only after the checks are complete and the required confirmations are valid.

The generator proposes explanations and repair actions. The verifier returns structured conflict categories, the constraints involved, and the permitted scope of repair. The agent then chooses an alternative according to its goals and enterprise policy. Designated interfaces adjudicate permissions and credential validity. Conflict reports respect disclosure boundaries and provide the information permitted for the decision.

When a proposal cannot be repaired lawfully, the runtime records rejection or termination of negotiation. When a repair path exists, it changes permitted fields and revalidates the proposal. A single repair addresses a local conflict; multiple negotiation rounds coordinate participants and combined constraints. Exhaustion of a search budget is recorded separately from the absence of a lawful plan, preserving the distinction between computation and business feasibility.

\subsection{Negotiation strategies, agreement, and participant autonomy}\label{subsec:5.3}\label{sec:5.3}

\paragraph{The effect of a negotiation contribution.}
Process records distinguish what a participant reports from what it authoritatively confirms. The following types refine the candidate/audit records in Table~\ref{tab:7}; they introduce no additional core object.

\begin{description}[leftmargin=1.4em,style=nextline]
\item[Judgment or disclosed fact.] ``The proposed date appears feasible'' supports a query or candidate decision. Its issuer, scope, validity, and evidential status determine how it can be used; it supplies no confirmation of the candidate.
\item[Conditional position or counterproposal.] ``We can expedite if the buyer pays the extra freight'' proposes a condition and may generate a revised candidate. It establishes no conditional commitment until activation or migration registers that commitment.
\item[Pending response or clarification.] A participant may report that its response is pending, request information, or state when it expects to reply. Negotiation policy sets a waiting deadline. Silence, timeout, and a coordinator's announcement that discussion has ended supply no affirmative confirmation.
\item[Confirmation or refusal.] A formal confirmation carries the authoritative issuer, represented enterprise and role, candidate digest, governed subject, scope, and valid time required by the activation or migration rules. A refusal leaves that requirement unsatisfied. Expired or revoked grounds are rechecked before a write; a later revocation does not erase an already established obligation.
\end{description}

These meanings apply to external exchanges. Each enterprise chooses how its people and software formulate a response; CC consumes the resulting permitted disclosure, position, or authoritative verdict. A candidate field claiming approval cannot promote a pending response into consent.

The normative state machine specifies which actions may take effect, and the agent chooses a strategy within that permitted set. Alternating offers, auctions, distributed constraint solving, LLM negotiation, mediator-assisted search, and human escalation can all generate candidates through the same typed actions. Strategy choice determines search, convergence, efficiency, and allocation behavior; the independent transition rules in Section~\ref{subsec:4.3} determine normative effect.

When a request exceeds action or disclosure authority, the agent may refuse it, narrow its scope, propose an alternative, or request additional authority through an enterprise interface. Enterprise goals, risk preferences, benefits, negotiation costs, and waiting costs influence the choice. The LLM generates candidate strategies; policy interfaces and adjudicators constrain their execution. A search budget may expire or offers may remain mutually unacceptable. Formation then ends without an effective CC; evolution retains the previously effective version and proceeds only through permitted failure, breach, release, or termination paths.

An approved extension of authority changes the enterprise's current action or disclosure permissions. A change to established rights or obligations beyond an already authorized option or remedy requires a successor version and the prescribed affected-party confirmations. The experiments use preset permissions to cover permitted and prohibited paths. Enterprise decision latency can be represented at the boundary as delayed, denied, expired, or granted responses, leaving the internal decision procedure outside the CC model.

Autonomy includes the right to refuse collaboration or exit it. Contract terms provide explicit dispositions for established obligations, including \state{Released}, \state{Excused}, and \state{Violated}. Withdrawal handles proposals that have not taken effect; release handles commitments that have already become effective. Each action follows its own conditions.

\subsection{Dynamic fulfillment and monitoring}\label{subsec:5.4}\label{sec:5.4}

After version activation, enterprise systems execute actions and designated sources return business states and events. The Monitor checks antecedents, deadlines, outcomes, and evidence requirements against the currently effective commitments. Its outputs include fulfilled, evidence pending, deviation detected, disposition pending, and permitted terminal states. Events are appended with their actual version references and do not overwrite earlier terms.

Task completion, path safety, and provable fulfillment are measured separately. Arrival of goods records a business outcome. Fulfillment also requires an authorized participant to perform the agreed obligation and an outcome consistent with the effective constraints. The corresponding verified state is written after the prescribed inspection, authority, and outcome evidence passes designated checks. Missing evidence leaves the judgment unresolved.

\paragraph{Routing a business change.}
The runtime first determines whether the event changes facts, exercises an existing entitlement, or requires new shared terms. The route follows the effective agreement and authoritative grounds; an agent cannot choose a less demanding route merely by naming it differently.

\begin{center}
\small
\begin{tabularx}{\linewidth}{@{}p{0.27\linewidth}YY@{}}
\toprule
\textbf{Business situation} & \textbf{Permitted route} & \textbf{Continuity requirement} \\
\midrule
New facts; terms still apply & Record admissible evidence and update monitoring & Preserve commitments and current version \\
Exercise of an agreed option or remedy & Check its trigger, authority, notices, and any confirmations already required & Apply the effective clause and retain its event grounds \\
Change beyond authorized terms & Enter EP and validate a successor & Retain the old version until the complete migration commits \\
No acceptable successor & Record failure to agree; use a permitted stopping or disposition path & Waiting or refusal alone neither suspends deadlines nor extinguishes obligations \\
\bottomrule
\end{tabularx}
\end{center}

\paragraph{Exercising an option without overwriting a version.}
The effective version fixes both the ordinary terms and any authorized conditional branches. The Monitor derives the currently applicable arrangement from those clauses together with validated option-exercise events. For example, a clause may allow a two-day extension after an attested transport disruption. The runtime checks the trigger, exercising party, affected commitments, notice and confirmation requirements, exercise deadline, and permitted number of uses. It then appends an event referencing the clause and its grounds and updates the runtime projection used for deadline checks. The original version and its ordinary deadline remain immutable; replay of the same exercise event cannot extend the deadline again. Pending or disputed grounds leave applicability unresolved under the clause's rules.

Selecting an already authorized execution branch can retain the version. Changing enterprise--role bindings, creating a new inter-enterprise commitment, or altering a term beyond the clause's permission requires the corresponding successor-version and confirmation procedure. This specifies the interpretation required by the business-routing rule; the present evaluations do not isolate a general option-exercise service.

The procurement and manufacturing domains impose different execution pressures. Procurement mainly exposes information and selection risks during formation. Manufacturing also involves queues, setup, changing production rates, equipment faults, inspections, and rework. Enterprise-system simulators or world interfaces provide these events, which monitoring and remedies use to handle changes during execution.

\subsection{Collaboration-contract evolution and renegotiation protocol}\label{subsec:5.5}\label{sec:5.5}

An enterprise change enters the evolution protocol when continuing collaboration requires terms beyond the rights, obligations, evidence conditions, or remedies already authorized by the effective CC. Internal substitutions that preserve the agreement remain enterprise execution choices, and pre-agreed options follow the route in Section~\ref{subsec:5.4}. When a successor is required, the protocol proceeds as follows:

\begin{enumerate}[leftmargin=2.3em,label=\textbf{EP\arabic*.}]
\item \textbf{Notify and ground the change.} An authorized source or participant reports the event with its relationship, version, time, and available grounds.
\item \textbf{Derive impact and affected parties.} The runtime starts from the effective version bindings, ledger, dependency rules, and enterprise policies to identify outstanding obligations, affected participants, and required adjudications. An Agent suggestion cannot shrink this set.
\item \textbf{Negotiate a successor.} Affected agents may propose repairs, counterpropose, reject, request additional information or authority, substitute a permitted participant, or choose a stopping path.
\item \textbf{Construct and validate migration.} A candidate $C^{v+1}$ supplies a disposition mapping $\mu$ for every member of $K_{\mathrm{out}}^v$, including conditional commitments whose antecedents have not yet become true. The runtime validates the new terms, grounds, and complete ledger-derived scope.
\item \textbf{Confirm the affected change.} Every typed requirement in the runtime-derived $\mathrm{ReqConf}_{\Delta}(C^v,C^{v+1},\mu,t_{\mathrm{commit}})$ is satisfied for the exact successor, version-specific bindings, and disposition mapping through the authoritative interface named by its enterprise, role, confirmation type, and subject. Rejection returns structured reasons and leaves $C^v$ effective.
\item \textbf{Commit or follow a lawful stopping path.} Successful migration atomically stores the successor, $\mu$, and migration event before changing the current-version pointer. If no successor is confirmed, the runtime retains the old obligations and applies permitted suspension, breach, excuse, release, or termination procedures.
\end{enumerate}

Replacing a participant, postponing delivery, reducing quantity, or changing an execution path may require different confirmations. Recoverability requires both a lawful remedy trajectory and the corresponding confirmations; a locally feasible repair is insufficient when it changes another enterprise's rights.

Old-version evidence continues to explain earlier events and the grounds for migration. New-version fulfillment is checked against the new obligations and any explicitly permitted lineage. If a concurrent or stale proposal loses a version race, the runtime returns the current reference and prevents silent overwriting.

\section{Evidence Closure and Lifecycle Extensions}\label{sec:6}

\subsection{Evidence registration, verification, and closure}\label{subsec:6.1}\label{sec:6.1}

The evidence chain connects terms, authority grounds, execution events, and outcomes. Enterprise systems or designated registries retain original credentials; the contract runtime retains their references, sources, and integrity-check results. Credential issuance is an independent world-side event. For example, a delivery system registers a delivery fact, and a payment interface registers a payment fact and the credential's origin. Evidence submission by an agent only queries existing registrations. Submission does not issue a credential. Logs separately record issuance, attachment, verification, and fulfillment adjudication for later responsibility analysis and review.

Responsibility analysis and experience consumption build on the core records using additional causal models, ownership rules, and applicability conditions.

Evidence verification checks source, relationship and commitment binding, version applicability, time, and purpose. The reference entry point queries credentials already issued by the world-side registry. An unregistered source remains unresolved and is rejected from the ledger; a credential bound to another relationship is also rejected. The implementation returns \code{origin_unattested} and \code{origin_mismatch}, respectively. An explicit source declaration must agree with registration. The submission destination cannot overwrite the original origin. Use of an old fact along a lineage requires an explicit rule; a new-version outcome obligation requires appropriate support. A content hash locates a registration record. Authenticity depends on source authentication and factual adjudication at the interface.

\paragraph{A legitimate use of one business report in several relationships.}
A quality report may cover a batch of 20 serialized units, of which eight fulfill contract A and twelve fulfill contract B. The original report remains at its authoritative source. An authorized issuer can provide separate applicability attestations that reference that report and identify each covered subset, specification version, permitted purpose, valid time, and target \rid/$v$/\kid{}. Each application retains its own evidence reference and verification result. The runtime checks coverage of the particular obligation and any restrictions on allocation or reuse; a complete reference alone does not establish coverage. Reusing the report as information does not authorize counting the same exclusive delivery twice. A credential already bound to A cannot be relabeled as B by its submitter.

This example specifies an enterprise evidence-interface extension: source facts remain unchanged, and independently authorized applications carry relationship-specific scope. The base registry's cross-relationship rejection remains in force; the reported checks do not implement or evaluate batch-subset attestation or shared allocation accounting.

$\field{VerifiedFulfilled}(\kappa)$ requires a satisfied performance antecedent, designated factual support for the outcome, a path consistent with the effective constraints, and satisfaction of $\field{evReq}$. $\field{VerifiedClosed}$ first requires an effective collaboration agreement, then checks lawful dispositions and corresponding evidence for every established and carried commitment in the relationship. Conditional commitments with unsatisfied antecedents remain in this scope. Excuse, breach disposition, or authorized release may yield permitted terminal states. Closure records preserve how termination occurred and distinguish it from the originally agreed delivery.

\code{close} is an action through which an authorized participant requests that a closure judgment be written into the normative record. It confirms that the specified matters in the relationship have lawful dispositions; enterprise acceptance, payment, and consent remain the responsibilities of their respective business interfaces. The reference implementation grants the supplier agent preset relationship-scoped \code{close} authority and checks the effective version, complete obligation domain, and evidence. An ineffective empty record cannot be written as $\field{VerifiedClosed}$. The base reference kernel has no multiparty closure-signature workflow. If a delegated runtime component triggers closure, the business protocol must specify its authority and retain the triggering grounds.

Evidence closure has both safety and coverage objectives. I4 prevents a relationship without the prescribed grounds from being written as provably closed. It does not issue credentials or ensure that every completed business case reaches that state. The runtime must also arrange credential issuance, permitted queries, version-specific uses, and closure requests to convert actual completion into a supported judgment. Section~\ref{subsec:10.2} reports business completion and positive fulfillment adjudication separately.

\subsection{Normative and causal grounds for responsibility analysis}\label{subsec:6.2}\label{sec:6.2}

Responsibility analysis first identifies the effective version and particular obligation, establishing who undertook which outcome, deadline, and risk. It then uses runtime events and a domain causal model to propose causes, applying checkpoint/fork operations to remove or replace events and check whether the outcome changes. Finally, it applies excuse, risk-transfer, and remedy rules to produce an auditable analysis. Causal influence and normative responsibility require separate judgments.

The same road closure may cause late transport, yet responsibility depends on whether the terms cover force majeure, whether valid event evidence exists, and whether other obligations could still be fulfilled. Joint faults may require simultaneous removal of several causes to restore the outcome. Attribution based only on the last log entry can miss such combinations. E5 and E6 test these mechanisms within specified causal models and fault families.

A responsibility claim is a derived result outside the four-object core. It must inherit the relevant \rid/$v$/\kid\ references and additionally cite applicable terms, authority and event sources, causal judgments, and excuse grounds for review by participants or other adjudicators. These are output requirements checked by the responsibility extension itself. Structural causal models and PROV provide causal and provenance foundations, respectively~\cite{r53,r54}. Complete references make grounds locatable; they do not establish model correctness, factual truth, or legal responsibility.

\subsection{Experience assets derived from concluded collaboration records}\label{subsec:6.3}\label{sec:6.3}

Collaboration experience uses concluded records with identified outcomes and evidential support. Sources include closed effective relationships and concluded negotiation episodes. A negotiation may yield a supported lesson after a correct refusal or failure to agree without ever activating a CC. Its source remains the provisional relationship and candidate/audit history, and it is not labeled $\field{VerifiedClosed}$. An experience repository may store
\[
\begin{aligned}
x=\langle &\field{xid},\field{sourceRef},\field{contextSig},\field{applicability},\field{lesson},
\\
&\field{evidenceRefs},\field{owner},\field{validityWindow},\field{riskBoundary}\rangle.
\end{aligned}
\]
The source reference identifies the particular outcome rather than only the relationship:
\[
\field{sourceRef}=\begin{cases}
(\rid,v,\field{outcomeEventId}), & \text{effective relationship},\\
(\rid,\field{candidateDigest},\field{auditEventId}), & \text{concluded negotiation}.
\end{cases}
\]
The second form can refer to a provisional relationship that never acquired an effective version. A lesson spanning several revisions retains a set of these references and their supersession links. Each cited event carries its outcome type and supporting grounds: refusal, withdrawal, timeout, and a verified business failure remain distinct. A participant's statement that a date is infeasible supports a record of that judgment; establishing actual infeasibility requires independent grounds. The source and evidence references permit provenance review. Applicability predicates, validity periods, and risk limits specify when a lesson may be used. The repository belongs to an enterprise or another party with lawful use rights. It is not automatically shared across enterprises.

Experience extraction converts outcome-grounded records with permitted uses into candidate assets supported by evidence. Structured retrieval and summary compression organize the content. Both fields and natural-language explanations require support checks. Uncertain analyses retain uncertainty markers or are excluded. E9 evaluates structured qualification and free-text support separately.

A process lesson can retain observed cues, permitted queries and their replies, candidates and exclusion reasons, conditional feedback, authority requests, and actual outcomes. For example, a buyer may learn to ask whether a promised date denotes arrival or release after inspection before negotiating a production start. Supported refusals and withdrawn candidates can identify unsuitable conditions. A counterparty's concession or an enterprise authorization is recorded as external support, so an outcome is not attributed wholly to the focal agent. Only observable, permitted records support these lessons; private deliberations are outside the shared trace.

Experience consumption has retrieval and qualification stages. Similarity retrieves candidates; a qualification gate checks applicability, ownership, evidence, validity, risk, and the required interpretation framework. Similar history may contain another enterprise's private information, expired rules, or a different minimum quantity. Similarity therefore does not determine permission to reuse. After qualified experience enters a proposal context, the new contract still undergoes independent validation.

\subsection{Experience benefits and source contamination}\label{subsec:6.4}\label{sec:6.4}

Experience evaluation reports incorrect reuse, coverage of qualified candidates, and subsequent construction outcomes. Actual reuse, coverage, and SafeReuseYield measure gate safety and usable output. Source isolation and independent cases test for direct answer sharing in the apparent experience benefit.

Task values, case identifiers, and repair actions in historical records may directly share answers with a current task. Source audits check identifiers, case summaries, contract fingerprints, exact numerical tuples, and near-duplicate text. Positive controls measure detection sensitivity. Effects of model pretraining sources are treated separately as an external-validity issue.

The \code{gated_v2} experiment evaluates a consumption method combining applicability predicates, ownership, evidence constraints, and framework semantics. The online experience family evaluates qualification gating under a fixed agent stack. These address construction benefits and reuse qualification; Section~\ref{subsubsec:11.4.4} sets out the separate evaluation needed for process learning.

\section{Contract Networks and Composition}\label{sec:7}

\subsection{Dependencies across relationships}\label{subsec:7.1}\label{sec:7.1}

Complex industrial tasks involve relationships for material supply, manufacturing, and transport. Each relationship maintains its own \rid, versions, and participant commitments. A contract dependency graph $G_C=(N,E_D)$ uses contract versions or commitments as nodes and resource, timing, authority, disclosure, remedy, or evidence dependencies as edges. Edges propagate effects between agreements. Changes in each relationship still require approval by its responsible participants.

Composition validation uses temporal and constraint-solving tools to check resource oversubscription, temporal cycles, permission mismatches, contradictory commitments, and conflicting evidence requirements. It identifies conflicts under the specified rules and facts. Participants subsequently confirm agreements and execute them, with outcomes entering the corresponding evidence chains.

\begin{figure}[htbp]
\centering
\includegraphics[width=\linewidth]{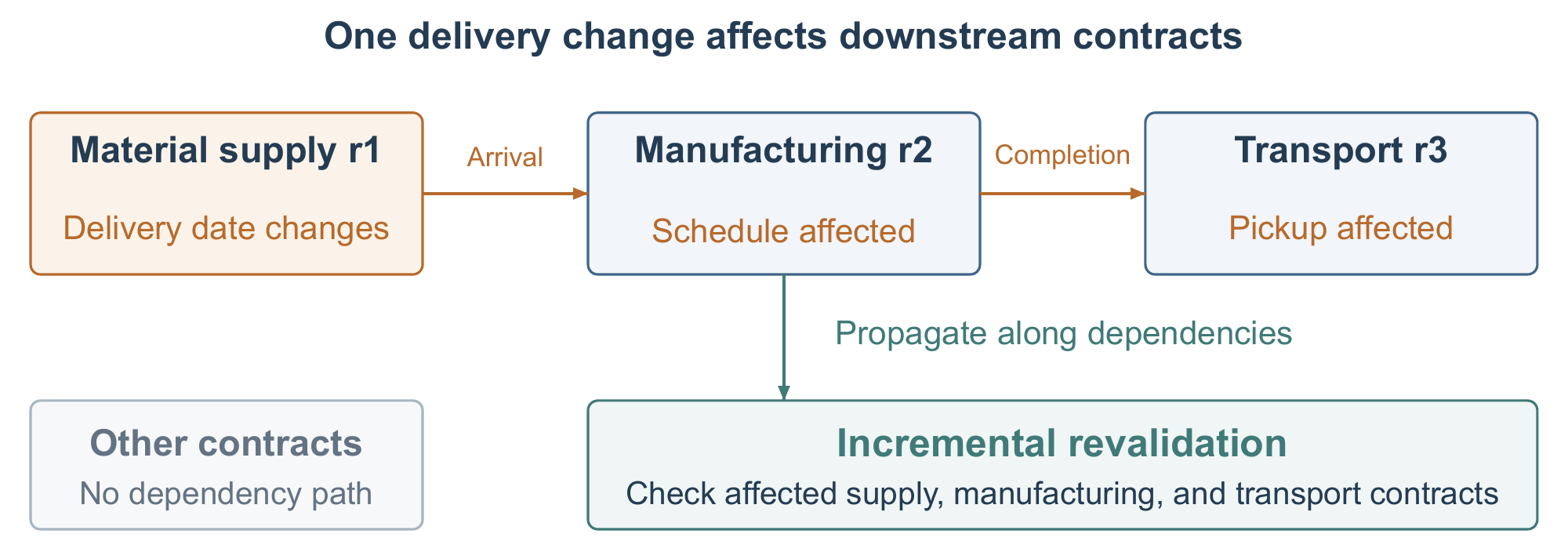}
\caption{Contract dependencies and incremental revalidation following a local change.}\label{fig:5}
\end{figure}

\subsection{Change impact and local revalidation}\label{subsec:7.2}\label{sec:7.2}

When a local version or resource fact changes by $\Delta$, the runtime follows dependency edges to compute $\field{Impact}(\Delta)$, then checks the relevant constraints and interface judgments in that set. Shared resources and transitive dependencies may propagate the effect to other relationships. Validation provides conflict reports to the affected participants. Obligation changes still follow each relationship's normal commitment or migration procedures.

The reference implementation uses a conservative closure that includes potentially affected nodes to reduce omissions. Evaluation measures conflict judgments and the tightness of the affected set separately. Lower precision indicates that more actually unaffected nodes were included. E10 reports missed impacts, precision, and validation workload to characterize this tradeoff.

Contract networks organize collaboration across multiple relationships. The current evaluation covers preset composition primitives and changes. Deeper dependencies, asynchronous cross-enterprise transactions, and conflicts between autonomous strategies lie outside the current evaluation scope. Composition validation and single-relationship normative-state checking have separate definitions and evaluations.

\section{Reference Implementation and Adoption Conditions}\label{sec:8}

\subsection{Runtime components and data interfaces}\label{subsec:8.1}\label{sec:8.1}

The reference architecture connects candidate generation and negotiation, domain-constraint validation, contract and event stores, monitoring, evidence and counterfactual analysis, experience repositories, and dependency validation. These components exchange agreements and grounds using relationship, version, and commitment references. The controlled platform implements the corresponding methods experimentally. Its base normative kernel supports registration of existing commitments, conditional performance, version dispositions, evidence attachment, and authorized closure. The running example describes requirements of the complete framework. Transactions for activating new obligations and obtaining multiparty closure signatures are not exposed by the base reference entry points.

The formation and evolution protocols cross three implementation regions. Agent software performs invitation, proposal search, counterproposal, and repair selection. Enterprise interfaces return permission, disclosure, confirmation, and factual verdicts. The CC runtime validates candidates and performs activation or migration writes. The base kernel receives identified participants and structured candidates at this boundary; it does not provide general peer discovery, dynamic coalition formation, a universal negotiation strategy, or a cross-replica consensus service. Those mechanisms can vary while preserving the protocol actions and normative preconditions in Sections~\ref{subsec:5.1} and~\ref{subsec:5.5}.

The full-lifecycle study in Section~\ref{subsec:10.8} uses an experimental extension that instantiates atomic new-obligation activation and multi-party closure as switchable guards. This configuration supports controlled ablation of those paths. It remains an experimental harness rather than a completed production interface, so its evidence is reported separately from the base-kernel regressions.

Section~\ref{subsec:10.10} evaluates two implementation configurations through public candidate, authorization, confirmation, activation, migration, and closure interfaces, with trusted simulated enterprise and world services. They share lower-level machinery and automate specified world-fulfillment transitions.

The later developmental implementation adds atomic successor establishment, business-scope checks before confirmation issuance, source-preserving performance allocations, and multi-party closure scheduling. It retains separate enterprise identities, observations, and conversation histories. Its read-only preparation tool translates explicit business choices into typed candidates; controlled disclosure, obligation-targeted payments, and shipment-balance checks support execution. Section~\ref{subsec:10.11} reports its evaluated configuration and online path.

Adapters receive A2A messages or tool results, map tasks and context to \rid, and normalize untrusted candidates into actions awaiting adjudication. They call $\delta$ after the necessary external judgments are available. Message transport, tool invocation, and physical enterprise execution are outside the relationship state machine. Production deployment also requires enterprise governance of identity and factual sources.

Adapters retain invitations, responses, judgments, candidates, confirmations, and invalidations in candidate or audit stores keyed by a provisional \rid{} and candidate digest. These records carry pending-version status until activation, as specified in Definition D5. The richer contribution types in Section~\ref{subsec:5.3} describe adapter responsibilities beyond the historical online stacks' typed negotiation support. A coordinator, transactional database, permissioned ledger, or distributed commit service can realize the logical atomic-write requirement across enterprises.

\begingroup
\small
\setlength{\tabcolsep}{4pt}
\renewcommand{\arraystretch}{1.18}
\setcounter{table}{6}
\begin{table}[htbp]
\caption{Core collections and protocol audit records in the runtime specification. Section~\ref{subsec:8.1} states implementation coverage.}\label{tab:7}
\centering
\small
\begin{tabular}{@{}>{\RaggedRight\arraybackslash}p{0.24\dimexpr\linewidth-4\tabcolsep\relax}>{\RaggedRight\arraybackslash}p{0.36\dimexpr\linewidth-4\tabcolsep\relax}>{\RaggedRight\arraybackslash}p{0.40\dimexpr\linewidth-4\tabcolsep\relax}@{}}
\toprule
\textbf{Collection and key} & \textbf{Stored content} & \textbf{Consistency requirement} \\
\midrule
Relationship: rid & Current version, q, revision, and broad participant scope & Stable task mapping; compare-and-swap version \\
ContractVersion: rid/v & Predecessor, frozen enterprise--role bindings, terms, commitments, rules, and $\mu$ & Atomic version/$\mu$ commit; no overwriting \\
Commitment: rid/v/kid & Frozen commitment, authority, and evidence requirements & State projected from appended events \\
EvidenceEvent: eid & Version/obligation binding and registration/verification references & Retain sources; separate attachment, verification, and closure \\
Candidate/audit record: rid/digest/event & Invitation, judgment, conditional position, pending reply, candidate, confirmation, refusal, and invalidation & Positions and pending replies supply no consent; retain superseded confirmations and bind activation to one digest \\
\bottomrule
\end{tabular}
\end{table}
\endgroup

\subsection{Persistence, idempotency, and rejection feedback}\label{subsec:8.2}\label{sec:8.2}

Normative events use idempotency keys composed of fields such as issuer, message ID, action, version, and candidate digest. Duplicate messages return the previous result. Stale replies and concurrent activation are adjudicated against the current version. A rejected confirmation request does not by itself exhaust a later confirmation for a corrected digest or version unless an enterprise policy says so. New versions, disposition mappings, and migration events are committed atomically. Recovery reconstructs active obligations and evidence mappings from immutable versions and appended events. Section~\ref{subsec:10.6} reports checks under changes of carrier.

Agent-facing action interfaces need types, permitted values, and repair hints consistent with disclosure policy. The adjacent-configuration interface primarily returns acceptance and rejection reasons. Some closure paths also return missing items. Its renegotiation schema names the fields and gives limited guidance on the five base disposition kinds and valid $\mu$ mappings.

The base kernel derives migration and closure scopes from the established ledger. Both \code{perform} and fulfillment dispositions in $\mu$ check antecedents, performance facts, and evidence; other terminal dispositions check their required sources, objects, and times. It rejects duplicate \kid{} values, retains pre-migration grounds, and queries world-side credential registrations. Its migration entry point returns \code{unsupported_new_commitment} for obligations outside the ledger: separately validated new-obligation activation is outside this base transaction set. Section~\ref{subsec:10.6} reports its regressions and controls.

The later successor transaction atomically combines old-claim disposition, new-\kid{} establishment, and version switching, using \code{superseded} records and predecessor links. Closure checks that the linked successor and disposition grounds are resolved. Prior delivery receives a separate allocation to the successor's total target: a 70-unit target with 20 units credited leaves 50 units to deliver. This extension implements the replacement semantics of Section~\ref{subsec:4.5} and is evaluated separately from the base interface.

Production integration must also make enterprise execution systems and counterparties process collaboration outcomes against the effective normative state. A message claiming that agreement has been reached must be checked against the current contract state. Enterprise infrastructure provides replica consistency, key governance, and source authentication. End-to-end message bypass is a separate adoption check.

\subsection{Controlled industrial worlds and reproducible experiments}\label{subsec:8.3}\label{sec:8.3}

The platform separates the business environment, method-visible state, and evaluator ground truth. Fault injection, dual clocks, checkpoint/fork, replay, and finite budgets control experiments. The procurement domain supplies events for supply, orders, certification, and logistics. MES/QMS simulators maintain manufacturing queues, rates, quality, and equipment events. World facts, participant observations, active disclosures, and runtime evidence have separate representations. Agents cannot read evaluator state. Preset policies and trusted simulated interfaces adjudicate action and disclosure authority. The reported experiments do not evaluate asynchronous enterprise decisions or their waiting costs; enterprise-internal approval procedures remain outside the model.

The online main panel retains original replies, candidate actions, world events, feedback, and state projections for offline recomputation of Tables~\ref{tab:17}--\ref{tab:20}. Kernel regressions, probes, bounded-path checks, and the small path diagnostic retain separate inputs and results. Section~\ref{sec:10.6} reports their endpoints, while Table~\ref{tab:audit-coverage} states which full-lifecycle properties the online records can adjudicate.

\subsection{Conditions for adopting the abstraction}\label{subsec:8.4}\label{sec:8.4}

Adopting collaboration contracts requires identifiable responsible participants, explicit action and disclosure policies, adjudicable commitment outcomes, accessible fact or evidence interfaces, and persistent versions and events. An enterprise can start with one cooperation process and map its terms to commitments, change conditions, and evidence requirements. Adapters can expose a production-capacity predicate from ERP or MES, a design-conformance verdict from computer-aided design or engineering (CAD/CAE), and a transport or inspection event from logistics or quality systems. Each adapter supplies the information authorized for the decision, identifies the relationship, and adds version and commitment references where they apply.

The interface contract must identify the issuer, governed subject, candidate or action scope, effective time, and verifiable grounds of each judgment. It must also represent pending and denied responses. Enterprises determine the people and procedures that supply these judgments. Deployment evaluation should measure interface cost, integration effort, and business benefit.

Table~\ref{tab:7a} gives an illustrative business mapping for adoption. Participating organizations retain the source records and determine issuer responsibilities and interface scope in their agreement.

\begingroup
\small
\setlength{\tabcolsep}{4pt}
\renewcommand{\arraystretch}{1.18}
\renewcommand{\thetable}{7a}
\renewcommand{\theHtable}{7a}
\begin{table}[htbp]
\caption{Illustrative integration of business objects with collaboration contracts.}\label{tab:7a}
\centering
\small
\begin{tabular}{@{}>{\RaggedRight\arraybackslash}p{0.27\dimexpr\linewidth-4\tabcolsep\relax}>{\RaggedRight\arraybackslash}p{0.35\dimexpr\linewidth-4\tabcolsep\relax}>{\RaggedRight\arraybackslash}p{0.38\dimexpr\linewidth-4\tabcolsep\relax}@{}}
\toprule
\textbf{Business object} & \textbf{CC object or external judgment} & \textbf{Illustrative authoritative source} \\
\midrule
Purchase order and outcome obligations & Contract version / commitment; order linked to rid & ERP/procurement system and contracting participants \\
Design or simulation result & Feasibility verdict or scoped evidence under an agreement version & CAD/CAE system and designated engineering issuer \\
Logistics arrival event & Evidence event referencing rid/v/kid & Logistics system and agreed issuer \\
Inspection report & Evidence event checked against evReq & QMS and agreed inspector \\
Enterprise action and disclosure permissions & External AuthorityVerdict and references & IAM/business authorization service \\
Rescheduling or supplier replacement & New version plus $\mu$; retain old-obligation lineage & Business amendment confirmed by authorized participants \\
\bottomrule
\end{tabular}
\end{table}
\endgroup
\setcounter{table}{7}

Table~\ref{tab:ops-spec} fixes the operational meaning of the five guards in the full-lifecycle study. O1 checks ledger-derived scope; O2 includes established conditional commitments before their performance antecedents are triggered (Section~\ref{subsec:4.5}).

\begingroup
\small
\setlength{\tabcolsep}{3.5pt}
\renewcommand{\arraystretch}{1.16}
\setcounter{table}{7}
\begin{table}[htbp]
\caption{Minimum operational specification of the experimental Full-CC guards.}\label{tab:ops-spec}
\centering
\small
\begin{tabular}{@{}>{\RaggedRight\arraybackslash}p{0.07\dimexpr\linewidth-8\tabcolsep\relax}>{\RaggedRight\arraybackslash}p{0.33\dimexpr\linewidth-8\tabcolsep\relax}>{\RaggedRight\arraybackslash}p{0.34\dimexpr\linewidth-8\tabcolsep\relax}>{\RaggedRight\arraybackslash}p{0.26\dimexpr\linewidth-8\tabcolsep\relax}@{}}
\toprule
\textbf{ID} & \textbf{Full-CC responsibility} & \textbf{Ablated behavior} & \textbf{Lifecycle point} \\
\midrule
O1 & \textbf{LedgerScope.} Derive the complete disposition and closure domain from the effective ledger & Permit candidate-supplied scope to omit established obligations & Migration and closure \\
O2 & \textbf{ConditionalInDomain.} Retain established conditional commitments before their antecedents trigger & Exclude an untriggered conditional commitment from the checked domain & Migration and closure \\
O3 & \textbf{AtomicNew.} Commit old-obligation dispositions and validated new obligations as one transaction, with rollback on failure & Apply stepwise writes without atomic rollback & Migration \\
O4 & \textbf{MultiSign.} Require affected parties to sign the relationship, version, and checked scope before closure & Permit closure to proceed with an incomplete signature set & Closure \\
O5 & \textbf{Composite.} Apply O1, O3, and O4 and accept evidence origins only from trusted registration & Remove O1, O3, and O4 and permit untrusted origin self-report & Migration, evidence, and closure \\
\bottomrule
\end{tabular}
\end{table}
\endgroup

\subsection{Threat model and security boundary}\label{subsec:8.5}\label{sec:8.5}

Cross-enterprise collaboration cannot assume an honest counterparty agent or treat fields generated by the local LLM as authority or facts. We treat messages, proposals, evidence submissions, and experience suggestions as potentially erroneous or malicious inputs. The trusted boundary comprises enterprise identity and permission interfaces, agreed factual issuers, and a normative kernel that executes rules and persists state. Attack objectives include changing obligations without consent, borrowing credentials from another collaboration, deleting prior responsibilities, and introducing unsupported experience into a new agreement.

Against credential fabrication and misbinding, issuance occurs in independent world or enterprise events. Submission queries registered credentials and checks source, object, version-specific use, and the complete obligation domain. Regressions and probes cover unregistered origins, cross-relationship credentials, and listed reference paths. They support rejection on these paths, without establishing the absence of bypasses across every encoding, adapter, cache, or concurrent entry point. Compromised issuers, leaked keys, hash collisions, and modified registry storage exceed the current trusted-input assumptions.

Selective disclosure permits enterprises to retain private information and also creates opportunities for strategic misrepresentation. Permission to withhold a reply and provision of a false guarantee require separate treatment. Missing necessary grounds leave a judgment unresolved. A decision-relevant guarantee should bind its issuer, scope, validity period, and responsibility. Mutual confirmation by two agents cannot substitute for required enterprise approval. If authorized parties or factual issuers collude to produce false grounds, complete references cannot expose the false fact; independent audits, dispute procedures, or other enterprise mechanisms are required.

Version checks, idempotency, and atomic commitment handle stale replies, repeated submissions, and competing proposals, but do not ensure availability. An attacker may repeatedly propose invalid changes, delay confirmation, or exhaust query budgets. Production deployments need resource limits, timeout and recovery rules per organization and relationship, and execution entry points that check effective state, preventing ordinary messages from bypassing rejected actions. Prompt injection, collusion, denial of service, and message bypass lie outside the current security evaluation scope.

\subsection{Connections to blockchain smart contracts and engineering tradeoffs}\label{subsec:8.6}\label{sec:8.6}

Blockchain smart contracts can host version registration, confirmation records, and state transitions. Collaboration contracts specify the business relationships, commitment lineage, individual dispositions, and evidence uses to be represented. The mechanisms can be combined, for example by placing version hashes and key confirmations on a chain while keeping private terms and original credentials in enterprise systems. On-chain execution still needs external facts about arrival, acceptance, and authority. Oracle authenticity, availability, and incentive compatibility are standard deployment concerns~\cite{r55}. A blockchain ledger therefore still requires explicit responsibilities for enterprise adjudication and external facts.

Deployment choices depend on enterprise trust and cooperation conditions. Shared databases and enterprise event stores require operational responsibilities, access controls, and dispute evidence procedures. Permissioned chains require membership governance and confirmation rules. Public chains also involve transaction fees and treatment of private business information. Ethereum, for example, meters computation and transactions using gas and charges fees~\cite{r56}. These are cost dimensions for evaluating implementations. We provide no on-chain comparison or evidence that a separately packaged contract runtime outperforms other implementations carrying equivalent rules.

The end-to-end cost of a collaboration includes proposal generation, domain solving, enterprise confirmation waits, fact or credential queries, persistent writes, and subsequent renegotiation. Integration costs also include term mapping, identity and authority adaptation, credential issuance, and dispute handling. The present measurements report model token consumption. Lifecycle latency and integration cost require measurements from the services that perform these tasks in a given deployment.

\section{Evaluation Design and Research Questions}\label{sec:9}

\subsection{Lifecycle questions and evaluation scope}\label{subsec:9.1}\label{sec:9.1}

\begingroup
\small
\setlength{\tabcolsep}{4pt}
\renewcommand{\arraystretch}{1.18}
\setcounter{table}{8}
\begin{table}[htbp]
\caption{Research questions, runtime mechanisms, and corresponding evaluations.}\label{tab:8}
\centering
\small
\begin{tabular}{@{}>{\RaggedRight\arraybackslash}p{0.27\dimexpr\linewidth-4\tabcolsep\relax}>{\RaggedRight\arraybackslash}p{0.35\dimexpr\linewidth-4\tabcolsep\relax}>{\RaggedRight\arraybackslash}p{0.38\dimexpr\linewidth-4\tabcolsep\relax}@{}}
\toprule
\textbf{Research question} & \textbf{Mechanisms to examine} & \textbf{Corresponding evaluation} \\
\midrule
MRQ-A: Formation and validity & Initiation, participation, selective disclosure, negotiation, validation, confirmation, activation, and effective commitments & Simulation E1/E2/E8 and T1; E0 N2/N6; online F1/F2/F3 \\
MRQ-B: Dynamic fulfillment and evolution & Monitoring, change impact, repair negotiation, version disposition, affected-party confirmation, and evidence coverage & Simulation E4/E7/E11; E0 N1/N3; online F4/F5 and full-lifecycle O1--O5; RA-NCT; targeted O1/O2 process audit and lawful-path diagnostics \\
MRQ-C: Evidence, accountability, and composition & Evidence-supported normative/causal analysis, contract dependencies, and impact closure & Simulation E5/E6/E10; E0 N4 \\
MRQ-D: Experience qualification and reuse & Source isolation, applicability, ownership, evidence, and usable yield & Simulation E9 and E8 $\times$ E9; E0 N5; online F6 \\
\bottomrule
\end{tabular}
\end{table}
\endgroup

MRQ-A--D in Section~\ref{subsec:2.3} address formation, fulfillment and evolution, evidence-supported responsibility and composition, and experience consumption. Table~\ref{tab:8} associates each question with mechanisms and evaluation materials. These check executability of specified rules, mechanism effects in controlled worlds, and effects of adjacent configurations on online outcomes. The formation and evolution protocols organize selected disclosures, validations, repairs, confirmations, migrations, and terminal checks across these studies. Each study retains its stated analysis unit; responsibility models, network changes, and experience consumption have their own units.

The evaluation examines simulated business effects, adjacent semantic configurations, individual runtime guards, process re-adjudicability, two implementations, and kernel write behavior. Table~\ref{tab:9} gives each study's scope and analysis unit. Replay and smoke checks supply path diagnostics.

A separate developmental question asks how three agents connect an entire formation--evolution--closure path through the public interfaces. Section~\ref{subsec:10.11} presents the adopted configuration and one observed path, with independently reconciled business facts. Development selection and the scope of this existence evidence are addressed in Section~\ref{subsec:12.2}.

\begingroup
\small
\setlength{\tabcolsep}{4pt}
\renewcommand{\arraystretch}{1.18}
\setcounter{table}{9}
\begin{table}[htbp]
\caption{Evaluation units and the questions they can address.}\label{tab:9}
\centering
\small
\begin{tabular}{@{}>{\RaggedRight\arraybackslash}p{0.24\dimexpr\linewidth-4\tabcolsep\relax}>{\RaggedRight\arraybackslash}p{0.34\dimexpr\linewidth-4\tabcolsep\relax}>{\RaggedRight\arraybackslash}p{0.42\dimexpr\linewidth-4\tabcolsep\relax}@{}}
\toprule
\textbf{Evidence source} & \textbf{Runtime and analysis unit} & \textbf{Question addressed} \\
\midrule
Two-domain simulation experiments & Cases, fault runs, or network changes of each experiment & Mechanism differences under specified worlds and rules \\
E0 minimal and bounded checks & Witnesses, transitions, and generated traces & Implementation behavior on generated inputs \\
Adjacent-configuration online panel (frozen Phase 3 same-stack configuration) & 380 cases; 760 online episodes & Successive semantic-configuration differences under that held-fixed stack and protocol; it predates the complete FP/EP specification \\
Full-lifecycle guard-ablation panel & Five paired 100-case panels; 1{,}000 online episodes & Full-CC experimental configuration versus guard-ablation differences in normatively correct terminal outcomes \\
Targeted O1/O2 process audit & 40 cases; 80 episodes; same-candidate re-adjudication & Process traceability, write integrity, and reachability diagnosis; no new effect estimate \\
Two-implementation comparison & 477 paired cases; 954 completed episodes; template-level primary contrast & Correct normative results and autonomous action reachability with corresponding public capabilities; no difference identified \\
Developmental online instance & One three-enterprise trajectory under a revised implementation & Formation through successor migration and closure, with route and deadline adjudication; existence evidence without a success-rate estimate \\
Current reference-kernel checks & 33 regressions; 11 probes/controls; 6 closure-precondition checks; current bounded paths & Specified normative actions, lawful scripted paths, and generator-reachable behavior \\
Interface-path diagnostics (frozen v2.5 reference interface) & 5 online samples; offline closure requests on saved terminal states & Action reachability under given events/budgets and acceptance of lawful terminal-state closure \\
\bottomrule
\end{tabular}
\end{table}
\endgroup

Table~\ref{tab:evidence-map} maps specification, implementation coverage, and online evidence. Its protocol rows summarize formation and evolution; the remaining rows identify individual mechanisms.

\begingroup
\raggedbottom
\footnotesize
\setlength{\tabcolsep}{3pt}
\renewcommand{\arraystretch}{1.08}
\setlength{\LTleft}{0pt}
\setlength{\LTright}{0pt}
\setcounter{table}{10}
\begin{longtable}{@{}>{\RaggedRight\arraybackslash}p{0.20\textwidth}>{\RaggedRight\arraybackslash}p{0.22\textwidth}>{\RaggedRight\arraybackslash}p{0.32\textwidth}>{\RaggedRight\arraybackslash}p{0.22\textwidth}@{}}
\multicolumn{4}{@{}p{\textwidth}@{}}{\label{tab:evidence-map}\small\textbf{Table \thetable:} Model scope, implemented subsets, and online evidence. F1--F6 refer to the frozen adjacent-configuration stack; O1--O5 refer to the experimental Full-CC configuration.}\\[6pt]
\toprule
\textbf{Object or semantics} & \textbf{Normative/design scope} & \textbf{Implementation coverage} & \textbf{Online evidence} \\
\midrule
\endfirsthead
\multicolumn{4}{@{}l}{\footnotesize\textit{Table \thetable{} continued}}\\
\toprule
\textbf{Object or semantics} & \textbf{Normative/design scope} & \textbf{Implementation coverage} & \textbf{Online evidence} \\
\midrule
\endhead
\midrule
\multicolumn{4}{r@{}}{\footnotesize\textit{Continued on next page}}\\
\endfoot
\bottomrule
\endlastfoot
Formation protocol & FP1--FP6: discovery, proposals, confirmation, activation & Candidate validation and activation in the kernel; separate enterprise sessions in the online instance; discovery is an extension point & F2/F3 conditional differences; 477-pair activation A:19/B:77; one joined path in Section~\ref{subsec:10.11} \\
Evolution protocol & EP1--EP6: complete disposition, consent, atomic migration & Base kernel continues existing obligations; successor implementation adds replacement, new obligations, and credit & O1 challenge contrast; one successor-to-closure path; broader convergence remains open \\
Relationship identity & D1/D2/D5: persistent relationship and party binding & N6 context/carrier checks; version-specific bindings & F1: corrected threshold not met \\
Commitments & D3: establishment, antecedents, fulfillment & Antecedent and ground checks; separate new-obligation extension & F2: conditional difference; O2: unresolved reversal; O3: floor \\
Current validity & I1: authority and applicable constraints & N2/N6 and authorization checks & F3: conditional difference \\
Version disposition & D2/$\mu$/I2: ledger-derived domain and lineage & N3; complete $\mu$ and grounded continuation & F4: no recovery increment; O1: 74\% versus 32\% on challenges \\
Evidence and closure & D4/I3/I4: provenance and scoped applications & Source checks and closure; successor credit allocation in later implementation & F5: nondiagnostic; O4: no corrected difference; online instance reconciles world facts \\
Experience & Applicability, ownership, evidence qualification & N5 and qualification gates & F6: conditional difference; process-learning extension remains open \\
Responsibility / networks & Derived attribution and dependency revalidation & N4 plus domain-specific simulation methods & E5/E6/E10 simulation results; no online family \\
Implementation comparison & Public-interface coordination and case-correct outcomes & Two internal configurations with shared lower-level machinery & 477 pairs; CNR 53:54; target attempts 27:38; Section~\ref{subsec:10.10} \\
Online lifecycle instance & Formation, adaptation, performance, closure & Successor implementation, enterprise boundary services, causal world & One agent-directed trace; 68 instantiated checks; configuration-specific existence evidence \\
\end{longtable}
\endgroup

\subsection{Baselines, metrics, and fairness}\label{subsec:9.2}\label{sec:9.2}

Metrics follow the business question. Table~\ref{tab:metrics} defines the recurrent endpoints and their interpretation boundaries.

\begingroup
\small
\setlength{\tabcolsep}{4pt}
\renewcommand{\arraystretch}{1.16}
\setcounter{table}{11}
\begin{table}[htbp]
\caption{Evaluation glossary: endpoints, abbreviations, and reading conventions.}\label{tab:metrics}
\centering
\small
\begin{tabular}{@{}>{\RaggedRight\arraybackslash}p{0.14\dimexpr\linewidth-4\tabcolsep\relax}>{\RaggedRight\arraybackslash}p{0.40\dimexpr\linewidth-4\tabcolsep\relax}>{\RaggedRight\arraybackslash}p{0.46\dimexpr\linewidth-4\tabcolsep\relax}@{}}
\toprule
\textbf{Term} & \textbf{Meaning} & \textbf{Reading convention} \\
\midrule
SCR & Safe completion rate & Violation-free completion divided by feasible episodes \\
EBCR & Evidence-backed completion rate & Positive fulfillment adjudications by the contract Monitor divided by feasible episodes; compare with SCR within that implementation \\
GT correctness & Agreement with evaluator ground truth & The specified case judgment defines correctness \\
NCT & Normatively correct terminal outcome & Combines agent progress, guard feedback, and adjudication under the evaluated protocol \\
RA-NCT & Record-audited NCT & Properties are pass, violation, or undetermined according to retained evidence \\
CNR & Correct normative result rate & Legal completion, justified exit, or illegal-target rejection, subject to audit; denominator: all retained cases \\
Target reachability & At least one attempt at the case-specific target action & Absence of that attempt is \code{NOT_ENGAGED}; proposals and other negotiation may still occur \\
CFS / WDC & Case-correct behavior over all cases / weighted disclosure cost & Feasible delivery or infeasible noncommitment; disclosure weights appear in Section~\ref{subsec:10.1} \\
Safe\-Reuse\-Yield & Qualified coverage multiplied by reuse precision & Actual reuse records; Section~\ref{subsec:10.7} \\
FP / EP & Formation protocol / evolution protocol & Step identifiers FP1--FP6 and EP1--EP6 \\
RL / RLA & Relationship-layer configuration / relationship-layer ablation & RL0--RL6 denote the adjacent semantic configurations \\
\bottomrule
\end{tabular}
\end{table}
\endgroup

Simulation experiments compare workflows, a constrained centralized approach, ordinary MAS, a policy guard, a static contract, and an adaptive runtime. They examine combinations of construction, prevention, monitoring, and remedy capabilities. The adjacent-configuration online panel holds the model, world, and budget fixed and compares successive semantic configurations in one stack. The full-lifecycle panel likewise holds the agent stack fixed and compares Full-CC with four single-guard ablations and one composite ablation. These two ablation studies estimate differences under their particular implementation and protocol. Targeted checks of the current kernel report their own inputs and endpoints.

The two-implementation comparison evaluates a CC runtime and a modular implementation with corresponding public capabilities, rather than switching individual semantics off. Section~\ref{subsec:10.10} specifies its frozen interface, sample, endpoint, and analysis. Shared project dependencies limit implementation independence.

The developmental trace uses the joint business and normative endpoint specified in Section~\ref{subsec:10.11}, including applicable route and deadline conditions.

Other experiment-specific metrics measure construction, disclosure, dangerous false acceptance, recovery, responsibility attribution, and experience consumption. Policy guard B4 has no contract Monitor, making its EBCR structurally zero. Within B6, SCR and EBCR describe the gap between completion and adjudication; cross-arm completion comparisons use SCR. The metrics retain their own denominators and are never pooled into a single effect.

\subsection{Online adjacent-configuration comparison}\label{subsec:9.3}\label{sec:9.3}

Relationship-layer ablation (RLA) compares adjacent configurations of the same agent stack. Starting from messages and tasks, RL0--RL6 successively add relationship identity, atomic commitments, current validity, version dispositions, evidence closure, and experience qualification. Table~\ref{tab:16} maps configurations to business capabilities. The model generates candidates online. Configurations determine visible state and feedback, so the comparison estimates their runtime differences under this protocol.

The panel contains 380 cases, each executed in two adjacent arms, yielding 760 online episodes. F1, F2, F3, F4, and F6 contain 60 cases each; F5 contains 80. Each 60-case family contains 30 cases from each simulated domain. F4 includes 20 recoverable, 20 unrecoverable, and 20 fault-free cases. F5 includes 20 cases each with complete, missing, fabricated, and stale evidence. The generator varies terms, deadlines, obligations, authority, and evidence content. Content fingerprints separate the evaluation panel from development samples.

The internal GPUStack service reports the identifier \code{qwen3.6-35b-a3b-fp8}. Both arms request temperature~0, \code{max_tokens=2048} per response, and \code{enable_thinking=False}. These fixed client settings define the operating point for the paired comparison. The requested temperature of 0 differs from the model card's recommended temperature of 0.7 for non-thinking mode~\cite{r61}. The client uses a 180-second timeout, at most three retries with linear $2(n+1)$-second backoff, and standard TLS verification. Each case permits at most 24 rounds and 30,000 tokens, with a global budget of three million tokens. Budgets are checked at runtime boundaries. The 760 records total 1,803,377 tokens. Arm order is randomized by \code{crc32(RUN_SEED:cid)}, and resume processing deduplicates by \code{(cid,arm)}.

\subsection{Online endpoints and statistical interpretation}\label{subsec:9.4}\label{sec:9.4}

For the adjacent-configuration panel, a correct case outcome requires the target event to occur and satisfy the specified judgment. Failure to reach it remains in the primary endpoint, with action attempts reported separately. Correctness therefore reflects both business progress and normative adjudication. F4 and F5 locate outcomes at \code{close}; other families use their corresponding target events. F5 excludes 11 pairs in which at least one arm lacks a stale-version trap, retaining 69 pairs within F5. The resulting analysis set contains 369 pairs across the six families. Defects in F5's version chain and scoring conditions make its results descriptive only. Appendix~\ref{app:online-analysis} gives the exclusion rule and analysis set.

Primary tests use exact two-sided McNemar tests and a six-family Holm correction. Table~\ref{tab:17} reports correct counts, favorable and reverse discordant pairs, differences, and interpretation. The subset in which both arms reach the target action describes adjudication conditional on that behavioral selection and cannot replace the full-case analysis. Experience qualification also reports qualified coverage, precision, and safe yield from actual reuse records.

Let $b$ count pairs in which the lower configuration is wrong and the higher configuration correct, $c$ count reverse pairs, and $n$ be the included pairs. The correctness difference is $(b-c)/n$. F5's test value remains in the six-family Holm calculation. Its interpretation is uniformly ``retained in the Holm family; no mechanism interpretation.''

The full-lifecycle guard comparison in Section~\ref{subsec:10.8} uses normatively correct terminal outcome as its endpoint. The main presentation uses two-sided exact McNemar tests with five-panel Holm correction within case class; the archived directional analysis applies its own correction to one-sided tests. Section~\ref{subsec:12.2} records the limitation on the test-sidedness provenance.

\subsection{Targeted O1/O2 process audit}\label{subsec:9.5}\label{sec:9.5}

The process audit uses 40 newly generated cases and 80 online episodes across O1 and O2, with lawful and challenge cases executed under Full-CC and the corresponding ablation. It retains complete candidate, effective-ledger, runtime-scope, adjudication, state-difference, and terminal records. Each parsable migration candidate is also applied, without another model call, to two deep copies of the same source-episode checkpoint. The copies share the same initial normative state, effective ledger, stored interface-derived approvals and attestations, and candidate; the only deliberate variation is the runtime configuration (Full-CC versus the corresponding ablation). The business world is not advanced during this paired re-adjudication. The audit evaluates process re-adjudicability, write integrity, and reachability; its terminal counts are descriptive, are not combined with the 1{,}000-episode study, and do not provide a new effect estimate.

\subsection{Implementation checks of the reference kernel}\label{subsec:9.6}\label{sec:9.6}

Targeted regressions use positive cases, rejection cases, and lawful controls to check performance antecedents, disposition grounds, complete commitment registration, credential origin, and normative state after rejection. Fixtures establish world judgments and issuance beforehand; submission only queries records. Probes satisfy upstream structural requirements before reaching the intended guard. A lawful control checks whether scripted actions can commit, separately from an agent's ability to construct the complete business chain autonomously. The adjacent-configuration stack and current reference kernel are evaluated as separate artifacts.

\section{Controlled Evaluation of Lifecycle Mechanisms}\label{sec:10}

Evaluation follows the collaboration-contract lifecycle. Formation tests disclosure, validation, and negotiation; fulfillment tests recovery and evidence coverage; subsequent analysis tests responsibility, composition, and experience. We then report reference-kernel checks, the adjacent-configuration online panel, interface-path diagnostics, and the full-lifecycle guard ablation. Each analysis retains its own unit and endpoint.

\subsection{Selective disclosure and contract construction}\label{subsec:10.1}\label{sec:10.1}

\begingroup
\small
\setlength{\tabcolsep}{4pt}
\renewcommand{\arraystretch}{1.18}
\setcounter{table}{12}
\begin{table}[htbp]
\caption{Simulated observations of selective disclosure and authoritative adjudication.}\label{tab:10}
\centering
\small
\begin{tabular}{@{}>{\RaggedRight\arraybackslash}p{0.26\dimexpr\linewidth-6\tabcolsep\relax}>{\RaggedRight\arraybackslash}p{0.21\dimexpr\linewidth-6\tabcolsep\relax}>{\RaggedRight\arraybackslash}p{0.22\dimexpr\linewidth-6\tabcolsep\relax}>{\RaggedRight\arraybackslash}p{0.31\dimexpr\linewidth-6\tabcolsep\relax}@{}}
\toprule
\textbf{Experiment and method} & \textbf{Case scope} & \textbf{Main readings} & \textbf{Interpretation} \\
\midrule
E1 exact disclosure D0 & 200 procurement cases & CFS 1.00; WDC 12 & Higher cost of disclosing raw values \\
E1 interval disclosure D1 & Same cases & CFS 0.93; WDC 7 & Boundary intervals retain uncertainty \\
E1 predicate statements D2 & Same cases & CFS 1.00; WDC 3 & Trusted interface returns decision predicates \\
E1 progressive refinement & Same cases & CFS 1.00; WDC 8.31 & Increase permitted granularity when needed \\
E2 planning/adjudication allocation & D1; LLM planning vs. interface adjudication & Selection agreement 0.435 / 0.965 & Language proposals with computable adjudication \\
\bottomrule
\end{tabular}
\end{table}
\endgroup

CFS measures case-correct behavior over all cases: delivery in a feasible case or no commitment in an infeasible case. WDC sums synthetic sensitivity weights for permitted disclosures, charging only incremental refinement per organization and field. The weights are 3 for exact price, 1.5 for a price band, 1 for exact lead time, 2 for exact capacity, 1 for a capacity bin, 0.25 for a feasibility predicate, and 1 for a coarse value band. They express the experiment's cost assumptions. Across 200 procurement cases, exact disclosure and predicate statements both achieve CFS 1.00, with WDC 12 and 3. Interval disclosure achieves CFS 0.93 and WDC 7; progressive refinement achieves CFS 1.00 and WDC 8.31. A trusted interface can supply decision-relevant predicates, while intervals spanning a judgment boundary require refinement. This supports choosing disclosure granularity according to formation constraints.

E2 compares LLM buyer planning with an allocation in which interfaces perform computable adjudication. Under coarse disclosure, selection agreement is 0.435 and 0.965, respectively; CFS judgment agreement is 0.485 and 0.995. Model-generated candidates and interface adjudication of permissions, resources, and timing allow negotiation under selective disclosure with explicit decision grounds.

\begingroup
\small
\setlength{\tabcolsep}{4pt}
\renewcommand{\arraystretch}{1.18}
\setcounter{table}{13}
\begin{table}[htbp]
\caption{Case outcomes under validation and negotiation without history.}\label{tab:11}
\centering
\small
\begin{tabular}{@{}>{\RaggedRight\arraybackslash}p{0.30\dimexpr\linewidth-6\tabcolsep\relax}>{\RaggedRight\arraybackslash}p{0.15\dimexpr\linewidth-6\tabcolsep\relax}>{\RaggedRight\arraybackslash}p{0.19\dimexpr\linewidth-6\tabcolsep\relax}>{\RaggedRight\arraybackslash}p{0.36\dimexpr\linewidth-6\tabcolsep\relax}@{}}
\toprule
\textbf{Method} & \textbf{Primary cases} & \textbf{Case-level GT correctness} & \textbf{Primary role} \\
\midrule
LLM one-shot generation & 300 & 0.653 & Construct candidates \\
Deductive validation after generation & Same cases & 0.653 & Dangerous false acceptance in unrepairable subset: 9/69 to 0/69 \\
One repair & Same cases & 0.757 & Repair local conflicts \\
Multiround negotiation without history & Same cases & 0.793 & Coordinate combined constraints and participant repairs \\
\bottomrule
\end{tabular}
\end{table}
\endgroup

The construction experiment compares one-shot generation, post-generation validation, one repair, and multiround negotiation on the same 300 cases. It covers quantity, timing, budget, authority, certification, disclosure, responsibility, evidence, and composition constraints. GT correctness measures final case outcomes: feasible or repairable cases must commit a valid contract, and genuinely unrepairable cases must produce a correct \code{INFEASIBLE_CONFIRMED} judgment. \code{REJECTED}, which rejects one invalid candidate, does not complete infeasibility adjudication. Validation reduces dangerous false acceptance in the unrepairable subset from 9/69 (13.04\%) to 0/69, while both arms retain 196/300 correct outcomes (0.653). One repair achieves 227/300 (0.757), and multiround negotiation 238/300 (0.793).

Case records show that validation changes 94 invalid \code{COMMITTED} outcomes to \code{REJECTED}: 85 remain repairable and nine are genuinely unrepairable. The 136 valid commitments and 60 correct infeasibility judgments remain unchanged, accounting for 196 correct outcomes. Ten incorrect infeasibility judgments also remain unchanged. Validation blocks dangerous commitments, but local rejection does not yet complete the repair or infeasibility judgment required by these cases. Overall correctness thus remains unchanged.

An independent 72-case T1 holdout varies where minimum-quantity and lawful \code{reduce_qty} semantics enter the process. Treating the target quantity as a hard minimum yields a first-round false-infeasibility rate of 0.569 and final correctness 0.431. Providing the minimum acceptable quantity and lawful reduction semantics during candidate construction reduces false infeasibility to zero and raises final correctness to 0.986. Providing the same hints only in the repair round leaves the baseline result unchanged. Formation-stage rule expression affects whether an agent enters later negotiation; later repair cannot recover a candidate chain terminated early.

\subsection{Dynamic fulfillment, recovery, and domain dynamics}\label{subsec:10.2}\label{sec:10.2}

\begingroup
\small
\setlength{\tabcolsep}{4pt}
\renewcommand{\arraystretch}{1.18}
\setcounter{table}{14}
\begin{table}[H]
\caption{Conditional mechanism results during dynamic execution.}\label{tab:12}
\centering
\small
\begin{tabular}{@{}>{\RaggedRight\arraybackslash}p{0.22\dimexpr\linewidth-6\tabcolsep\relax}>{\RaggedRight\arraybackslash}p{0.30\dimexpr\linewidth-6\tabcolsep\relax}>{\RaggedRight\arraybackslash}p{0.21\dimexpr\linewidth-6\tabcolsep\relax}>{\RaggedRight\arraybackslash}p{0.27\dimexpr\linewidth-6\tabcolsep\relax}@{}}
\toprule
\textbf{Experiment} & \textbf{Comparison and scope} & \textbf{Readings} & \textbf{Result scope} \\
\midrule
E4 fault recovery & 200 cases; 17,964 primary runs; static vs. adaptive & Recovery 0.522 / 0.726 & Benefits depend on fault type \\
E7 high-complexity completion & Procurement; policy guard B4 vs. adaptive B6 & SCR 0.715 / 0.524 & Policy guard has higher completion here \\
E7 completion and adjudication & Within B6 & SCR 0.524; EBCR 0.304 & Evidence-adjudication gap on completion paths \\
E11 manufacturing HIGH & Manufacturing; B4 vs. B6 & SCR 0.473 / 0.747 & Ranking depends on simulated dynamics \\
\bottomrule
\end{tabular}
\end{table}
\endgroup

The recovery experiment contains 200 cases and 17,964 primary runs. Recovery is 0.726 for the adaptive runtime, with a descriptive 95\% Wilson interval [0.717, 0.735], and 0.522 for the static contract, with interval [0.511, 0.532]. These run-level intervals summarize the simulated case--fault combinations; repeated conditions on the same 200 cases limit their interpretation as independent population samples. Differences are larger for participant exit and quality failure; capacity reduction and authority revocation show no difference. Recovery requires exception terms, resources, and necessary consent to jointly provide a lawful remedy path.

In high-complexity procurement, policy guard B4 has SCR 0.715 and adaptive runtime B6 has SCR 0.524. B6's positive fulfillment adjudication rate, EBCR, is 0.304, leaving a 22.0-percentage-point gap. The available aggregates establish a gap between business completion and the prescribed adjudication. They do not identify whether it arises from rejection of applicable old evidence, issuance timing, evidence requirements, or Monitor logic. B4 has no such Monitor and is excluded from EBCR quality ranking.

The 22.0-point adjudication gap motivates a process-level question: when are applicable credentials issued, attached, and recognized under the effective version? Completion events, version changes, evidence requirements, and closure opportunities need to be observed jointly to distinguish these contributions.

\begin{figure}[!htbp]
\centering
\includegraphics[width=\linewidth]{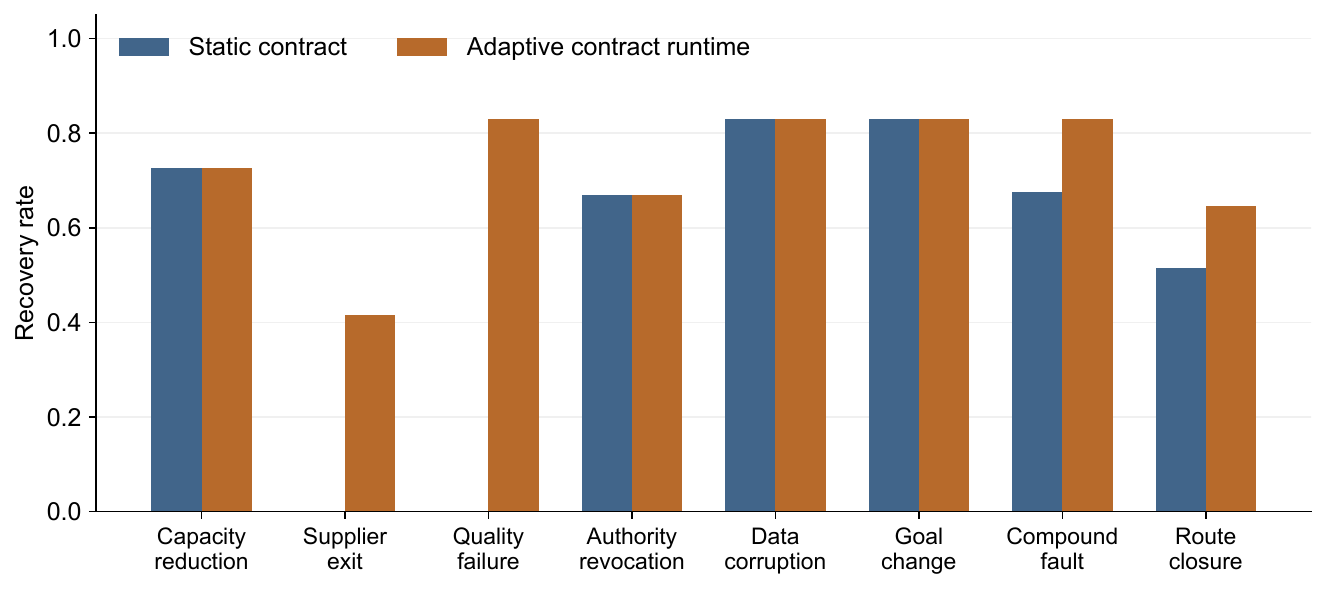}
\caption{E4 recovery rates by fault type for static and adaptive contract runtimes.}\label{fig:6}
\end{figure}

In high-complexity manufacturing, SCR is 0.747 for B6 and 0.473 for B4. At medium complexity, B4's safety margins and budget screening also lead to late or \code{NO_PLAN} paths. The reversal relative to procurement can be interpreted through business pressures: prevention is valuable when formation mistakes or irreversible authority risks dominate; runtime remedies have more opportunities when queues, setup, rates, and equipment conditions can be adjusted. This interpretation follows the simulated rules and does not isolate the causal contribution of each factor. Figure~\ref{fig:6} further shows that remedy benefits concentrate in particular fault types.

\FloatBarrier
\subsection{Evidence-supported responsibility analysis}\label{subsec:10.3}\label{sec:10.3}

\begingroup
\small
\setlength{\tabcolsep}{4pt}
\renewcommand{\arraystretch}{1.18}
\setcounter{table}{15}
\begin{table}[htbp]
\caption{Evidence-supported responsibility analysis and external-shock diagnostics.}\label{tab:13}
\centering
\small
\begin{tabular}{@{}>{\RaggedRight\arraybackslash}p{0.18\dimexpr\linewidth-6\tabcolsep\relax}>{\RaggedRight\arraybackslash}p{0.28\dimexpr\linewidth-6\tabcolsep\relax}>{\RaggedRight\arraybackslash}p{0.28\dimexpr\linewidth-6\tabcolsep\relax}>{\RaggedRight\arraybackslash}p{0.26\dimexpr\linewidth-6\tabcolsep\relax}@{}}
\toprule
\textbf{Experiment} & \textbf{Scenario and comparison} & \textbf{Observation} & \textbf{Interpretation boundary} \\
\midrule
E5 missing evidence & Log interpretation vs. contract evidence/counterfactual pipeline & Missing 0/10/30\%: logs 1.00/0.80/0.65; pipeline 1.00/1.00/1.00 & Specified faults, causal model, and evidence paths \\
E5 joint causes & Jointly necessary F04/F12 scenarios & Last-log attribution 0; counterfactual analysis finds joint causes & Joint removal changes the outcome \\
E6 external shock & Force majeure and excuse rules under a specified strong-shock slice & Strong-shock misattribution 0; recovery 0.83 & Targeted regression of the specified force-majeure and excuse rules \\
\bottomrule
\end{tabular}
\end{table}
\endgroup

E5 compares log interpretation with an evidence-and-counterfactual pipeline under specified causal scenarios and normative interpretations. At evidence-missing rates of 0\%, 10\%, and 30\%, log attribution accuracy is 1.00, 0.80, and 0.65. The pipeline achieves 1.00 on each of these scenario sets. The log baseline attributes by the recorded event pattern, whereas the pipeline uses the specified causal model and obligation references; this contrast evaluates those defined procedures. Cases with jointly necessary causes require removing a combination of causes to change the outcome. Responsibility analysis must therefore retrieve effective obligations and exception rules as well as outcome-level causal conditions. The domain model determines the scope of this judgment.

External-shock diagnostics combine force-majeure event checks with an \code{external-no-liability} interpretation. E6 evaluates the adopted force-majeure and excuse rule on a targeted strong-shock slice. In the specified strong-shock slice, supplier misattribution is zero and recovery is 0.83. It checks the resulting interpretation on that slice rather than external generalization. The domain model determines the causal effect of events such as road closures; risk and excuse terms determine who bears their consequences.

\subsection{Contract networks and incremental validation}\label{subsec:10.4}\label{sec:10.4}

\begingroup
\small
\setlength{\tabcolsep}{4pt}
\renewcommand{\arraystretch}{1.18}
\setcounter{table}{16}
\begin{table}[htbp]
\caption{E10 impact-closure metrics under successive changes.}\label{tab:14}
\centering
\small
\begin{tabular}{@{}>{\RaggedRight\arraybackslash}p{0.27\dimexpr\linewidth-4\tabcolsep\relax}>{\RaggedRight\arraybackslash}p{0.35\dimexpr\linewidth-4\tabcolsep\relax}>{\RaggedRight\arraybackslash}p{0.38\dimexpr\linewidth-4\tabcolsep\relax}@{}}
\toprule
\textbf{Scope or metric} & \textbf{Reading} & \textbf{Meaning} \\
\midrule
Successive random changes & 500 mutations & Preset relationship primitives and conflict families \\
Affected-set precision & 0.39 & Conservative expansion includes unaffected nodes \\
Validation work saved & WorkSaved 0.28 & Relative to full revalidation \\
Missed impacts and recall & FN 0; recall 1.00 & No missed nodes under complete GT in this test \\
Additional included nodes & 937 & Transitive neighbors and shared-resource closure \\
\bottomrule
\end{tabular}
\end{table}
\endgroup

The composition experiment applies 500 successive random changes involving resource, timing, authority, commitment, disclosure, responsibility/remedy, and evidence dependencies. Multilabel truth records concurrent effects of each change. The runtime computes a conservative closure along dependency edges and revalidates related nodes, comparing affected sets and workload with full validation.

Impact closure yields FN 0, recall 1.00, precision 0.39, 937 additional nodes, and WorkSaved 0.28. Conservative expansion retains affected nodes in this test set while adding checks. Precision 0.39 means that many selected nodes are conservatively included; WorkSaved 0.28 quantifies the resulting 28\% workload reduction relative to full validation. Coverage and selectivity therefore need to be assessed together. These results characterize coverage and workload under the preset dependencies and changes.

\subsection{Qualified experience and construction benefits}\label{subsec:10.5}\label{sec:10.5}

Experience extraction produces candidate assets from 240 historical traces and checks structured fields, reference existence, ownership, and free-text support separately. Structured checks pass at rate 1.0 in the preset set; 10.4\% of free-text explanations lack complete support. An eight-arm reuse comparison shows distinct roles for structured retrieval, owner constraints, and evidence constraints in ownership identification and qualified reuse. The no-memory configuration has zero coverage.

The consumption experiment uses seed 31337 to generate 300 independent cases with zero content-fingerprint overlap with E8 Main. It compares no history, known-contaminated synthetic history, ungated retrieval of real experience, and \code{gated_v2}. The last combines applicability predicates, ownership, evidence qualification, and an interpretation framework to evaluate the overall benefit of qualified experience in candidate construction. Synthetic history is retained as a direct-answer-sharing condition to identify the high scores contamination can produce.

\begingroup
\small
\setlength{\tabcolsep}{4pt}
\renewcommand{\arraystretch}{1.18}
\setcounter{table}{17}
\begin{table}[htbp]
\caption{Experience consumption and GT correctness on 300 independent cases.}\label{tab:15}
\centering
\small
\begin{tabular}{@{}>{\RaggedRight\arraybackslash}p{0.38\dimexpr\linewidth-4\tabcolsep\relax}>{\RaggedRight\arraybackslash}p{0.15\dimexpr\linewidth-4\tabcolsep\relax}>{\RaggedRight\arraybackslash}p{0.47\dimexpr\linewidth-4\tabcolsep\relax}@{}}
\toprule
\textbf{Experience configuration} & \textbf{GT correctness} & \textbf{Interpretation} \\
\midrule
No history & 0.833 & Construction baseline on the new test set \\
Known-contaminated synthetic history & 0.950 & Direct-answer-sharing condition \\
Ungated retrieval of real experience & 0.683 & Relevance does not establish applicable conditions \\
gated\_v2: qualification and framework semantics & 0.970 & Combined predicates, ownership, evidence, and interpretation \\
\bottomrule
\end{tabular}
\end{table}
\endgroup

GT correctness is 0.970 for \code{gated_v2}, 0.683 for ungated real experience, and 0.833 without history. The gated method improves on ungated retrieval by 0.287, with a paired normal-approximation 95\% interval [0.236, 0.338], and on no history by 0.137, with interval [0.096, 0.178]; both exact two-sided McNemar tests have $p<0.001$. The intervals use $\widehat\Delta\pm1.96\sqrt{(b+c)-(b-c)^2/n}/n$, where $b,c$ are discordant counts and $n=300$ paired cases. Known-contaminated synthetic history achieves 0.950. The gated difference relative to that condition is 0.020, with lower interval bound $-0.008$, above the stated noninferiority margin $-0.05$. Its role as a contamination control remains distinct from the clean experience-benefit comparisons.

\begin{figure}[htbp]
\centering
\includegraphics[width=\linewidth]{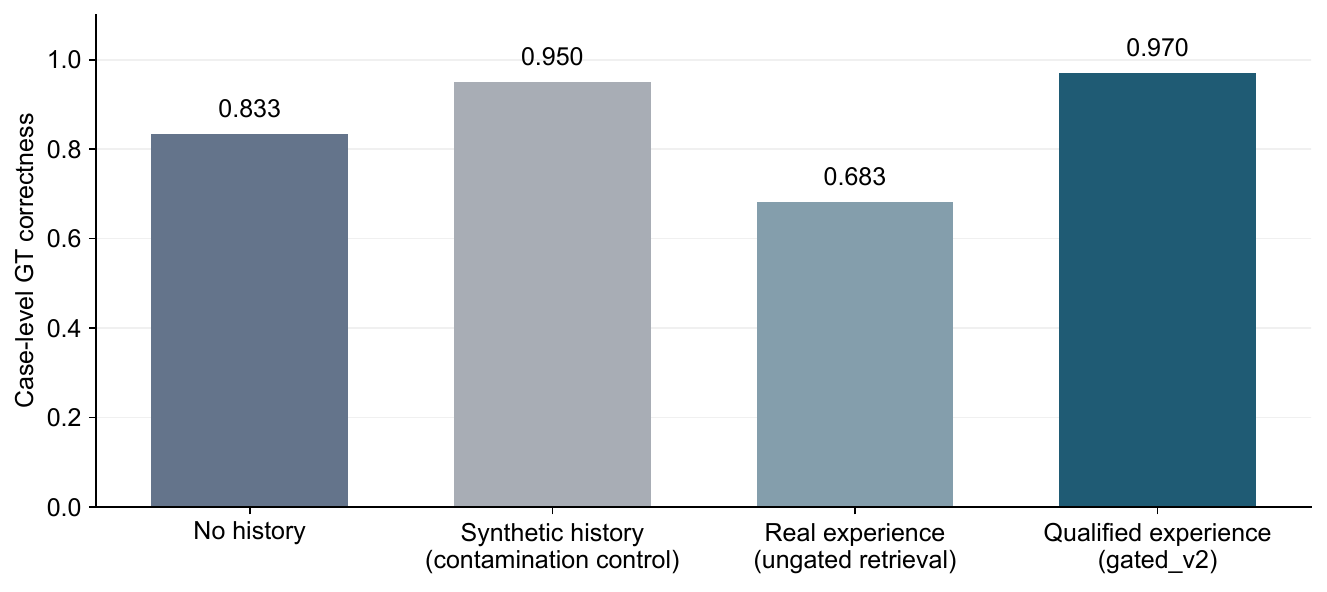}
\caption{History and gated experience on the source-isolated 300-case holdout. Known-contaminated synthetic history is shown as a contamination control.}\label{fig:7}
\end{figure}

Source-isolation checks cover identifiers, summaries, contract fingerprints, exact numerical tuples, and near-duplicate text. No checked direct-answer trace is found in the 300 cases. A known 5\% leakage positive control has recall 1.0 in all five checks. Figure~\ref{fig:7} compares consumption conditions, showing that the way experience is used affects construction outcomes. The observation is restricted to this audited experimental library. Pretraining sources and open historical distributions are discussed in Section~\ref{sec:12}.

\FloatBarrier

\subsection{Minimal witnesses and bounded implementation checks}\label{subsec:10.6}\label{sec:10.6}

Six executable witnesses, N1--N6, use positive, counterexample, and reversal controls to distinguish judgments and test carrier changes. The current reference kernel passes these specified inputs. Four guard mutants trigger detection markers at the entry point. Actual signals include semantic assertions, target returns, and an input-incompatibility exception in a reference mutant. These implementation checks do not replace a full state-space proof.

A branching generator with two preset commitments, versions v1/v2, and maximum depth six visits 17 projected states of the current kernel. It attempts 220 transitions, accepts 16, rejects 204, and finds no violation as defined by that checker. Five \code{carry} operations preserve both old obligations. Evidence-submission acceptance and verification outcomes are counted separately. The generator includes no conditional-performance or excuse candidates, and all its closure candidates are rejected; targeted lawful controls provide closure evidence. The 33 regressions, 11 adapter probes, and six additional closure-precondition checks pass.

A lawful scripted control through the current kernel and runner attaches an issued payment credential, disposes of an existing payment obligation, submits new-version delivery and payment evidence, and closes with authority. This establishes executability of the specified input chain. Closure-precondition checks separately distinguish ineffective empty records from effective relationships without outstanding obligations. Five online path diagnostics achieve autonomous combined outcomes in 0/5 cases: only one action opportunity remains after necessary credentials are accepted, and no case submits closure. Offline submission of closure to the saved terminal state is accepted. This small diagnostic cannot separate event timing, action opportunity, and strategy.

\subsection{Same-stack online comparison: outcomes and action reachability}\label{subsec:10.7}\label{sec:10.7}

\begingroup
\small
\setlength{\tabcolsep}{4pt}
\renewcommand{\arraystretch}{1.18}
\setcounter{table}{18}
\begin{table}[htbp]
\caption{Adjacent online configurations and preset capability gaps.}\label{tab:16}
\centering
\small
\begin{tabular}{@{}>{\RaggedRight\arraybackslash}p{0.23\dimexpr\linewidth-4\tabcolsep\relax}>{\RaggedRight\arraybackslash}p{0.34\dimexpr\linewidth-4\tabcolsep\relax}>{\RaggedRight\arraybackslash}p{0.43\dimexpr\linewidth-4\tabcolsep\relax}@{}}
\toprule
\textbf{Family and comparison} & \textbf{Added semantics} & \textbf{Preset target gap} \\
\midrule
F1: RL0 vs. RL1 & Relationship identity and binding & Cross-relationship misbinding \\
F2: RL1 vs. RL2 & Atomic commitments and active obligations & Omitted outstanding obligations \\
F3: RL2 vs. RL3 & Current authority and validity adjudication & Invalid activation or rejection of lawful actions \\
F4: RL3 vs. RL4 & Immutable versions and $\mu$ dispositions & Old-obligation loss and failed lawful remedies \\
F5: RL4 vs. RL5 & Evidence references and closure guards & False provable closure \\
F6: RL5 vs. RL6 & Experience qualification gate & Unsupported reuse \\
\bottomrule
\end{tabular}
\end{table}
\endgroup

\begingroup
\small
\setlength{\tabcolsep}{4pt}
\renewcommand{\arraystretch}{1.18}
\setcounter{table}{19}
\begin{table}[htbp]
\caption{Case outcomes in the same-stack online main panel.}\label{tab:17}
\centering
\small
\begin{tabular}{@{}>{\RaggedRight\arraybackslash}p{0.23\dimexpr\linewidth-10\tabcolsep\relax}>{\RaggedRight\arraybackslash}p{0.07\dimexpr\linewidth-10\tabcolsep\relax}>{\RaggedRight\arraybackslash}p{0.15\dimexpr\linewidth-10\tabcolsep\relax}>{\RaggedRight\arraybackslash}p{0.12\dimexpr\linewidth-10\tabcolsep\relax}>{\RaggedRight\arraybackslash}p{0.13\dimexpr\linewidth-10\tabcolsep\relax}>{\RaggedRight\arraybackslash}p{0.30\dimexpr\linewidth-10\tabcolsep\relax}@{}}
\toprule
\textbf{Mechanism} & \textbf{Pairs n} & \textbf{Correct: low / high} & \textbf{b : c} & \textbf{Difference (pp)} & \textbf{Holm and interpretation} \\
\midrule
Identity (F1) & 60 & 40 / 52 & 19 : 7 & +20.00 & p=0.0869; threshold not met \\
Commitments (F2) & 60 & 0 / 23 & 23 : 0 & +38.33 & p<0.0001; conditional difference \\
Current validity (F3) & 60 & 30 / 45 & 15 : 0 & +25.00 & p=0.0002; conditional difference \\
Version disposition (F4) & 60 & 3 / 2 & 0 : 1 & -1.67 & p=1; no increment identified \\
Evidence closure (F5) & 69 & 17 / 17 & 4 : 4 & Descriptive only & Retained in Holm family; no mechanism interpretation \\
Experience qualification (F6) & 60 & 12 / 58 & 46 : 0 & +76.67 & p<0.0001; conditional difference \\
\bottomrule
\end{tabular}
\end{table}
\endgroup

Table~\ref{tab:17} reports correctness differences as higher minus lower configuration. Commitment recognition, current validity, and experience qualification meet the preset direction and six-family corrected threshold on the targeted cases, with differences of +38.33, +25.00, and +76.67 percentage points. Relationship identity has a +20.00-point difference (Holm $p=0.0869$), and version disposition has one discordant pair and a $-1.67$-point difference. Both F5 arms record 17 correct outcomes across 69 pairs. Defects in the version chain and scoring restrict F5 to descriptive reporting.

\begingroup
\small
\setlength{\tabcolsep}{4pt}
\renewcommand{\arraystretch}{1.18}
\setcounter{table}{20}
\begin{table}[htbp]
\caption{Action attempts and adjudicable outcomes at the original family-level targets.}\label{tab:18}
\centering
\small
\begin{tabular}{@{}>{\RaggedRight\arraybackslash}p{0.12\dimexpr\linewidth-8\tabcolsep\relax}>{\RaggedRight\arraybackslash}p{0.20\dimexpr\linewidth-8\tabcolsep\relax}>{\RaggedRight\arraybackslash}p{0.16\dimexpr\linewidth-8\tabcolsep\relax}>{\RaggedRight\arraybackslash}p{0.26\dimexpr\linewidth-8\tabcolsep\relax}>{\RaggedRight\arraybackslash}p{0.26\dimexpr\linewidth-8\tabcolsep\relax}@{}}
\toprule
\textbf{Family} & \textbf{Attempts: low / high} & \textbf{Both attempted} & \textbf{Only low / only high / neither} & \textbf{Correct among both: low / high} \\
\midrule
F1 & 60/52 & 52 & 8/0/0 & 33/52 \\
F2 & 42/42 & 42 & 0/0/18 & 0/23 \\
F3 & 60/60 & 60 & 0/0/0 & 30/45 \\
F4 & 4/9 & 3 & 1/6/50 & 2/2 \\
F5 & 45/22 & 20 & 25/2/22 & 16/17 \\
F6 & 60/58 & 58 & 2/0/0 & 12/58 \\
\bottomrule
\end{tabular}
\end{table}
\endgroup

\FloatBarrier

Both arms attempt closure in three F4 pairs and 20 F5 pairs. Within those subsets, the lower and higher arms have two correct outcomes each in F4 and 16 and 17, respectively, in F5. Table~\ref{tab:18} also gives the one-arm and neither-arm attempt counts.

In F4, 107 arm records end with \code{inform} after earlier renegotiation attempts. Among F5 records that stop before the target, 58 end with \code{inform} and 13 reach the budget limit.

\begingroup
\small
\setlength{\tabcolsep}{4pt}
\renewcommand{\arraystretch}{1.18}
\setcounter{table}{21}
\begin{table}[htbp]
\caption{Distinct denominators for F4 renegotiation and closure.}\label{tab:19}
\centering
\small
\begin{tabular}{@{}>{\RaggedRight\arraybackslash}p{0.68\dimexpr\linewidth-4\tabcolsep\relax}>{\RaggedRight\arraybackslash}p{0.16\dimexpr\linewidth-4\tabcolsep\relax}>{\RaggedRight\arraybackslash}p{0.16\dimexpr\linewidth-4\tabcolsep\relax}@{}}
\toprule
\textbf{Action metric} & \textbf{RL3} & \textbf{RL4} \\
\midrule
Cases & 60 & 60 \\
Cases attempting renegotiate & 60 & 60 \\
Cases with at least one accepted renegotiate & 0 & 1 \\
Cases attempting close & 4 & 9 \\
Pairs in which both attempt close & 3 & 3 \\
\bottomrule
\end{tabular}
\end{table}
\endgroup

Both F4 arms attempt renegotiation in 60/60 cases. The action is available only in the higher configuration, which accepts one of 62 submissions. Of the remaining submissions, 45 use invalid dispositions and 16 contain malformed mappings. A typical candidate names an undefined postponement disposition. The schema gives limited guidance on disposition values and repair (Section~\ref{subsec:8.2}), leaving interface design and candidate construction intertwined. Table~\ref{tab:19} separates these submissions from subsequent closure attempts. The scripted controls in Section~\ref{subsec:10.6} show how selected lawful paths reach closure.

\begingroup
\small
\setlength{\tabcolsep}{4pt}
\renewcommand{\arraystretch}{1.18}
\setcounter{table}{22}
\begin{table}[htbp]
\caption{Qualified coverage and safe yield of actual F6 reuse.}\label{tab:20}
\centering
\small
\begin{tabular}{@{}>{\RaggedRight\arraybackslash}p{0.60\dimexpr\linewidth-4\tabcolsep\relax}>{\RaggedRight\arraybackslash}p{0.20\dimexpr\linewidth-4\tabcolsep\relax}>{\RaggedRight\arraybackslash}p{0.20\dimexpr\linewidth-4\tabcolsep\relax}@{}}
\toprule
\textbf{Metric and denominator} & \textbf{RL5} & \textbf{RL6} \\
\midrule
Actual reuse among eligible candidates & 12/12 & 12/12 \\
Actual reuse of ineligible candidates & 48/48 & 0/48 \\
Qualified coverage & 1.00 & 1.00 \\
Actual reuse precision & 12/60 = 0.20 & 12/12 = 1.00 \\
SafeReuseYield & 0.20 & 1.00 \\
\bottomrule
\end{tabular}
\end{table}
\endgroup

Both F6 arms reuse all 12 eligible candidates. Among the 48 ineligible cases, RL5 reuses every candidate and RL6 rejects 46 submissions. The other two RL6 cases make no attempt and remain unsuccessful in the primary endpoint. Table~\ref{tab:20} reports qualified coverage (reuse among eligible cases), precision (the eligible share of actual reuse), and SafeReuseYield (their product). Each case supplies one candidate.

Across 52 normal controls, outcomes agree in 50/52 pairs. The differences are failure to advance to \code{perform} in the higher identity configuration and failure to reach \code{close} in the higher version configuration. F4 normal controls reach the original outcome in 3/20 and 2/20 cases, respectively. Outcome agreement and actual completion therefore require separate reporting.

\subsection{Full-lifecycle online evaluation by semantic guard ablation}\label{subsec:10.8}\label{sec:10.8}

This study compares semantic guards within an experimental Full-CC configuration. Within each final panel, model, industrial world, action budget, client path, and case set are held fixed while a guard is removed. The endpoint is \emph{normatively correct terminal outcome} (NCT) for lifecycles involving migration, new obligations, evidence, and multi-party closure. Table~\ref{tab:ops-spec} specifies each guard; protocol and statistical provenance are consolidated in Section~\ref{subsec:12.2}.

Five panels (O1--O5) each contain 100 paired cases: 50 lawful cases and 50 challenge cases containing a targeted trap such as omission of a conditional commitment, stale or foreign credentials, or a concurrent revision. O1--O4 remove one component at a time: ledger-derived scope, retention of conditional commitments in that scope, atomic activation of new obligations, and multi-party closure signatures. O5 combines removal of the O1, O3, and O4 guards with untrusted evidence-origin self-report. The experiment does not include a separate evidence-use ablation. Every episode permits at most 37 rounds. The model, decoding settings, retry policy, and business world match the controlled online setup in Section~\ref{subsec:9.3}.

NCT requires the terminal ledger state to match the case expectation, including the disposition required for every obligation. Challenge and lawful cases are analyzed separately. Table~\ref{tab:full-lifecycle} reports two-sided exact McNemar tests, with Holm correction across the five panels within each case class. The archived directional analysis tests Full-CC outperforming the ablation with a separate one-sided correction; Appendix~\ref{app:protocol-provenance} records its provenance. In Table~\ref{tab:full-lifecycle}, $b$ counts pairs correct only under Full-CC and $c$ counts pairs correct only under the ablation. Scripting checks establish that the 40 designated lawful action chains are executable. Within the finite defect generator, each ablated configuration is also exposed to its targeted defect family; Full-CC accepts none of those defects in 6{,}912 offline runs. These generated checks establish path behavior for the implemented defect set and do not substitute for online outcomes.

\begingroup
\small
\setlength{\tabcolsep}{4pt}
\renewcommand{\arraystretch}{1.18}
\setcounter{table}{23}
\begin{table}[htbp]
\caption{Full-lifecycle online guard-ablation results by case class (50 paired cases per row). Exact two-sided tests use Holm correction separately within each case class.}\label{tab:full-lifecycle}
\centering
\small
\begin{tabular}{@{}>{\RaggedRight\arraybackslash}p{0.20\dimexpr\linewidth-10\tabcolsep\relax}>{\RaggedRight\arraybackslash}p{0.09\dimexpr\linewidth-10\tabcolsep\relax}>{\RaggedRight\arraybackslash}p{0.17\dimexpr\linewidth-10\tabcolsep\relax}>{\RaggedRight\arraybackslash}p{0.09\dimexpr\linewidth-10\tabcolsep\relax}>{\RaggedRight\arraybackslash}p{0.13\dimexpr\linewidth-10\tabcolsep\relax}>{\RaggedRight\arraybackslash}p{0.32\dimexpr\linewidth-10\tabcolsep\relax}@{}}
\toprule
\textbf{Panel (ablated component)} & \textbf{Class} & \textbf{NCT: Full-CC / ablation} & \textbf{b : c} & \textbf{Difference (pp)} & \textbf{Two-sided Holm $p$ and interpretation} \\
\midrule
O1 (ledger scope) & challenge & 37/50 : 16/50 & 24 : 3 & +42.0 & $0.00025$; Full-CC higher \\
O1 (ledger scope) & lawful & 30/50 : 30/50 & 3 : 3 & +0.0 & $1.000$; no difference \\
O2 (conditional in domain) & challenge & 1/50 : 10/50 & 0 : 9 & $-$18.0 & $0.0156$; exploratory reverse difference \\
O2 (conditional in domain) & lawful & 29/50 : 21/50 & 9 : 1 & +16.0 & $0.107$; no corrected difference \\
O3 (atomic new obligations) & challenge & 0/50 : 0/50 & 0 : 0 & +0.0 & $1.000$; floor-limited \\
O3 (atomic new obligations) & lawful & 31/50 : 35/50 & 0 : 4 & $-$8.0 & $0.500$; no corrected difference \\
O4 (multi-signature close) & challenge & 31/50 : 37/50 & 4 : 10 & $-$12.0 & $0.359$; no corrected difference \\
O4 (multi-signature close) & lawful & 30/50 : 30/50 & 2 : 2 & +0.0 & $1.000$; no difference \\
O5 (composite ablation) & challenge & 30/50 : 41/50 & 5 : 16 & $-$22.0 & $0.080$; no corrected difference \\
O5 (composite ablation) & lawful & 31/50 : 35/50 & 4 : 8 & $-$8.0 & $1.000$; no corrected difference \\
\bottomrule
\end{tabular}
\end{table}
\endgroup

On O1's 50 targeted omission challenges, Full-CC reaches NCT in 37 cases (74\%) compared with 16 (32\%) after removal of the ledger-derived scope guard (adjusted two-sided $p=0.00025$). Both arms reach NCT in 30 of 50 lawful O1 cases. Of the 24 pairs favoring Full-CC, seven ablated episodes stop at closure or terminal-state checks and 17 diverge earlier through structural loops or early stops. Timing gates block Full-CC in the three pairs favoring the ablation. The measured contrast includes terminal checks and agents' responses to guard feedback during execution.

O2 runs in the reverse direction on challenges. Full-CC reaches NCT in 1 of 50 cases and the ablated configuration in 10 of 50 (exploratory two-sided $p_{\mathrm{adj}}=0.0156$). The traces give tool outcomes but omit the action arguments needed to distinguish blocked lawful progress from unsafe shortcuts.

Both arms score 0/50 on O3 challenges within the 37-round limit under this model, request configuration, and action interface. This floor leaves the online effect of the atomic-new-obligation guard unresolved. O4 and O5 fall short of the corrected threshold. O5 reaches NCT in 30/50 challenge cases with Full-CC and 41/50 with the composite ablation ($p_{\mathrm{adj}}=0.080$, exploratory). All lawful-case comparisons fall short after correction.

The arm-independent \emph{record-audited NCT} (RA-NCT) combines NCT with the absence of detected violations among properties the records can adjudicate. The audit assigns each property pass, violation, or undetermined, with missing arguments leaving the relevant checks undetermined. All 1{,}000 episodes attempt migration, 867 submit evidence, and 546 record an accepted agent-initiated close. Table~\ref{tab:audit-coverage} gives the number assessable for each property. The audit identifies an O5 challenge episode that closes with incomplete signatures. Reclassifying that success gives exploratory two-sided $p_{\mathrm{adj}}=0.124$ for O5. The separate post-window closure probe records 275 liveness failures.

\begingroup
\scriptsize
\setlength{\tabcolsep}{3pt}
\renewcommand{\arraystretch}{1.14}
\setcounter{table}{24}
\begin{table}[htbp]
\caption{Actual record-audit coverage in the 1{,}000-episode full-lifecycle study.}\label{tab:audit-coverage}
\centering
\scriptsize
\begin{tabular}{@{}>{\RaggedRight\arraybackslash}p{0.22\dimexpr\linewidth-8\tabcolsep\relax}>{\RaggedRight\arraybackslash}p{0.12\dimexpr\linewidth-8\tabcolsep\relax}>{\RaggedRight\arraybackslash}p{0.14\dimexpr\linewidth-8\tabcolsep\relax}>{\RaggedRight\arraybackslash}p{0.25\dimexpr\linewidth-8\tabcolsep\relax}>{\RaggedRight\arraybackslash}p{0.27\dimexpr\linewidth-8\tabcolsep\relax}@{}}
\toprule
\textbf{Property} & \textbf{Screened; reached} & \textbf{Adjudicable} & \textbf{Observed result} & \textbf{Boundary} \\
\midrule
Terminal normative correctness (NCT) & 1{,}000; 1{,}000 & 1{,}000 & 505 pass; 495 nonpass & Includes action reachability and terminal-state matching \\
Agent-close signer completeness & 1{,}000; 546 & 433 & 432 pass; 1 violation; 113 undetermined & Reached count is accepted agent-initiated closure; dynamic closure domain is not recoverable \\
Accepted migration before agent close when a new obligation is required & 1{,}000; 151 & 151 & 151 pass; 0 violation & Checks migration presence, not its arguments or atomic rollback \\
Evidence identity and version binding & 1{,}000; 867 & 0 & 867 undetermined & Reached count is evidence submission; credential identity/version arguments are absent \\
Disposition-map content & 1{,}000; 1{,}000 & 0 & 1{,}000 undetermined & Every episode attempted migration; candidate $\mu$ arguments are absent \\
Candidate-declared scope narrowing & 1{,}000; 1{,}000 & 0 & 1{,}000 undetermined & Every episode attempted migration; scope arguments are absent \\
Post-window closure probe & 1{,}000; 333 & 333 & 58 pass; 275 liveness failures & Separate liveness diagnostic; not a safety violation \\
\bottomrule
\end{tabular}
\end{table}
\endgroup

Repeated approval and confirmation requests and premature evidence submissions consume action opportunities across arms. Section~\ref{subsec:10.9} examines candidate and confirmation records in a separate process audit.

\subsection{Process-audit diagnosis of O1/O2 reachability}\label{subsec:10.9}\label{sec:10.9}

The audit retains all 80 episodes and the migration candidates needed for independent re-adjudication. Rejected actions preserve normative status, version, and signatures. The audit registry records 137 rejected evidence submissions. Accepted migrations commit the version returned by the guard, and all specified lawful controls pass (Table~\ref{tab:process-audit}).

Both arms reach NCT in 9/10 O1 challenges in this diagnostic sample. The comparative estimate comes from the O1 panel (100 pairs, 200 episodes) of the 1{,}000-episode study. The audit log label \code{closed} records accepted lifecycle termination. NCT and \code{VerifiedClosed} each require their own checks.

All 40 O2 episodes fail NCT. Across the two audited panels, 35 of 560 migration records have mismatched candidate $\mu$ keys and runtime checking domains. Full-CC and the ablation derive different scopes in 185 of 560 same-candidate forks, all in O2. Every O2 episode names the buyer in its first confirmation request for the new freight obligation, while most candidates name the logistics provider as creditor. The confirmation check rejects the mismatch, and the world's one-shot issuance rule blocks replacement with a corrected token. This shared path keeps the scope differences from reaching the terminal outcome in the audit sample.

\begingroup
\small
\setlength{\tabcolsep}{4pt}
\renewcommand{\arraystretch}{1.16}
\setcounter{table}{25}
\begin{table}[htbp]
\caption{Claim-relevant results of the targeted O1/O2 process audit.}\label{tab:process-audit}
\centering
\begin{tabular}{@{}>{\RaggedRight\arraybackslash}p{0.27\dimexpr\linewidth-4\tabcolsep\relax}>{\RaggedRight\arraybackslash}p{0.25\dimexpr\linewidth-4\tabcolsep\relax}>{\RaggedRight\arraybackslash}p{0.48\dimexpr\linewidth-4\tabcolsep\relax}@{}}
\toprule
\textbf{Audit question} & \textbf{Observed result} & \textbf{Interpretation} \\
\midrule
Can migration decisions be re-adjudicated? & 560/560 candidate records complete & Candidate, effective ledger, runtime scope, guard decision, and state difference are jointly inspectable \\
Do rejected or accepted writes remain consistent? & 0/818 normative mutations after rejection; 0/99 version mismatches after acceptance & The recorded transition paths preserve reject immutability and accepted-version consistency \\
Are specified lawful paths executable? & 24/24 scripted controls pass & The listed control paths reach their intended normative writes \\
Does the O1 audit sample reproduce the main-panel difference? & Challenge NCT: 9/10 versus 9/10 & Descriptive process sample; no second O1 effect estimate \\
Where does O2 fail? & NCT 0/40 & Shared one-shot confirmation mismatch prevents observed candidate-level conditional-scope differences from affecting terminal NCT \\
\bottomrule
\end{tabular}
\end{table}
\endgroup

\subsection{Two implementation configurations: normative outcomes and autonomous coordination}\label{subsec:10.10}

\subsubsection{Comparison purpose and frozen configuration}

Arm A uses the CC runtime, and Arm B assembles session, commitment, authorization, versioning, evidence, recovery, and experience functions in separate modules. The study compares these implementations of the public collaboration interface on the same generated cases. Arm B was developed in a separate coding-agent context within the same project. Both arms use shared lower-level code and evaluation infrastructure.

\begin{center}
\begin{minipage}{\linewidth}
\small
\textbf{Implementation scope in the paired comparison}\par\smallskip
\begin{tabularx}{\linewidth}{@{}p{0.20\linewidth}YY@{}}
\toprule
Dimension & Arm A & Arm B and shared basis \\
\midrule
Candidate orchestration & CC runtime with candidate and enterprise-service stores & Separate session, commitment, and compensation modules; common public tool contract \\
Authority and confirmation & Enterprise-service decision and issued-confirmation records checked at the runtime boundary & Separate authorization book and candidate-local decisions; both map authority into the shared kernel grant representation \\
Activation & Checks required enterprise confirmations; resolves a kernel acceptor from role bindings & Additionally requires the submitting actor to be the proposer; both call the shared acceptance machinery \\
State and migration & Uses the Phase 5 state representation and migration kernel with CC prechecks & Uses the same state foundation and migration function, with separate prechecks and result translation \\
Agent observations & Role identifiers with enterprise/identity mapping and candidate digests & Enterprise-name representation and candidate digests; corresponding fields do not imply identical adapter behavior \\
World and evaluation & Frozen generated cases, service client, target definitions, audit and scoring rules & Same case inputs, model client and evaluator; separate world/session adapters and event handling \\
\bottomrule
\end{tabularx}
\end{minipage}
\end{center}

The interfaces differ in adapter and orchestration behavior. B requires the candidate's proposer to submit activation. A validates enterprise confirmations and selects the low-level acceptor from the candidate's role bindings. Both call the shared acceptance machinery.

The frozen panel contains 477 paired cases across six families and 48 structural templates in the procurement and manufacturing worlds. The families cover relationship binding, authority and disclosure, version disposition, evidence, recovery, and experience qualification. Three online development instances were excluded from the original 480-case panel before the main run; 45 templates retain ten cases and three retain nine. The retained panel shares template structures with development cases and is described as the existing panel after development-instance exclusion. Both arms execute exactly the same retained case set and the frozen run order. The main run contains 954 completed episodes, with no parameter adjustment, case deletion, or sample extension based on their outcomes.

Enterprise agents receive their identities, permitted commitments, pending candidates and digests, world observations, and tool feedback. The frozen tool contract defines eight actions and their parameters: authorization request, proposal, enterprise decision and confirmation, activation, migration, closure, query, and termination of participation. The two implementations expose corresponding business information through different role encodings and state representations. The harness supplies business events and automatic fulfillment transitions. Enterprise approvals are simulated at the interface.

Both arms use the same GPUStack service identifier, \code{qwen3.6-35b-a3b-fp8}, and the frozen common client. The client sends temperature 0 and a 2,048-token response limit, requests \code{enable_thinking=False}, and retains its documented fallback for a service rejecting that option. Each episode has a shared 20-decision limit, a 40-call cap, and boundary checks against 100,000 prompt and 10,000 completion tokens. Queries and malformed dispatched actions consume decisions. The action \code{done} ends participation without charging a decision. Reaching the target on the twentieth decision takes precedence over budget termination.

The correct normative result rate (CNR) counts legal completion of a case-specific target, justified exit, or correct rejection of an illegal target, subject to independent state auditing. Its denominator is 477 cases per arm. The label \code{NOT_ENGAGED} marks cases without a target attempt, including justified exits that score CNR=1. We report full closure separately.

The evaluator associates each audit with its episode identifier and checks state changes, rejection integrity, time-indexed authorization, and basic closure legality independently of the runtime's claimed success. Missing or insufficient audit records block formal scoring. The primary contrast averages the arm difference equally across the six families and, within each family, across templates. It uses 100,000 template-level sign permutations and 20,000 stratified template bootstrap samples. Exact two-sided case-paired McNemar results are secondary; six family-level tests use Holm correction. These analysis units account for the designed template structure without treating the panel as a sample of all enterprise collaborations.

\subsubsection{Main result and CNR composition}

CNR is 53/477 (11.1\%) for A and 54/477 (11.3\%) for B. The primary A-minus-B difference is $-0.21$ percentage points, with a 95\% bootstrap interval of $[-2.08,+1.67]$ points and permutation $p=1.0$. The case-level paired counts are 38 both correct, 15 A-only correct, 16 B-only correct, and 408 neither correct (exact McNemar $p=1.0$). The panel detects no CNR difference, and no family reaches the corrected threshold (Table~\ref{tab:phase6-results}).

\begin{table}[htbp]
\setcounter{table}{26}
\caption{Two-implementation comparison on 477 paired cases. CNR counts legal completion of case-specific targets, justified exits, and correct rejections.}\label{tab:phase6-results}
\centering\small
\begin{tabularx}{\linewidth}{@{}Xrr@{}}
\toprule
Outcome or diagnostic & Arm A & Arm B \\
\midrule
CNR=1 & 53/477 (11.1\%) & 54/477 (11.3\%) \\
\quad Case-correct \code{legal_completion} & 21 & 20 \\
\quad Case-correct \code{justified_exit} & 32 & 24 \\
\quad Case-correct \code{guard_reject_illegal} & 0 & 10 \\
Target action attempted & 27/477 (5.7\%) & 38/477 (8.0\%) \\
Target action not attempted (\code{NOT_ENGAGED}) & 450/477 (94.3\%) & 439/477 (92.0\%) \\
Audit-sufficient completed episodes & 477 & 477 \\
Violations detected within audit coverage & 0 & 0 \\
\bottomrule
\end{tabularx}
\par\smallskip\raggedright\footnotesize
Primary template-weighted difference (A--B): $-0.21$ pp; 95\% CI $[-2.08,+1.67]$ pp; template sign-permutation $p=1.0$. Case-level discordant pairs: 15:16, exact two-sided McNemar $p=1.0$. Correct exits overlap with target nonparticipation in 32 A and 24 B episodes.
\end{table}

Correct exits account for 32 of A's 53 correct outcomes and 24 of B's 54. These episodes have label \code{NOT_ENGAGED}. The legal-completion counts are 21 and 20 for the case-specific targets. Six A episodes and three B episodes reach accepted closure.

The specified audit covers all 954 completed episodes and detects no safety violation within its rule set.

\subsubsection{Action reachability and coordination failures}

Every episode registers a candidate. Table~\ref{tab:phase6-stages} counts episodes with at least one corresponding action and distinguishes attempts from acceptances. The eligible episodes can differ by action. An accepted confirmation records one enterprise's decision on a particular digest.

\begin{table}[htbp]
\setcounter{table}{27}
\caption{Action reachability in the two-implementation study. Each count denotes episodes with at least one corresponding action; all denominators are 477. Accepted confirmations can concern different candidates.}\label{tab:phase6-stages}
\centering\small
\begin{tabularx}{\linewidth}{@{}Xrr@{}}
\toprule
Action and counting rule & Arm A & Arm B \\
\midrule
Candidate submission registered & 477 & 477 \\
Authorization request accepted & 410 & 471 \\
Enterprise decision/confirmation accepted & 339 & 472 \\
Activation attempted / accepted & 294 / 19 & 467 / 77 \\
Migration attempted / accepted & 0 / 0 & 1 / 0 \\
Closure attempted / accepted & 11 / 6 & 24 / 3 \\
\bottomrule
\end{tabularx}
\end{table}

\Needspace{8\baselineskip}
Activation is attempted in 294 A episodes and 467 B episodes, but accepted in only 19 and 77. Missing confirmations account for 449 rejected attempts in 280 A episodes and 657 in 396 B episodes. B additionally returns \code{unauthorized} for 593 activation attempts in 323 episodes; A has no such activation rejection in these records. The different caller rules described above help explain why the rejection profiles differ, without isolating their causal effect on final outcomes.

Accepted decisions or confirmations refer to at least two recorded digests in 212 A episodes and 312 B episodes. Sequential revisions can produce this pattern. In B, 91 activation rejections list a placeholder buyer identifier among the missing confirmers, exposing a candidate-construction problem in those records. Appendix~\ref{app:phase6-process} reports the event definitions and family-level counts.

\begin{table}[htbp]
\setcounter{table}{28}
\caption{Final outcomes among episodes containing an accepted activation. The subset is selected by behavior and is reported descriptively.}\label{tab:phase6-activation-cross}
\centering\small
\begin{tabularx}{\linewidth}{@{}Xrr@{}}
\toprule
Final classification within activated episodes & A & B \\
\midrule
Activated episodes & 19 & 77 \\
\quad Case-correct legal completion & 5 & 2 \\
\quad Case-correct justified exit & 0 & 3 \\
\quad Case-correct rejection of illegal target & 0 & 10 \\
\quad CNR=0 & 14 & 62 \\
\midrule
Target action not attempted (overlapping diagnostic) & 10 & 58 \\
\bottomrule
\end{tabularx}
\par\smallskip\raggedright\footnotesize
The four indented outcome rows partition each activated subset. Target nonparticipation overlaps them. Paired activation membership: both arms 3 cases, A only 16, B only 74. This table reports accepted activation, not an independently established complete business lifecycle.
\end{table}

\Needspace{7\baselineskip}
Table~\ref{tab:phase6-activation-cross} explains why more accepted activations need not increase CNR. Only three paired cases have an accepted activation in both arms; 16 do so only in A and 74 only in B. The net difference of 58 is not a set of 58 additional paired successes. Among B's 77 activated episodes, 58 never attempt their case target: 32 end as \code{other}, 23 exhaust the budget, and three correctly exit. The other case-correct outcomes comprise two legal completions and ten correct rejections. All ten B correct rejections occur in activated episodes, but this association does not establish that activation caused them or that A lacked the corresponding guard. A contributes more correct exits overall (Table~\ref{tab:phase6-results}); its activated subset has five case-correct outcomes. Activation, target exposure, and correct final judgments consequently describe different parts of the process.

The case-specific target is attempted in 27 A episodes and 38 B episodes, including 19 pairs in which both arms attempt it. The primary comparison uses all 477 pairs to include cases that stop before the target. The two arms register some activations and closures, with zero accepted migrations.

\subsubsection{Scripted executability, autonomous behavior, and cost}

In one excluded development case, scripted witnesses use public observations and tool feedback to reach activation and closure in 13 A decisions and 19 B decisions, within the 20-decision budget. These paths establish executability for that case through the shared interface. The paired online panel counts correct agent outcomes under the same decision limit.

The 954 completed records contain 21,917,378 tokens for A and 16,792,034 for B, totaling 38,709,412. Decision totals are 6,663 and 7,806, or 14.0 and 16.4 per completed episode. No completed episode has a run error. Interrupted attempts and their unavailable consumption are documented separately in Section~\ref{subsec:12.4}.

The next comparison will retain matched business information and authority conditions while broadening exposure to the target transactions. It will distinguish end-to-end lawful completion, transition-level adjudication, recovery after refusal, and coordination cost, and compare independently implemented workflows across more models and enterprise scenarios. The present 477-pair result supplies a baseline for that programme.

\subsection{An agent-directed online formation--evolution--closure instance}\label{subsec:10.11}

\subsubsection{Purpose, implementation, and operating conditions}

This instance connects the protocol stages through the developmental implementation in Section~\ref{subsec:8.1}. Three independently scheduled agents represent the buyer, supplier, and logistics enterprise, each with its own conversation history, action identity, and permitted observations. Agents choose business terms and enterprise responses; \code{prepare_successor} translates explicit choices into candidate structures and reports missing prerequisites.

The configuration uses implementation commit \code{c7d8a22}, a 40-file freeze manifest, and the GPUStack service identifier \code{qwen3.6-27b-fp8}. Requests specify temperature~0, \code{max_tokens=2048}, and \code{enable_thinking=False}. The budget is 180 decisions, 240 requests, and two million tokens, with fixed heuristic admission reservations and service-reported settlement. The configuration and joint endpoint are fixed for this execution.

The order requests 100 units at a CNY~47,000 negotiation baseline. The accepted initial agreement is for 75 units at that total price, route R1, and absolute delivery deadline~65. This experimental case permits initial quantity negotiation; it is distinct from the illustrative minimum-80 requirement in Section~\ref{sec:2}. After the first 20 units are inspected, the world discloses capacity reduction and closure of R1. Successor quantity and price ranges and a separate freight-fee range then apply. Shipments, payments, evidence registration, and signatures require agent actions; transit takes two simulated ticks.

\subsubsection{Joint endpoint and observed process}

\begin{figure}[tbp]
\centering
\includegraphics[width=\linewidth]{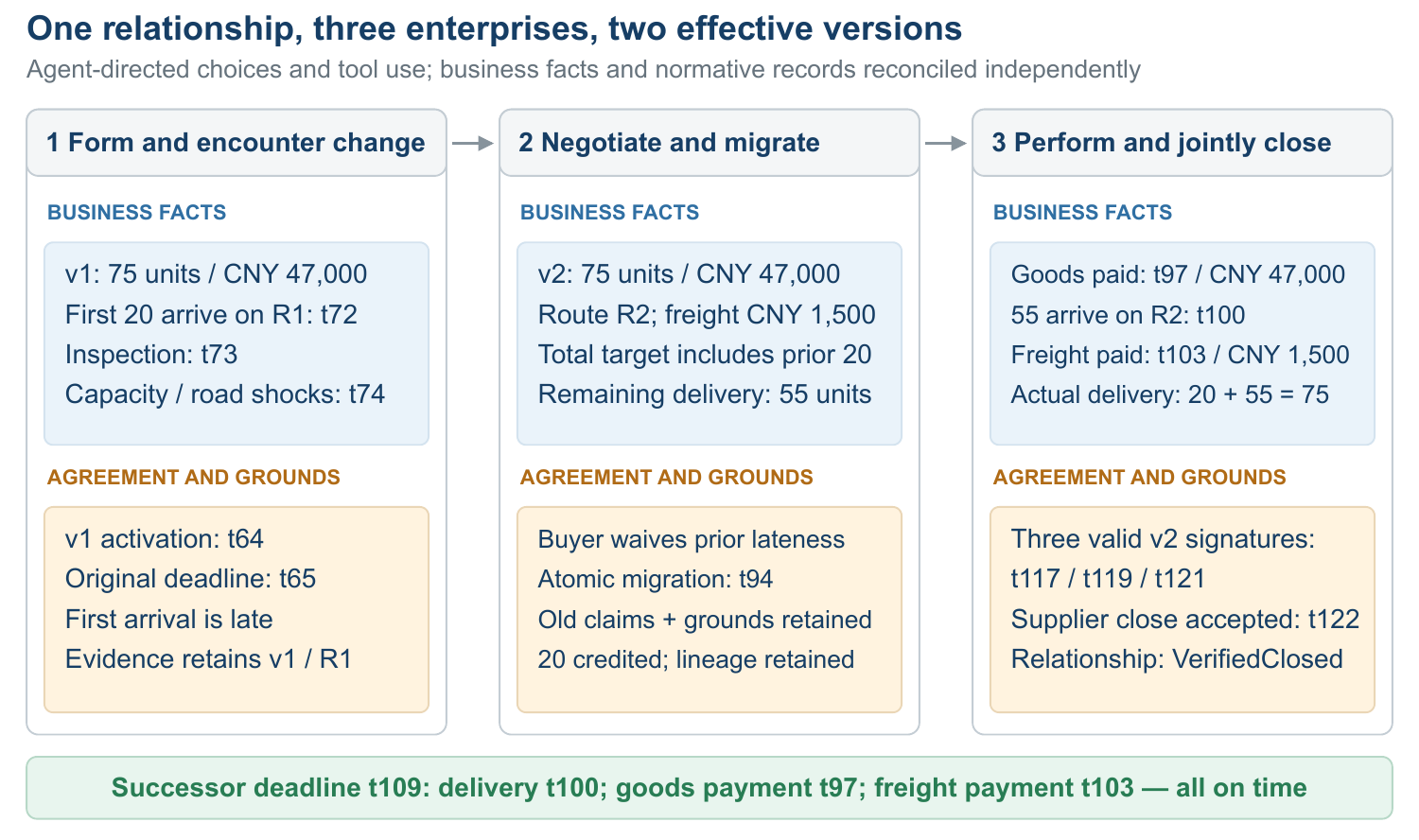}
\caption{Business execution and normative continuity in the agent-directed online instance. Ticks denote simulated world events; the panels summarize an observed path, rather than a prescribed agent script.}\label{fig:online-instance}
\end{figure}

The joint endpoint requires initial activation; authorized successor migration with complete old-obligation treatment; reconciled prior-performance credit and delivery; correct payments by purpose; and effective multi-party signatures followed by accepted closure. Applicable business terms are checked against accepted candidates, version events, world facts, and evidence allocations. Figure~\ref{fig:online-instance} joins the business and normative records of the adopted run, \code{cfgB-r11}.

Version~1 activates at tick~64 after nine initial candidates. Its deadline of 65 is later than activation but shorter than the two-tick transport horizon for a new shipment. The activation check in this implementation tests a future deadline, not full execution-horizon feasibility. First-batch arrival at~72 is consequently late. At migration tick~94, the buyer explicitly waives the prior lateness through a formal decision bound to the candidate. The successor retains 75 units and CNY~47,000, changes the route to R2, adds CNY~1,500 freight, and sets each successor deadline to 15 ticks after migration: tick~109. Quantity and goods-payment reductions are zero; the waiver concerns lateness.

The transaction preserves the old version, records the old obligations as \code{superseded} with their disposition grounds, and establishes linked successor obligations. It allocates the first 20 delivered units once to the successor's total target. A further 55 units arrive on R2 at tick~100, giving 75 units in total. The original R1 batch keeps its version~1 provenance; the new batch matches version~2. World payment records contain exactly CNY~47,000 to the supplier at~97 and CNY~1,500 to logistics at~103.

\begin{table}[htbp]
\centering
\small
\setlength{\tabcolsep}{4pt}
\caption{Active deadline evidence in \code{cfgB-r11}. Successor deadlines are independently derived as migration tick $94+15=109$.}\label{tab:online-deadlines}
\begin{tabularx}{\linewidth}{@{}P{0.24\linewidth}P{0.15\linewidth}P{0.21\linewidth}Y@{}}
\toprule
\textbf{Obligation} & \textbf{Deadline} & \textbf{World fact} & \textbf{Adjudication} \\
\midrule
Original delivery \code{kd9} & Absolute 65 & First 20 arrive at 72 & Late; buyer explicitly waives prior lateness in the accepted migration \\
Successor delivery \code{kd92} & 109 & Remaining 55 arrive at 100; prior 20 credited & Total target 75 satisfied on time \\
Goods payment \code{kq92} & 109 & CNY~47,000 paid at 97 & On time; no excess payment \\
Freight payment \code{kf2} & 109 & CNY~1,500 paid at 103 & On time; registration at 115 does not change the factual payment time \\
\bottomrule
\end{tabularx}
\end{table}

The case's payment antecedents are satisfied by first-batch acceptance and arrival, so goods payment at~97 can precede the final batch. Deadlines apply to factual arrival and payment times; subsequent evidence registration records those facts. Thus the fee paid at~103 remains timely when acknowledged at~115. A premature closure-signature request at~110 is rejected while fee disposition is incomplete. Logistics, supplier, and buyer sign at~117, 119, and~121; the qualified signatures bind version~2 and \code{all_terms}. The supplier's \code{close} is accepted at~122.

\subsubsection{Independent adjudication and evidence level}

The independent evaluator returns \code{VERIFIED_SUCCESS}: all 68 instantiated checks pass, including route provenance, authorized prior-lateness disposition, quantity and payment reconciliation, actual successor deadlines, and qualified closure signatures. The rule families are fixed with the evaluator; the instantiated count reflects the obligations and events in this trace. Archive-based recomputation reproduces the checks, the 128 settled-and-executed requests, and 1,410,796 tokens (1,401,245 input and 9,551 output). All returned service identifiers match the requested model, with no parameter fallback. The run finishes naturally within its three resource limits.

The trace provides configuration-specific existence evidence for the joined lifecycle. Section~\ref{subsec:12.2} states its development-selection and generalization limits.

\section{System Implications and Evidence Boundaries}\label{sec:11}

\subsection{An integrated software framework for the collaboration lifecycle}\label{subsec:11.1}\label{sec:11.1}

CC gives maintainers separate checks for a change to a service implementation and a change to the agreement. A replacement adapter can be tested against the same interface contract for permitted disclosures, authority, and factual evidence. Reassigning a delivery obligation to another enterprise requires a confirmed successor version and a lawful disposition of the original obligation. The explicit dependencies identify checks to rerun across affected relationships. The composition experiment in Section~\ref{subsec:10.4} measures the resulting coverage and workload on preset dependencies. Table~\ref{tab:evidence-map} links these methods to their evaluated scope.

\subsection{Responsibilities of trusted interfaces and autonomous agents}\label{subsec:11.2}\label{sec:11.2}

The separation of agent strategy from enterprise adjudication makes failures attributable at different boundaries: a poor proposal, an unauthorized acceptance, and an unsuccessful business execution demand different remedies. Delegation contracts govern how an agent acts for its enterprise; CC governs commitments among independently accountable parties~\cite{r60}. Strategy comparisons can therefore vary candidate search while holding authority and normative acceptance fixed.

Deployment must make the adjudicating sources dependable. The integration and threat boundaries in Sections~\ref{subsec:8.4}--\ref{subsec:8.5} identify the required issuer, scope, time, and challenge information. Source failures and organizational disputes call for evaluation alongside agent behavior.

\subsection{Constructing normative actions and adjudicating them at runtime}\label{subsec:11.3}\label{sec:11.3}

\paragraph{Finding A: ledger-derived checking scope has conditional empirical support.}
The O1 comparison provides the clearest controlled evidence: deriving disposition scope from the effective ledger improves the terminal endpoint on targeted omission challenges. This supports an operational rule with a direct business interpretation---a proposed amendment must account for the outstanding agreement even when its author omits an obligation. The discordant pairs span terminal blocks and earlier guard-mediated behavior, so the measured contrast belongs to the evaluated runtime and agent configuration.

Further comparisons of conditional-obligation retention need action records that distinguish rejected migrations from failures to obtain confirmation. The O3 result motivates cases that reach the new-obligation transaction often enough to evaluate it.

\paragraph{Finding B: normative adjudication and autonomous coordination require complementary support.}
Agents must construct an action, obtain the right enterprise responses, satisfy the runtime's checks, and retain opportunities for business execution and closure. Typed schemas, visible prerequisites, repair feedback, controlled cross-enterprise requests, and participant re-engagement make that chain operational. Independent records then distinguish incorrect acceptance from lawful but incomplete progress.

The 477-pair comparison locates coordination attrition behind similar CNR totals. The online instance adds a positive path through the joined protocols, while its candidate revisions and context consumption expose convergence cost. The matched-exposure comparisons proposed in Section~\ref{subsec:10.10} can assess that cost alongside correctness.

\subsection{Implementation boundaries and evaluation criteria}\label{subsec:11.4}\label{sec:11.4}

Four areas organize the next stage of the research programme. They extend the shared model into broader implementation and operating conditions, with evaluation criteria that retain both business progress and normative correctness.

\subsubsection{Enterprise interfaces and dependable adjudication}\label{subsubsec:11.4.1}\label{sec:11.4.1}

The open interface challenge is to sustain shared agreements under selective disclosure, asynchronous information, and conflicting adjudications while retaining unknown, stale, and disputed states.

Next steps include agreements on adjudication scope and effective time, permitted queries and refinement, and confirmation invalidation across asynchronous services. Evaluation should measure wrongful acceptance, blocked lawful collaboration, disclosure, adjudication latency, and dispute cost.

\subsubsection{Contract evolution and obligation continuity}\label{subsubsec:11.4.2}\label{sec:11.4.2}

Concurrent amendments raise the next continuity problem: multiple enterprises may add, continue, or terminate obligations while approvals and facts arrive at different times. The research task is to preserve earlier responsibilities and identify the effective agreement throughout that concurrency.

Evaluation should extend to version races, delayed approvals, cross-relationship amendments, independently developed adapters, and alternative negotiation strategies. It should jointly check obligation omission, erroneous activation, and reachability of lawful changes, recording confirmation cost, completion latency, and candidate revisions.

The adopted online path provides a concrete joint endpoint for independent-case evaluations varying organizational policies and disturbance timing.

\subsubsection{Contract networks and local revalidation}\label{subsubsec:11.4.3}\label{sec:11.4.3}

For contract networks, the open challenge is to identify sufficient dependency knowledge under selective disclosure and to represent unknown impacts without omitting required checks.

Methods to examine include permitted dependency summaries, composition-conflict explanations, and incremental validation. Evaluation should measure omitted dependencies, additional checks, computation, communication, and renegotiation cost. Tighter affected sets must retain necessary coverage.

E10's conservative coverage--workload trade-off motivates tests with deep dependencies and competing changes, examining which additional dependency knowledge narrows the affected set.

\subsubsection{Constrained strategies and qualified experience}\label{subsubsec:11.4.4}\label{sec:11.4.4}

The next strategy studies should compare prevention and remedy under different risks, approval delays, and costs of querying or waiting. They should separate wrongful acceptance, recovery, negotiation cost, and successful business progress.

Process-learning evaluation should compare candidate quality under the same complete information, then information acquisition under the same initial view and query budget. Failed negotiations and correctly declined requests should remain in the data, with external assistance controlled. Reuse measures should include ineligible use, qualified coverage, safe yield, benefit, and disclosure cost.

Enterprise approval, open historical distributions, and long-term cooperation also require testing whether qualified experience continues to help as conditions and permissions change.

The Agentic Living Lab methodology~\cite{agenticlivinglab} treats the service episode as a test unit and documents how participants and execution environments are realized and grounded. A future CC study could use it to compare simulated episodes with episodes involving enterprise systems and processes. Participating enterprises would help define scenarios and acceptance criteria, validate human roles or their proxies, and review recorded incidents for later tests.

A companion open benchmark could supply synthetic cases, change requests, expected outcomes, and replayable traces. Under the same business rules, permissions, and exposed capabilities, the benchmark could compare a CC implementation and a workflow-based assembly. Measures would include the effort to implement a change, the components and tests it affects, and consistency across agreement versions, outstanding obligations, and evidence.

\section{Threats to Validity and Auditability}\label{sec:12}

\subsection{Construct and internal validity}\label{subsec:12.1}\label{sec:12.1}

The testbed uses preset rules, faults, and method-invisible world truth. Responsibility accuracy depends on the specified causal models, while comparisons across paradigms differ in several runtime components. EBCR depends on contract-Monitor adjudication vocabulary and describes evidence coverage in this implementation. Online configurations jointly supply normative projections and feedback, so estimated outcome differences combine semantic guards, visible information, repair feedback, and behavioral participation.

Low exposure to target actions makes online endpoints sensitive to coordination as well as adjudication. The developmental instance uses predetermined participants and the business rules in Section~\ref{subsec:10.11}, including first-batch payment antecedents and authorized disposition of prior lateness.

The adjacent-configuration generators target specified semantic gaps. Content-fingerprint checks separate development inputs. The contrasts in F2, F3, and F6 therefore characterize targeted challenge cases. Independently authored case distributions and cross-mechanism challenges would broaden the evaluation.

NCT measures terminal ledger correctness. The transition audit in Table~\ref{tab:audit-coverage} identifies which process judgments the retained records support. The earlier traces leave evidence identity and version, migration-map content, and candidate-declared scope undetermined. Section~\ref{subsec:10.9} diagnoses the confirmation behavior of a separately recorded sample.

\subsection{Statistical conclusions and evidence levels}\label{subsec:12.2}\label{sec:12.2}

The studies use the analysis units in Table~\ref{tab:9}. Content fingerprints limit direct repeats, while cases generated from the same business template can remain correlated.

The full-lifecycle plan specifies exact McNemar tests and Holm correction and leaves test sidedness unspecified. The results use the two-sided calculation. Appendix~\ref{app:protocol-provenance} documents the archived directional analysis.

The final full-lifecycle panels used the interface fixed after initial O1/O2 execution and reused the initial O1/O2 business seeds. Transport handling changed between panel blocks. Aggregate token use exceeded the declared budget, while all final pairs completed under matched limits within their panels. Appendix~\ref{app:protocol-provenance} details these deviations.

Candidate forks in the 40-case process audit are repeated actions within episodes. The 477-pair comparison uses template-level inference on the retained cases and includes structural templates represented in development. The two implementations share the acceptance and migration machinery and differ in confirmation handling, activation-caller rules, and adapters. These differences jointly define the comparison. Few episodes enter late-stage checks, and conditioning on jointly attempted targets selects on agent behavior. The estimated CNR contrast leaves implementation equivalence open.

The developmental instance uses a known case, with the preparation tools, information views, scheduling support, and enlarged budget fixed before the reported execution. Development stopped after an observed success. The 68 predicates characterize that trace. The archive retains unsuccessful and interrupted runs with their original configurations.

E0's coverage depends on its action generator, search depth, and projection deduplication. Separate controls exercise conditional performance, excuse, and closure prerequisites. The base kernel uses same-\kid{} continuation. The 6{,}912 generated defect runs test the guard configuration in Section~\ref{subsec:10.8}, which includes new-obligation activation.

\subsection{External validity and experience sources}\label{subsec:12.3}\label{sec:12.3}

Evaluation uses two parameterized industrial domains, preset authority policies, trusted simulated interfaces, and an author-built runner. The comparative panels hold the requested decoding settings fixed within each pair. The developmental instance uses the separate service identified in Section~\ref{subsec:10.11}. Low completion can arise from model capacity, the served FP8 configuration, action schemas, confirmation rules, feedback, and action opportunities. Cross-model and cross-decoding studies are needed to distinguish these effects.

Field evaluation would need to measure human approval delays, partner discovery, negotiation strategies, legacy ERP/MES integration, replica consistency, and organizational disputes. Section~\ref{subsec:8.5} sets out the risks from source failures, malicious participants, and bypass of the runtime. Section~\ref{subsec:8.6} identifies the components of end-to-end latency and integration cost that remain to be measured.

Source-isolation checks control direct answer sharing between the experimental experience library and evaluation set. Model pretraining sources may still influence outcomes. The online experience family contains only 12 eligible controls, limiting coverage and precision conclusions. Free-text grounds remain incomplete in some records, and consumption benefits combine qualification constraints with the interpretation framework.

\subsection{Record retention and auditability}\label{subsec:12.4}\label{sec:12.4}

The public materials for this preprint comprise the manuscript source and figures. The authors retain the experimental records and analysis code in their research archive.

Archived inputs, episode records, and analysis code support recomputation of the reported tables. The earlier full-lifecycle records preserve normalized tool trajectories and terminal fields. Reconstructing the path from replies to parsed actions would require the complete assistant replies omitted from those records. The separate 80-episode audit preserves requests, replies, actions, candidate forks, and outcomes. The online archives record the request settings and returned service identifiers. Exact backend weight hashes and date-locked snapshots were unavailable from the service.

The two-implementation archive retains the frozen case set, per-episode scores, 954 completed traces, authority and state records, scripts, and SHA-256 manifests. The analysis selects the formal case identifiers. Four interrupted attempts lack completed records and token totals. For the developmental instance, the archive preserves source, the request and reply ledger, checkpoint, and world and normative events. Its offline entry point verifies the manifest and uses the bundled source to recompute the checks and token totals.

\section{Conclusion}\label{sec:13}

Collaboration contracts give a persistent cross-enterprise relationship explicit state and operations. The model defines how authorized decisions change versions and commitments, how evidence supports fulfillment, and how relationships depend on one another. The engineering methods use these semantics to validate candidates, commit amendments, verify fulfillment, and recheck dependencies after change.

Reference implementations and controlled evaluations provide initial executable evidence. Ledger-derived disposition scope has the clearest comparative support on targeted challenge cases; the other panels identify coordination and coverage questions for further study. A three-enterprise online instance joins the protocols through independently recomputable business execution and closure.

Experimental coverage remains selective across the lifecycle model. The continuing programme will broaden mechanism and scenario comparisons and test long-running collaboration with enterprise authority, execution, and evidence systems.

\appendix
\section{Interpretation of the Online Main Panel}\label{sec:panel-interpretation}\label{app:online-analysis}

The 380 main-panel cases each run two arms, producing 760 episodes. Relationship identity, commitment recognition, current validity, version disposition, and experience qualification contain 60 pairs each. Evidence closure contains 80. The final scoring rule excludes 11 F5 pairs if either arm lacks a stale-version trap. The archived pre-registration and frozen analysis commit \code{9761ff4} specify this rule at 12:40 UTC+8 on 1 October 2026, before the documented main-run start at 12:44. Applying the rule to the completed pairs gives suffixes 061, 062, 064, 066, 067, 069, 073, 074, 075, 078, and 079, with common prefix \code{F5-rla-evidence-v8}. This leaves 69 descriptive F5 pairs and 369 pairs in total.

Favorable discordant pairs $b$ have an incorrect lower configuration and correct higher configuration; $c$ denotes the reverse. The difference is $(b-c)/n$. Primary judgments use exact two-sided McNemar tests with six-family Holm correction. The original F5 test value is 1 and remains in that family, although we do not display it as an interpretable mechanism test. Unadjusted descriptive intervals, behaviorally selected action subsets, and later exploratory samples do not replace the primary analysis.

Attempts are located at target events, with F4 and F5 using \code{close}. Nonattempts remain unsuccessful in the primary endpoint. After an attempt, unmet obligations or an incorrect judgment may still prevent success. Table~\ref{tab:18} retains one-arm, both-arm, and neither-arm counts; Table~\ref{tab:19} separates renegotiation expression and acceptance; Table~\ref{tab:20} reports actual experience reuse. \code{inform} and \code{budget_hit} are recorded stop labels and do not establish the internal agent motives. These denominators and the identifiers above define the analysis set used by the main-text tables.

\FloatBarrier
\section{Process Analysis of the Two-Implementation Study}\label{app:phase6-process}

This descriptive analysis joins the 477 formal case identifiers per arm to episode scores and trajectories. Development traces co-located in the log directory are excluded. A stage is reached if at least one corresponding action has an explicit successful tool return. Candidate registration, an accepted decision/confirmation call, activation, and closure are counted separately. Only final scores determine CNR.

\begingroup
\renewcommand{\thetable}{B\arabic{table}}
\setcounter{table}{0}
\begin{table}[H]
\caption{Family-level process counts. Every A/B cell gives counts over the same per-arm family denominator. Stages and outcome classes can overlap.}\label{tab:phase6-families}
\centering\small
\begin{tabular}{@{}lrrrrrr@{}}
\toprule
Family & Cases/arm & Activated & Target attempted & Correct exit & Correct rejection & CNR=1 \\
\midrule
S1: binding & 79 & 2/22 & 0/3 & 6/4 & 0/1 & 6/5 \\
S2: authority & 79 & 3/7 & 0/4 & 4/5 & 0/1 & 4/7 \\
S3: versions & 80 & 4/10 & 4/2 & 6/4 & 0/1 & 8/5 \\
S4: evidence & 80 & 3/12 & 0/4 & 6/2 & 0/3 & 6/6 \\
S5: recovery & 79 & 3/6 & 20/21 & 5/5 & 0/0 & 23/23 \\
S6: experience & 80 & 4/20 & 3/4 & 5/4 & 0/4 & 6/8 \\
\midrule
All & 477 & 19/77 & 27/38 & 32/24 & 0/10 & 53/54 \\
\bottomrule
\end{tabular}
\end{table}

Rejected-action counts include dispatched actions whose trace explicitly records \code{ok=false}. The classification uses the tool and the first segment of the returned reason or error. Parsing failures without a dispatched action are excluded. Repeated rejections count as separate attempts; an episode is counted once per category. A record exposing only an unspecified error remains unspecified. Under this rule, A has 567 rejected attempts and B has 1,442. Table~\ref{tab:phase6-rejections} presents the main categories. Its episode columns overlap, so their sum is not a count of distinct episodes.

\begin{table}[H]
\caption{Recorded rejection attempts and affected episodes. Episode denominators are 477 per arm.}\label{tab:phase6-rejections}
\centering\small
\begin{tabularx}{\linewidth}{@{}Xrrrr@{}}
\toprule
Tool and rejection category & A attempts & A episodes & B attempts & B episodes \\
\midrule
Activation: missing confirmation & 449 & 280 & 657 & 396 \\
Activation: unauthorized caller & 0 & 0 & 593 & 323 \\
Decision: not a required confirmer & 85 & 66 & 132 & 113 \\
Closure: missing authorization & 9 & 5 & 21 & 19 \\
Activation: unknown predecessor & 0 & 0 & 15 & 15 \\
Other dispatched rejections & 24 & -- & 24 & -- \\
\midrule
Total rejected attempts & 567 & -- & 1,442 & -- \\
\bottomrule
\end{tabularx}
\end{table}

The observed missing-confirmation share is 79.2\% of A's rejected attempts and 45.6\% of B's; unauthorized activation contributes another 41.1\% in B. These are distributions of observed return codes, not independent causes. Ninety-one B missing-confirmation records name a placeholder buyer identifier, illustrating how a construction error can surface as a confirmation rejection. B checks the submitting actor against the proposer before checking missing confirmations, so the reason returned also depends on guard order.

Accepted decision/confirmation calls contain at least two different recorded candidate-digest values in 212/477 A episodes and 312/477 B episodes. The means are 1.41 and 2.00 distinct values per episode. This mechanical count does not adjudicate the temporal validity of confirmations or distinguish legitimate candidate replacement from unresolved competition. It should not be interpreted as a convergence-failure rate. Actor identities, digest supersession, and confirmation validity must be considered when diagnosing a particular negotiation.

Table~\ref{tab:phase6-activation-cross} uses the activated subset. In A, its 14 CNR=0 records comprise 11 budget stops, one \code{other} stop, one legal-completion-labelled record scored CNR=0, and one rejection-labelled record scored CNR=0. In B, the 62 comprise 26 budget stops, 32 \code{other} stops, one legal-completion-labelled record scored CNR=0, and three rejection-labelled records scored CNR=0. These terminal labels describe where records ended and are not inferred psychological reasons. The primary comparison continues to use all 477 paired cases; no effect test is claimed for these behaviorally selected subsets.
\FloatBarrier
\endgroup

\section{Protocol provenance of the full-lifecycle comparison}\label{app:protocol-provenance}

The body reports the adopted final panels. This appendix retains the material execution and analysis facts needed to interpret that evidence.

The full-lifecycle study has material protocol deviations. The final procedure was frozen after an initial O1/O2 run had begun. Its 400 balanced episodes were excluded because the agent interface did not expose the authorization-request tool required by the intended protocol. Final O1/O2 cases reused the initial case-generation seeds, preserving business parameters and world timelines, while Agent behavior was generated again under the final interface; 395 of 400 paired tool sequences differ between the excluded and final runs. The runner enforced the 37-round limit but contained no runtime token-budget check. The final main run consumed 377{,}563{,}216 tokens, and accumulated initial-run, pilot, and final-run use was approximately $5.67\times10^8$, 13.4\% above the $5.0\times10^8$ declared global budget (not enforced as a runtime hard cap). All 1{,}000 final episodes completed, with identical case sets and world timelines across arms, so the overrun did not truncate or asymmetrically extend either arm. A transport repair disabling inherited proxy settings entered after O1/O2 and before O3--O5. It left request content and model parameters unchanged, and both arms within each panel used the same transport path, but it prevents treating all five panels as one identically configured run. Panels are therefore analyzed separately and combined only through the specified multiple-testing correction. The final-panel counts remain those of the adopted configuration.

The retained \code{PRE_REGISTRATION.md} specifies case-paired exact McNemar testing and five-panel Holm correction by case class. It does not explicitly specify one-sided testing. The final report and analysis code use the Full-CC-greater alternative and also provide independently corrected two-sided results. The main text foregrounds the two-sided calculation. For reference, the separately corrected directional values are $p_{\mathrm{adj}}=0.00012$ for O1 challenge cases and $0.054$ for O2 lawful cases, compared with two-sided values of $0.00025$ and $0.107$, respectively. Directionality is treated as reported, rather than independently established as fixed before results were inspected.

The targeted 80-episode audit uses a separate 40-case sample and complete action records. Its driver-only time-box setting leaves the semantic files unchanged and does not truncate episodes. It diagnoses the confirmation behavior in that sample; it cannot reconstruct arguments omitted from the earlier records.

\FloatBarrier
\begingroup
\small
\setlength{\parskip}{0pt}
\setlength{\itemsep}{2pt}

\endgroup
\end{document}